\documentclass[11pt,a4paper]{article}
\pdfoutput=1
\usepackage{jcappub}
\usepackage{graphicx}
\usepackage{dcolumn}
\usepackage{bm}
\usepackage{color}
\usepackage{calrsfs}
\usepackage{hyperref}
\usepackage{multirow}
\usepackage{float}
\input epsf

\newcommand{\be}{\begin{equation}}
\newcommand{\e}{\end{equation}}
\newcommand{\bear}{\begin{eqnarray}}
\newcommand{\ear}{\end{eqnarray}}

\begin{document}

\title{Modeling the neutral hydrogen distribution in the post-reionization Universe: intensity mapping}

\author[a,b]{Francisco Villaescusa-Navarro,} \author[a,b]{Matteo Viel,} 
\author[c]{Kanan K. Datta,} \author[c]{T. Roy Choudhury} 

\affiliation[a]{INAF - Osservatorio Astronomico di Trieste, Via Tiepolo 11, 34143, Trieste, Italy}
\affiliation[b]{INFN sez. Trieste, Via Valerio 2, 34127 Trieste, Italy}
\affiliation[c]{National Centre for Radio Astrophysics, TIFR, Post Bag 3, Ganeshkhind, Pune 411007, India}

\emailAdd{villaescusa@oats.inaf.it}
\emailAdd{viel@oats.inaf.it}
\emailAdd{kanan@ncra.tifr.res.in}
\emailAdd{tirth@ncra.tifr.res.in}

\abstract{

We model the distribution of neutral hydrogen (HI) in the
post-reionization era and investigate its detectability in 21 cm
intensity mapping with future radio telescopes like the Square
Kilometer array (SKA).  We rely on high resolution hydrodynamical
N-body simulations that have a state-of-the-art treatment of the low
density photoionized gas in the inter-galactic medium (IGM). 
The HI is assigned \textit{a-posteriori} to the gas particles
following two different approaches: a \textit{halo-based method} in which HI is
assigned only to gas particles residing within dark matter halos;
a \textit{particle-based method} that assigns HI to all gas particles
using a prescription based on the physical properties of the
particles.

The HI statistical properties are then compared to the observational properties 
of Damped Lyman-$\alpha$ Absorbers (DLAs) and of lower column density systems
and reasonable good agreement is found for all the cases.  Among
the \textit{halo-based method}, we further consider two different
schemes that aim at reproducing the observed properties of DLAs by
distributing HI inside halos: one of
this results in a much higher bias for DLAs, in agreement with recent observations,
which boosts the 21 cm power spectrum by a factor $\sim 4$ with respect
to the other recipe. Furthermore, we quantify the contribution of HI in the diffuse IGM to
both $\Omega_{\rm HI}$ and the HI power spectrum finding to be
subdominant in both cases.

We compute the 21 cm power spectrum from the simulated HI distribution
and calculate the expected signal for both SKA1-mid and SKA1-low
configurations at $2.4 \leq z \leq 4$. We find that SKA will be able to detect the 21 cm
power spectrum, in the non-linear regime, up to $k\sim 1\,h$/Mpc for SKA1-mid
and $k\sim 5\,h$/Mpc for SKA1-low with 100 hours of observations. 

We also investigate the perspective of  imaging the HI distribution.  Our findings indicate 
that SKA1-low could detect the most massive HI
peaks with a signal to noise ratio (SNR) higher than 5 for an
observation time of about 1000 hours at $z=4$, for a synthesized beam width of $2'$. 
Detection at redshifts $z\geqslant2.4$ with SKA1-mid would instead require a much longer
observation time to achieve a comparable SNR level.}

\maketitle

\section{Introduction}

The cosmology standard model is able to reproduce reasonably well a variety of observables from large to small scales like the fluctuations in the cosmic microwave background (CMB), the spatial distribution of galaxies, the weak lensing statistical properties, the clustering of the cosmic-web filaments as seen in Lyman-$\alpha$ , the cluster number counts, etc.
Among the different tracers of the large scale structure, Hydrogen is the most abundant element with a contribution of $\sim75\%$ to the total baryonic mass. The Lyman-$\alpha$ forest, in the post-reionization era, is a powerful probe of the diffuse hydrogen atoms
that are  ionized due to the presence of a rather strong ultraviolet (UV) background radiation flux. However, it is expected that the UV background radiation would not be able to penetrate  dense hydrogen environments associated to galaxies. Thus, the neutral hydrogen within massive galaxies is believed to be self-shielded agains the external UV radiation. It is also thought that the neutral hydrogen within galaxies represents an intermediate phase between the diffuse inter-galactic medium and the high density molecular hydrogen, the fuel for the formation of stars. Therefore, any successful theory of galaxy formation/evolution should predict not only the galactic HI mass but also the interplay between the galactic properties and the HI content and distribution.

The amount of neutral hydrogen can be parametrized with the parameter $\Omega_{\rm HI}(z)=\bar{\rho}_{\rm HI}(z)/\rho_{\rm c}^0$, with $\bar{\rho}_{\rm HI}(z)$ being the average comoving density of neutral hydrogen (HI) at redshift $z$ and $\rho_{\rm c}^0$ the Universe critical density at $z=0$. Observations of the abundance and properties of the DLAs have been used to infer a value of $\Omega_{\rm HI}(z)\simeq10^{-3}$, almost independently of redshift \cite{Peroux_2003, Zwaan_2005, Rao_2006, Lah_2007, Martin_2010, Braun_2012, Noterdaeme_2012, Zafar_2013}. 

Neutral hydrogen can be detected through its redshifted 21-cm emission (see \cite{Furlanetto_2006, Morales_2009} for a recent reviews). Even though galaxies can not
be detected individually at high redshift, variations in the 21 cm signal
on large scales can be used to measure the power spectrum: a technique which is known as intensity mapping \cite{Bharadwaj_2001A, Bharadwaj_2001B,Chang_2008,Loeb_Wyithe_2008}. Moreover, the fluctuations in the 21 cm intensity emission are believed to be one of the most powerful probes of the epoch of reoinization (EoR) (see for instance \cite{Furlanetto_2006, Zaldarriaga_2004, Datta_2007, Morales_2009, Pritchard_2011} and references therein). In the post-reionization era, the much larger volume available for 21 cm observations, with respect to those probed by galaxy surveys, is expected to allow us to measure the matter power spectrum with an unprecedented precision \cite{Loeb_Wyithe_2008, Wyithe_2008, Camera_2013,Bull_2014}. 

The current radio-telescope technology only allows us to detect the 21 cm emission from galaxies only at relatively low redshift ($z\lesssim0.3$). On the other hand,
existing radio interferometers such as the Giant Meterwave Radio Telescope (GMRT)\footnote{http://gmrt.ncra.tifr.res.in/}, the Ooty Radio Telescope (ORT) \cite{Ooty}, the Low-Frequency Array (LOFAR)\footnote{http://www.lofar.org/}, the Murchison Wide-field Array (MWA)\footnote{http://www.mwatelescope.org/} or upcoming like ASKAP (The Australian Square Kilometer Array Pathfinder)\footnote{http://www.atnf.csiro.au/projects/askap/index.html}, MeerKAT (The South African Square Kilometer Array Pathfinder)\footnote{http://www.ska.ac.za/meerkat/} and SKA (The Square Kilometer Array)\footnote{https://www.skatelescope.org/} are designed to detect the fluctuations in the emission of 21 cm neutral hydrogen. We notice that the 21 cm signal has recently been detected at $z\sim1$ by cross-correlating intensity maps and the large scale structure (via catalogues of optically selected galaxies) \cite{Chang_2010,Masui_2013} and also forecasting (see for instance \cite{Ghosh_2010,Ghosh_2011}). Therefore, a deep understanding of the HI distribution is required from the theory/numerical side in order to extract the maximum possible information from the data. 

In this paper we aim at modeling the distribution of HI using high resolution hydrodynamical N-body simulations. The HI distribution has been previously studied in several works \cite{Bharadwaj_2004,Nagamine_2004, Kohler_2006, Pontzen_2008, Razoumov_2008, Tescari_2009, Altay_2010, Popping_2009, Bagla_2010, Nagamine_2010, McQuinn_2011, Fumagalli_2011,Duffy_2012, Bird_2013, Rahmati_2013, Dave_2013, Marinacci_2014, Bird_2014}. Here, we use two different approaches to simulate the distribution of HI. With the first one, the so-called \textit{halo-based method}, we only assign HI to the gas particles belonging to dark matter halos. This approach is based on Bagla et al. 2010 \cite{Bagla_2010} method. We will then use two different schemes: one it is just the Bagla et al. method whereas with the other we try to reproduce the observational measurements of both the DLAs column density distribution and their bias as obtained recently by the SDSS-III/BOSS collaboration \cite{Font_2012}. The two schemes are quite different and by exploring both of them in terms of 21 cm power spectrum we are conservatively exploring the expected signal.

Our second approach, consists of assigning HI to the N-body gas particles using a particle-by-particle basis and make use of the Dav\'e et al. 2013 \cite{Dave_2013} method, which accounts for the most important effects that set the hydrogen phase: photo-ionization equilibrium in low density regions and self-shielding and HI-H$_2$ conversion in high density zones. Ref.~\cite{Dave_2013} showed that this method, applied to their simulations, 
is able to reproduce many of the low-redshift HI observations \cite{Meyer_2004, Giovanelli_2005, Catinella_2010}. Moreover, this method allows us to quantify the contribution of HI residing outside halos to the overall Universe HI content and to the HI power spectrum. 

We also investigate the sensitivity of future radio telescopes such as the SKA to the 21 cm signal arising from the spatial distribution of neutral hydrogen. Furthermore, we study the prospect of direct imaging the largest peaks in the HI distribution.

The paper is organized as follows. In Sec. \ref{Simulations} we present the hydrodynamical N-body suite of simulations  used in this paper. The methods employed to model the distribution of neutral hydrogen are described in Sections \ref{HI_halos} and \ref{HI_particles}. In the first, we focus on the halo-based method, whereas in the latter we use the particle-based method. In Sec. \ref{HI_outside_halos_section} we quantify the contributions of HI outside dark matter halos to both the overall amount of HI in the Universe and the HI power spectrum. The 21 cm power spectrum from our simulated HI distribution is presented in Sec. \ref{21cm_section}. In that section we also discuss its detectability for future radio surveys as SKA. In Sec. \ref{imaging} we investigate the possibility, for the SKA, to detect bright, isolated, HI peaks. Finally, a summary and the main conclusions of this paper are given in Sec. \ref{Conclusions}.

\section{N-body simulations}
\label{Simulations}

We run a set of five high-resolution hydrodynamical N-body simulations
including both cold dark matter (CDM) and baryon (gas + star) particles. We
use the TreePM-SPH N-body code {\sc GADGET-III}, which is an improved and
faster version of the code {\sc GADGET-II} \cite{Springel_2005}. 

The starting redshift of the simulations is set to $z=99$. At this
redshift, the N-body initial conditions are generated by displacing
the particle positions from a cubic grid using the Zel'dovich
approximation. We make use of CAMB \cite{CAMB} to compute the CDM and
baryon transfer functions together with the matter power spectrum at
the starting redshift.

The simulations contain $512^3$ CDM and $512^3$ baryon particles
within a periodic box of linear comoving size 15, 30, 60 and 120
$h^{-1}$ Mpc. Thus, the resolution of the 15 $h^{-1}$ Mpc simulation
is 512 times higher than the one of the simulation with box size equal to
120 $h^{-1}$ Mpc. The reason of having different simulations with
different resolutions is because we will perform convergence tests
along the paper, investigating the stability of our results.

The gravitational softening of the particles (for both CDM and
baryons) is set to $1/30$ of the mean inter-particle linear spacing,
corresponding to $\sim1,2,4$ and 8 $h^{-1}$ com. kpc for the simulations
with boxes of 15, 30, 60 and 120 $h^{-1}$ com. Mpc, respectively. The values
of the cosmological parameters, in roughly good agreement with the latest
Planck results \cite{Planck_2013}, are the same for all the
simulations: $\Omega_{\rm m}=\Omega_{\rm cdm}+\Omega_{\rm b}=0.2742$,
$\Omega_{\rm b}=0.046$, $\Omega_\Lambda=0.7258$, $h=0.7$, $n_s=0.968$
and $\sigma_8=0.816$.

The code includes radiative cooling by hydrogen and helium and heating
by a uniform Ultra Violet (UV) background, while star formation is
implemented using the effective multiphase sub-grid model of Springel
\& Hernquist \cite{Springel-Hernquist_2003}.  The UV background and
the cooling routine have been modified in order to achieve a wanted
thermal history which corresponds to the reference model of
\cite{viel13} which has been shown to provide a good fit to the
statistical properties of the transmitted Lyman-$\alpha$ flux. In this
model, hydrogen reionization takes place at $z\sim 12$ and the
temperature-density relation for the low-density IGM
$T=T_0(z)(1+\delta)^{\gamma(z)-1}$ has $\gamma(z)=1.3$ and
$T_0(z=2.4,3,4)=(16500,15000,10000)$ K.

One of the simulations is run with feedback in the form of energy
driven galactic wind as implemented in Ref. \cite{Springel-Hernquist_2003}:
star forming particles receive 'a kick' of $\sim480$ km/s according to
a probabilistic criterion. Hydrodynamical forces are then turned off
for a fixed time interval or until the density of the wind particle
reaches a value of $0.1\rho_{\rm th}$, whichever happen earlier, with
$\rho_{\rm th}$ is the density above which star formation takes place.
For further details we refer the reader to Refs.
\cite{Springel-Hernquist_2003, Barai_2013}.
A summary of the simulation suite is presented in table
\ref{tab_sims}.

\begin{table}
\begin{center}
\resizebox{14cm}{!}{
\begin{tabular}{|c|c|c|c|c|c|}
\hline
Name & Box & $m_{\rm CDM}$ & $m_{\rm b}$ & wind model & $z_{\rm end}$ \\
 & ($h^{-1}\rm{Mpc}$) & ($h^{-1}M_\odot$) & ($h^{-1}M_\odot$) & & \\
\hline
\hline 
$\mathcal{B}120$ & 120 & $8.16\times10^{8}$ & $1.64\times10^{8}$ & no winds & 2.4\\ 
\hline
$\mathcal{B}60$W & 60 & $1.02\times10^{8}$ & $2.05\times10^{7}$ & constant velocity winds & 3.0\\ 
\hline
$\mathcal{B}60$ & 60 & $1.02\times10^{8}$ & $2.05\times10^{7}$ & no winds & 2.4 \\ 
\hline
$\mathcal{B}30$ & 30 & $1.28\times10^{7}$ & $2.56\times10^{6}$ & no winds & 2.4 \\ 
\hline
$\mathcal{B}15$ & 15 & $1.59\times10^{6}$ & $3.20\times10^{5}$ & no winds & 2.4\\ 
\hline
\end{tabular}
}
\end{center} 
\caption{Summary of the simulations. The value of the cosmological
  parameters is the same for all simulations: $\Omega_{\rm
    m}=\Omega_{\rm cdm}+\Omega_{\rm b}=0.2742$, $\Omega_{\rm
    b}=0.046$, $\Omega_\Lambda=0.7258$, $h=0.7$, $n_s=0.968$ and
  $\sigma_8=0.816$. Each simulation contains $512^3$ CDM and $512^3$
  baryon particles.}
\label{tab_sims}
\end{table}

The simulations without galactic winds have been run until $z=2.4$,
whereas the simulation with galactic winds has been run only until
$z=3$. We analyzed snapshots at redshifts $z=2.4,3,4,6$ (except for
the simulation with galactic winds where we only keep snapshots at
$z=3,4,6$), on top of which we have run the Friends-of-Friends (FoF)
\cite{FoF} algorithm with a value of the linking length parameter
equal to $b=0.2$. 

For each gas particle, we consider its mass, SPH smoothing length and
HI/H fraction. The neutral hydrogen fraction, together with the
abundance of other ionization states of hydrogen and helium, is
computed by the code assuming optically thin gas and photo-ionization
equilibrium with the external UV background using the formalism
described in \cite{Katz_1996}.  In the version used in the present
work, we stress that the code does not take into account neither
HI self-shielding effects nor the formation of molecular
hydrogen. However, the HI content outside halos and the IGM physical
state is realistic in the sense that it fits all the observational
constraints.

\section{HI modeling: halo-based}
\label{HI_halos}

In this paper we model the neutral hydrogen distribution using two
different techniques. In the first one, we assume that all the HI
resides in dark matter halos and we thus assign HI only to gas
particles belonging to dark matter halos. We refer to this method as
the \textit{halo-based} method and we describe it in detail in this
section. With the second technique we assign instead HI to all the gas
particles in the simulation, according to the physical properties of
the particles themselves. We label this second scheme the
\textit{particle-based} method and we shall depict it in section
\ref{HI_particles}.

The HI assignment in the halo-based model proceeds as follows: given
an N-body simulation snapshot we first identify all the dark matter
halos within it (FoF halos); then we compute, for each dark matter halo, the total
HI residing in it; finally we split the total HI mass among the gas
particles inside the halo.

In order to carry out the HI assignment using the halo-based method we
need two ingredients: 1) the HI mass within a halo as a function of
its total mass $M_{\rm HI}(M)$; 2) the scheme according to which the
halo HI mass is split among its gas particles. We neglect environment
effects and therefore, the HI mass that a halo host is only a function
of its mass.  The function $M_{\rm HI}(M)$ completely determines the
value of the parameter $\Omega_{\rm HI}$:
\begin{equation}
\Omega_{\rm HI}(z)=\frac{1}{\rho_c}\int_0^\infty n(M,z)M_{\rm HI}(M)dM\,
\label{Omega_HI_constrain}
\end{equation}
where $n(M,z)$ is the halo mass function and $\rho_{\rm c}$ is the
Universe critical density at $z=0$. We have explicitly checked that
the abundance of halos in our simulations is well reproduced by the
Sheth \& Tormen halo mass function \cite{Sheth-Tormen}, that we use
through the paper. Notice that, even if our simulations do include gas cooling, we assume that the effects of baryons on
the halo mass function are negligible (see however \cite{Weiguang_2014}) and will not impact on our results.

We now investigate how the spatial properties of the neutral hydrogen
depend on the function $M_{\rm HI}(M)$ and on the HI distribution
within the dark matter halos. For that purpose we consider two
different models which use a different $M_{\rm HI}(M)$ function. We
model the spatial distribution of HI within halos to reproduce the
observed distribution of column densities.

\subsection{Model 1}
\label{Model_1}

In this model we follow the work of Bagla et al. 2010
\cite{Bagla_2010}, where three different schemes for the function
$M_{\rm HI}(M)$ were presented and applied to N-body dark matter only simulations. 
In particular, we focus on their model number 3 which parametrizes the $M_{\rm HI}(M)$ function as:
  
\begin{equation}  
M_{\rm HI}(M) = \left\{ 
  \begin{array}{l l}	
  
    f_3\frac{M}{1+\left(\frac{M}{M_{\rm max}}\right)} & \quad \text{if $M_{\rm min}\leqslant M$}
    \\
    0 & \quad \text{otherwise}\\
  \end{array} \right.
\label{M_HI_Bagla3}
\end{equation}  
where $f_3$ is a free parameter describing the amplitude
of the function $M_{\rm HI}(M)$, $M_{\rm min}$ represents the minimum
mass of a dark matter halo able to contain neutral hydrogen and
$M_{\rm max}$ corresponds to the mass scale above which the fraction
of HI in a halo becomes suppressed.

We have chosen the Bagla et al. model number 3 to characterize the $M_{\rm HI}(M)$ 
function
since that function is physically better justified than the other two\footnote{We notice that 
authors in \cite{Bagla_2010} also demonstrated that, on large scales,
 the HI power spectrum obtained by using any of their three $M_{\rm HI}(M)$ functions is 
 almost the same and we have also verified their findings.} (see \cite{Bagla_2010} for 
 details). However, we have explicitly checked that our results do not significantly change 
if we use the Bagla et al. model number 2.
  Following \cite{Bagla_2010}, the value of $M_{\rm min}$ and
$M_{\rm max}$ can be obtained assuming that dark matter halos with
circular velocities below 30 km/s does not contain any HI and that
halos with circular velocities higher than 200 km/s will present a
suppressed HI mass fraction. The relationship between the halo
circular velocity ($v_c$) and its mass can be approximated by
\begin{equation}
M=10^{10}M_\odot\left(\frac{v_c}{60~{\rm km/s}}\right)^3\left(\frac{1+z}{4}\right)^{-3/2}~.
\end{equation}
The value of the free parameter $f_3$ is obtained by requiring that
 $\Omega_{\rm HI}(z)=10^{-3}$, as suggested by observations. 
Once the total HI mass within a given dark matter halo is
computed, we split it equally among all the gas particles belonging to
the halo. From now on we label this method as the \textit{halo-based model 1}.
We stress that unlike the original Bagla et al. implementation we rely on 
hydrodynamical simulations and thus the HI is assigned to gas particles.

The values of both $\Omega_{\rm HI}$ and the DLAs line density,
$dN/dX$, that we obtain by assigning HI to the gas particles using the
halo-based model 1 are shown in table \ref{Omega_HI_tab_model1}.
 For a particular simulation and redshift the value of
$\Omega_{\rm HI}$ is obtained by summing the HI content assigned
to all the gas particles, whereas the value of $dN/dX$ is computed
from the HI column density distribution (see below).

Notice that differently from \cite{Bagla_2010}, we obtain
the amplitude of the function $M_{\rm HI}(M)$, $f_3$, by using the
whole mass function range, not just the range that the simulations can
resolve. In this way we make sure that halos of the same mass contain
the same HI mass, independently of the resolution of the simulation. On the
other hand, low mass halos can not be resolved in the simulations with
the large box sizes. Thus, the value of $\Omega_{\rm HI}$ that we
obtain from the simulation $\mathcal{B}60$ is below the one from
observations, and has its origin on the fact that low mass dark matter
halos are not properly resolved in that simulation.
  
\begin{table}
\begin{center}
\resizebox{12cm}{!}{ 
\begin{tabular}{|c|c|c|c|c|}
\cline{1-5}
z & quantity & $\mathcal{B}60$ & $\mathcal{B}30$ & $\mathcal{B}15$ \\ \cline{1-5}

\multicolumn{1}{ |c| }{\multirow{2}{*}{2.4} } &
\multicolumn{1}{ c| }{$\Omega_{\rm HI}$} & $0.738\times10^{-3}$ & $0.937\times10^{-3}$ & $0.988\times10^{-3}$  \\ \cline{2-5}
\multicolumn{1}{ |c|  }{}   &
\multicolumn{1}{ c| }{$dN/dX$} & $0.026$ & $0.033$ & $0.037$ \\ \cline{1-5}

\multicolumn{1}{ |c| }{\multirow{2}{*}{3.0} } &
\multicolumn{1}{ c| }{$\Omega_{\rm HI}$} & $0.692\times10^{-3}$ & $0.960\times10^{-3}$ & $1.012\times10^{-3}$  \\ \cline{2-5}
\multicolumn{1}{ |c|  }{}   &
\multicolumn{1}{ c| }{$dN/dX$} & $0.028$ & $0.037$ & $0.043$ \\ \cline{1-5}

\multicolumn{1}{ |c| }{\multirow{2}{*}{4.0} } &
\multicolumn{1}{ c| }{$\Omega_{\rm HI}$} & $0.585\times10^{-3}$ & $0.953\times10^{-3}$ & $1.006\times10^{-3}$ \\ \cline{2-5}
\multicolumn{1}{ |c|  }{}   &
\multicolumn{1}{ c| }{$dN/dX$} & $0.027$ & $0.040$ & $0.047$ \\ \cline{1-5}
\end{tabular}
}
\end{center}
\caption{Values of the DLAs line density, $dN/dX$, and $\Omega_{\rm
    HI}$ obtained by assigning the HI to the gas particles according
  to the halo-based model 1.}
\label{Omega_HI_tab_model1}
\end{table}

\begin{figure}
\begin{center}
\includegraphics[width=1.0\textwidth]{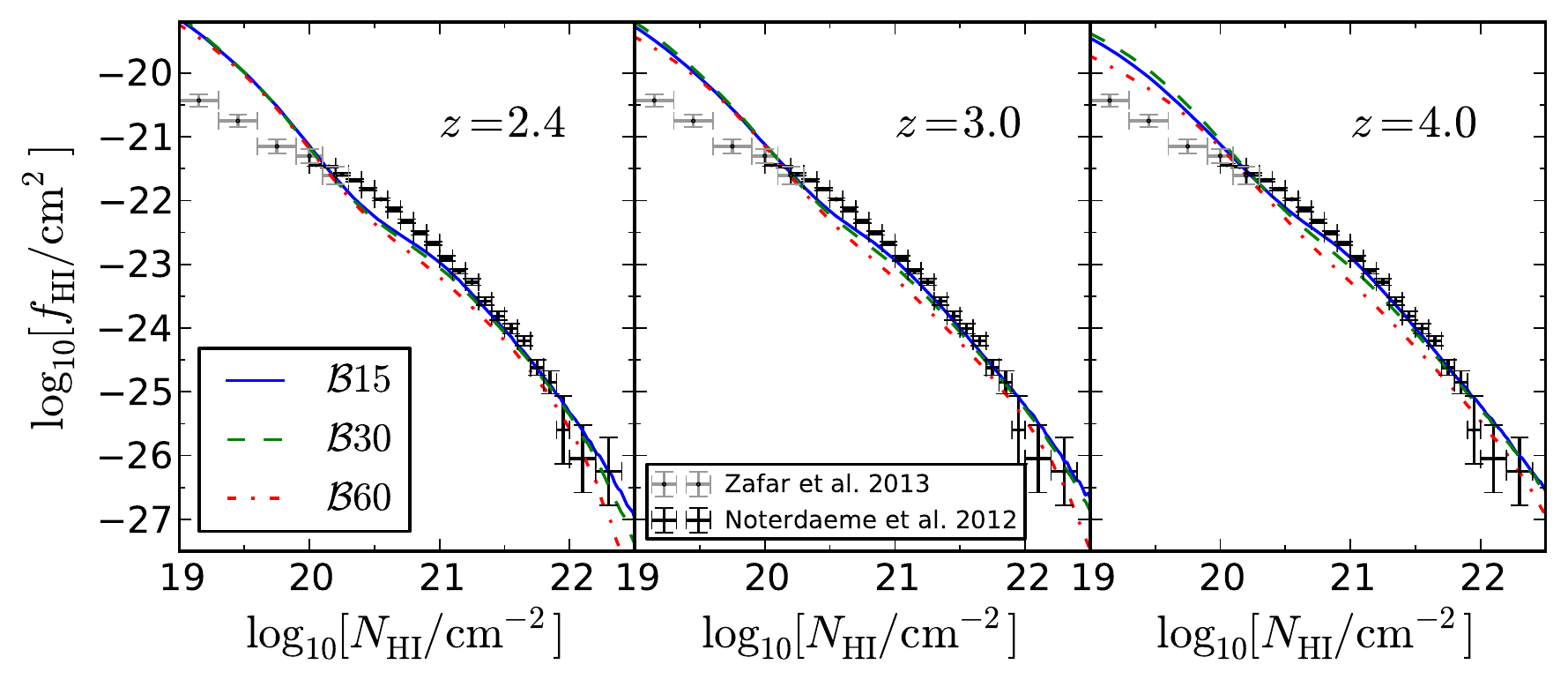}\\
\end{center}
\caption{Column density distribution obtained by using the halo-based
  model 1 at $z=2.4$ (left), $z=3.0$ (middle) and $z=4$ (right) for
  the simulations $\mathcal{B}15$ (solid blue), $\mathcal{B}30$
  (dashed green) and $\mathcal{B}60$ (dot-dashed red). The
  observational measurements of Noterdaeme et al. 2012
  \cite{Noterdaeme_2012} are shown in black whereas those by Zafar
  et al. 2013 \cite{Zafar_2013} are displayed in gray.}
\label{f_HI_Bagla}
\end{figure}

We compute (see appendix \ref{column_density_appendix} for details)
the column density distribution function, $f_{\rm HI}$, for the simulations
$\mathcal{B}15$, $\mathcal{B}30$ and $\mathcal{B}60$ at redshifts
$z=2.4$, $z=3$ and $z=4$ and show the results in
Fig.~\ref{f_HI_Bagla}, together with the measurements of
Noterdaeme\footnote{These results are obtained for DLAs in the
redshift range $z\in[2,3.5]$, with a mean redshift value of $\langle
z \rangle=2.5$.} et al. 2012 \cite{Noterdaeme_2012} and Zafar et al.
2013 \cite{Zafar_2013}. For simplicity
we assume that the HI column density distribution from observations
does not vary with redshift, i.e. we consider that the observed abundance 
of DLAs and Lyman Limit Systems (LLS) at the redshift studied in this paper 
($z=2.4,3,4$) are given by the measurements by Noterdaeme et al. (2012) and 
Zafar et al. (2013).

Overall we find a good agreement between our HI distribution and
the observational measurements for absorbers with column densities
larger than $10^{21}~{\rm cm}^{-2}$. However,
we find that this model under-predicts the abundance of absorbers
with column densities in the range $[10^{20}-10^{21}]~{\rm cm^{-2}}$, 
which gives rise to a lower value of the
DLA line density that the one observed (see table
\ref{Omega_HI_tab_model1}),
and over-predicts the abundance of Lyman Limit
Systems (LLS). The spatial distribution of the column densities, 
obtained by assigning HI to the gas particles of the simulation $\mathcal{B}15$
at $z=3$ using the three methods investigated in this paper, is shown in the appendix
\ref{HI_distribution_appendix}. The column density distributions of the simulations
$\mathcal{B}15$ and $\mathcal{B}30$ are in good agreement among
themselves even though the resolution of the former is eight times larger
than the one of the latter. The column density distribution of the
simulation $\mathcal{B}$60 is below those obtained from the other two
simulations, pointing out the lack of the DLAs residing in low mass
halos that the simulation is not able to resolve. We find that the
column density distribution exhibits a very weak redshift dependence.
We notice that we obtain almost the same HI column density distribution
if we assign HI to the halos gas particles according to the Bagla et al. model 2. 

\begin{figure}
\begin{center}
\includegraphics[width=1.0\textwidth]{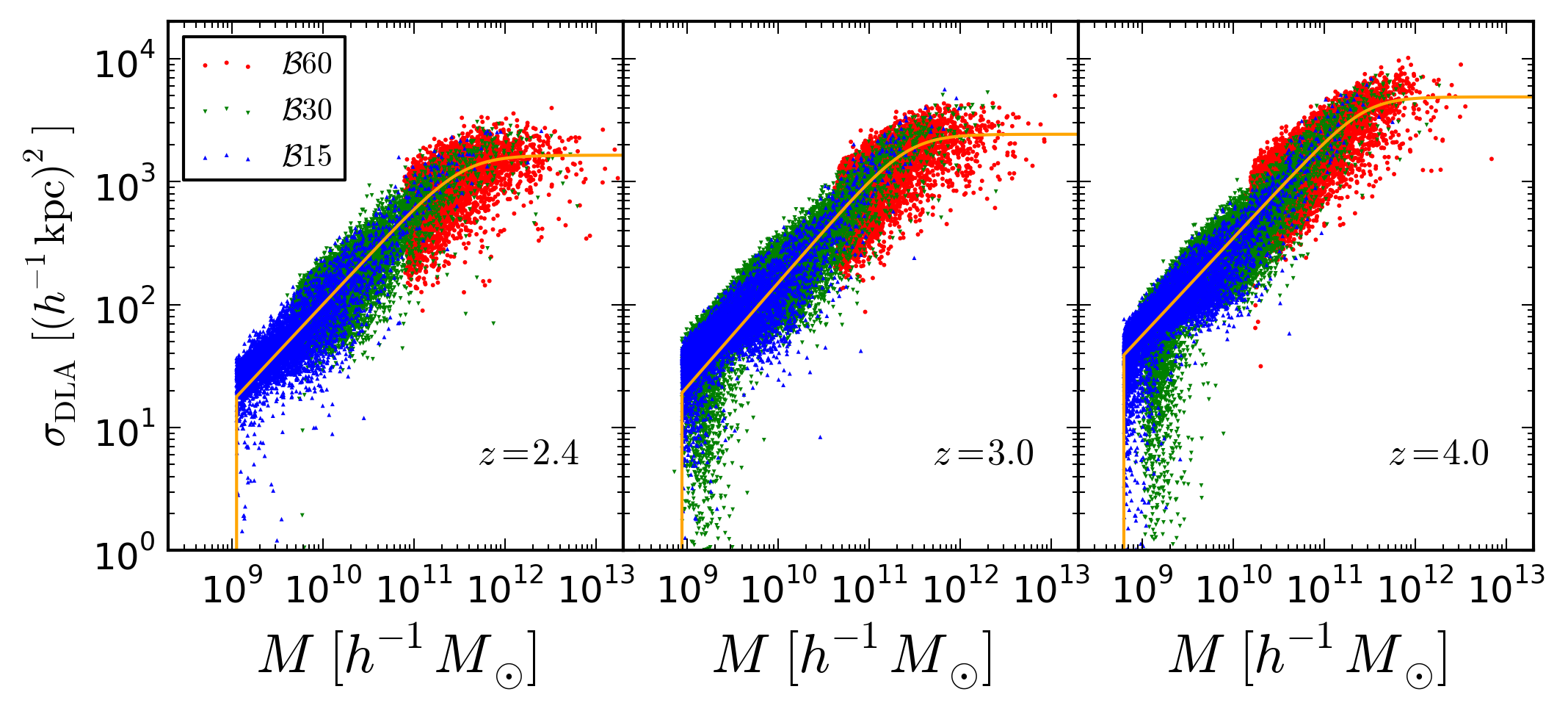}\\
\end{center}
\caption{DLA comoving cross-section for a subsample of halos at $z=2.4$ (left), $z=3$ (middle) and $z=4$ (right) from the simulations $\mathcal{B}60$ (red), $\mathcal{B}30$ (green) and $\mathcal{B}15$ (blue) when the neutral hydrogen is assigned to the gas particles using the halo-based model 1. A fit to the results is shown with a solid orange line.}
\label{cross_section_Bagla}
\end{figure}

For each dark matter halo we compute the DLA cross-section,
$\sigma_{\rm DLA}$, as follows: we select the gas particles belonging
to the halo and we throw random lines of sight within the halo virial
radius. For each line of sight we then compute its column density and we
estimate the DLA cross section as \cite{Pontzen_2008}:

\begin{equation}
\sigma_{\rm DLA}=\pi R_{\rm halo}^2\left(\frac{n_{\rm DLA}}{n_{\rm total}}\right)~,
\end{equation}
where $R_{\rm halo}$ is the halo virial radius, $n_{\rm DLA}$ is the
number of lines of sights with column densities higher than
$10^{20.3}~{\rm cm}^{-2}$ and $n_{\rm total}$ is the total number of
lines of sight used. We note that since we are using FoF halos the
quantity $R_{\rm halo}$ is not well defined. We use the value of the
radius, centered on the position of the particle with the minimum
energy, within which the mean density is 200 times the mean density of
the Universe as definition for $R_{\rm halo}$. We have explicitly
checked that our results do not change if we use a different
definition for the halo radius. We use 5000 random LOSs for each dark
matter halo; we have checked that our results do not significantly
change if we increase the former number. In
Fig. \ref{cross_section_Bagla} we show the comoving cross-section of
the DLAs for the simulations $\mathcal{B}60$, $\mathcal{B}30$ and
$\mathcal{B}15$ at redshifts 2.4, 3 and 4. We find that the DLA
cross-section increases with the halo mass, although for halos more
massive than $\sim 5\times10^{11}~h^{-1}{\rm M}_\odot$ the cross
section remains constant, almost independent of redshift. This is a
consequence of the $M_{\rm HI}(M)$ function used, which assigns, for
halos with masses above $M_{\rm max}$, a decreasing $M_{\rm HI}/M$
fraction. The mean DLA cross-section is well reproduced
by a function of the form:
\begin{equation}  
\sigma_{\rm DLA}(M) = \left\{ 
  \begin{array}{l l}	
  
    \sigma_0\left[\frac{M}{M_0}\right]^{\alpha}\left[1+\left(\frac{M}{M_0}\right)^{\beta}\right]^{-\alpha/\beta} & \quad \text{if $M_{\rm min}\leqslant M$}\\
    \\
    0 & \quad \text{otherwise}~.\\
  \end{array} \right.
  \label{cross_section_Bagla_fit}
\end{equation}

\begin{table}
\begin{center}
\resizebox{10cm}{!}{
\begin{tabular}{|c|c|c|c|c|c|}
\hline
$z$ & $\sigma_0$ & $M_0$ & $\alpha$ & $\beta$ & $M_{\rm min}$ \\ 
 & $(h^{-1}{\rm kpc})^2$ & $(h^{-1}M_\odot)$ & & & $(h^{-1}M_\odot)$ \\
\hline
$2.4$ & $1641$ & $3.41\times10^{11}$  & $0.79$ & 1.81 & $1.12\times10^9$\\
\hline
$3.0$ & $2429$ & $2.67\times10^{11}$  & $0.85$ & 1.89 & $8.75\times10^8$\\
\hline
$4.0$ & $4886$ & $2.81\times10^{11}$  & $0.79$ & 2.04 & $6.26\times10^8$ \\
\hline
\end{tabular}
}
\end{center}
\caption{Best fit values of the fitting formula \ref{cross_section_Bagla_fit} used to reproduce the average comoving cross-section obtained by assigning HI to the gas particles belonging to dark matter halos using the halo-based model 1.}
\label{cross-section_Bagla_table}
\end{table}

In table \ref{cross-section_Bagla_table} we present the best fit
values of the parameters of the above fitting function as a function of
redshift. Those fitting are also shown with a solid orange line in
Fig. \ref{cross_section_Bagla}. Next, we use mean DLAs cross-section to
estimate the DLAs bias as:
\begin{equation}
b_{\rm DLA}(z)=\frac{\int_0^\infty b(M,z)n(M,z)\sigma_{\rm DLA}(M,z)dM}{\int_0^\infty n(M,z)\sigma_{\rm DLA}(M,z)}~\, ,
\label{biaseq}
\end{equation}
with $n(M,z)$ being the halo mass function and $b(M,z)$ the halo
bias. For the former quantity we use the Sheth \& Tormen halo mass
function whereas for the latter we employ the Sheth, Mo \& Tormen halo
bias formula \cite{SMT}. We find the following values for the DLAs
bias: $b_{\rm DLA}=1.47$ at $z=2.4$, $b_{\rm DLA}=1.74$ at $z=3$ and
$b_{\rm DLA}=2.10$ at $z=4$. At $z=2.4$ our results are in strong
tension with recent observational measurements obtained with SDSS-III/BOSS
survey, which find a value of $b_{\rm DLA}=(2.17\pm0.20)\beta_F^{0.22}$, with
$\beta_F\sim 1-1.5$, at $\langle z \rangle=2.3$ \cite{Font_2012}. We notice that the values of the DLAs bias barely change by using the Bagla et al. model 2 to characterize the function $M_{\rm HI}$.

We try to simulate different HI density profiles within dark matter halos by distributing the neutral hydrogen according to the physical properties of the gas particles (see Eq. \ref{M_HI_Paco_sim}) to investigate whether we can fit at the same time the DLAs column density distribution and their bias. Our results indicate that this model is unable to fit simultaneously both quantities. The reason arises from the $M_{\rm HI}(M)$ used in this model, which assigns a constant HI mass to the most massive halos. In order to reproduce the DLAs bias we would need a HI density profile which only allocate DLAs to the most massive halos. On the other hand we find that such density profile can not reproduce the observations of the DLAs column density distribution function.

\begin{figure}
\begin{center}
\includegraphics[width=1.0\textwidth]{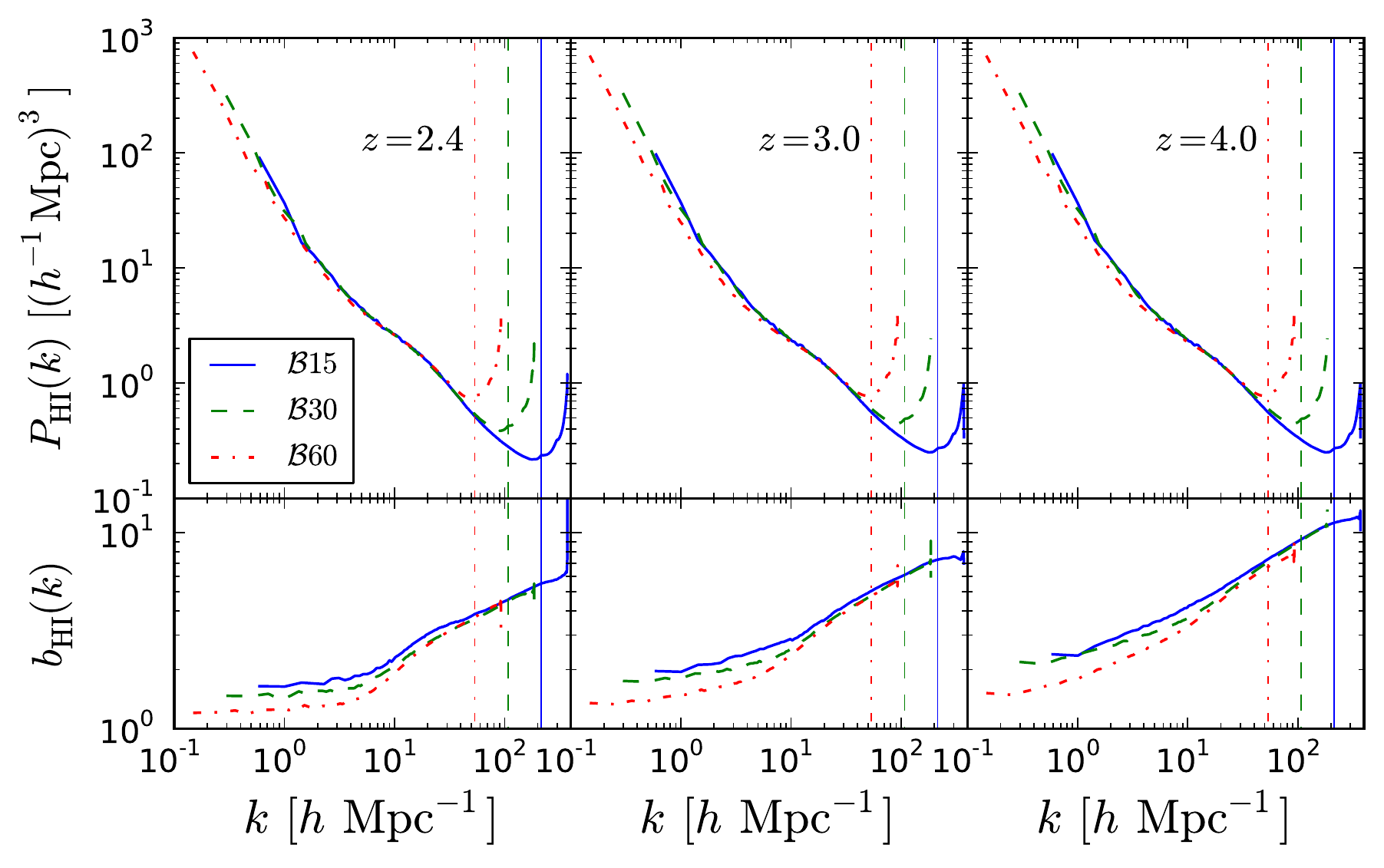}\\
\end{center}
\caption{HI power spectrum obtained by assigning the HI to the gas
  particles belonging to dark matter halos using the halo-based model
  1. The results are displayed at $z=2.4$ (left), $z=3$ (middle) and
  $z=4$ (right) for the simulations $\mathcal{B}60$ (dot-dashed red),
  $\mathcal{B}30$ (dashed green) and $\mathcal{B}15$ (solid blue). The
  value of the Nyquist frequency is displayed for each simulation with
  a vertical line. We show the bias between the distributions of HI and 
  matter, $b_{\rm HI}^2(k)=P_{\rm HI}(k)/P_{\rm m}(k)$, in the bottom panels.}
\label{Pk_HI_Bagla}
\end{figure}

In Fig. \ref{Pk_HI_Bagla} we show the results of computing the HI
power spectrum at redshifts $z=2.4$, $z=3$ and $z=4$ for the different
simulations. The values of the HI over-density are computed on 
a regular cubic grid with $1024^3$ points
using the Cloud-in-Cell (CIC) interpolation technique. The mean HI
mass per grid cell is computed using $\Omega_{\rm HI}=10^{-3}$
instead of the value of $\Omega_{\rm HI}$ from the simulation itself.
The reason for doing this arises because some simulations (for instance 
$\mathcal{B}60$) do not have resolution enough to resolve the smallest
halos that host HI and therefore, if we use the value of $\Omega_{\rm HI}$
from those simulations, the HI power spectrum from will be artificially above
the real one. The values of the neutral hydrogen mass over-densities are then
Fourier transformed using the Fast Fourier Transform algorithm and the
HI power spectrum is estimated through:
\begin{equation}
P_{\rm HI}(k)=\frac{1}{N_{\rm modes}}\sum_{\vec{k}\in k}\delta_{\rm HI}(\vec{k})\delta^*_{\rm HI}(\vec{k})
\end{equation}
with $N_{\rm modes}$ being the number of modes lying within the
spherical shell in $k$ considered. For concreteness, we compute the HI
power spectrum within spherical shells of width $\delta k=2\pi/L$,
with $L$ being the simulation box size.

We have explicitly checked that our results are the same on
intermediate and large scales if we compute the HI power spectrum
taking into account the SPH kernel of the gas particles (see appendix
\ref{power_spectrum_method}). Whereas there are some differences on
small scales, a proper computation requires a mode amplitude
correction to take into account the mass assignment scheme (MAS),
which is beyond the scope of this paper. We therefore use the CIC
technique, correcting the modes amplitude to account for the
MAS, since the calculation is much faster.

We find that the HI power spectrum from the simulations
$\mathcal{B}15$, $\mathcal{B}30$ and $\mathcal{B}60$ are in very good
agreement among themselves at all redshifts. On large scales, the
amplitude of the HI power spectrum is slightly lower in the simulation
$\mathcal{B}60$ than in the other two. This is because this simulation can not resolve
the smallest dark matter halos that host HI, and therefore, the
contribution of those halos to the HI power spectrum is not accounted
for. Even though the value of $\Omega_{\rm HI}$ is substantially
below $10^{-3}$ for the simulation $\mathcal{B}60$, the HI power
spectrum is almost converged, pointing out that the amplitude of the
HI power spectrum is set by the most massive, and biased, dark matter
halos.

In the bottom panels of Fig. \ref{Pk_HI_Bagla} we display the bias
between the distribution of neutral hydrogen and the one of the underlying matter:
$b^2_{\rm HI}(k)=P_{\rm HI}(k)/P_{\rm m}(k)$. We note that even though the HI
power spectra from the different simulations are in good agreement with each other,
the HI bias do not show such good convergence. This is due to the matter power
spectrum, whose amplitude decreases slightly with resolution due to cosmic variance
and also because small box size simulations do not account for the large scale modes 
that influence the structure growth. We expect this problem not to be present in large box-size
simulations, having resolution high enough to resolve the smallest halos capable of hosting HI.
As expected, we find that the HI bias increases with redshift. Our estimates for $b_{\rm HI}(k)$ agree quite well with previous works which use a prescription similar to ours
\cite{Bagla_2010,Guha_2012}.

\subsection{Model 2}
\label{Model_2}

In this model we follow the work of \cite{Barnes_2010, Barnes_2014}
and use a different $M_{\rm HI}(M)$ function, which assigns a constant
$M_{\rm HI}/M$ fraction for very massive halos:
\begin{equation}
M_{\rm HI}(M)=\alpha f_{\rm H,c}\exp{\left[ -\left(\frac{v_c^0}{v_c}\right)^\beta\right]}M~,
\label{M_HI_Paco}
\end{equation}
where $f_{\rm H,c}=0.76\Omega_{\rm b}/\Omega_{\rm m}$ is the cosmic
hydrogen mass fraction, $v_c$ is the halo circular velocity and
$\alpha$, $\beta$ and $v_c^0$ are the model free parameters. Our aim
is to reproduce the most important properties of the
DLAs (their bias, line density and column density distribution). We
first present a simple analytic model that is able to reproduce the
DLAs properties and then we use this as a framework to assign HI to the
gas particles belonging to the dark matter halos.

Our simple analytic model, based on the work of \cite{Barnes_2014},
consists of two ingredients: 1) the HI mass within a dark matter halo
of mass $M$; 2) the density profile of the neutral hydrogen within
it. For the former we use Eq. \ref{M_HI_Paco} whereas we
phenomenologically model the HI density profile within any dark matter
halo as:
\begin{equation}
\rho_{\rm HI}(r)=\frac{\rho_0}{\left( \frac{r}{r_s}+0.06\right)^2\left(\frac{r}{r_s}+1\right)^3}~\, ,
\label{HI_profile_Paco}
\end{equation}
in order to match observations.
We assume that the HI profile extends out to the halo virial radius,
$R$ and we take the value of the parameter $r_s$ to be $r_s=Rc$, with $c$
being the halo concentration following the law:
\cite{Bullock_2001,Maccio_2007}
\begin{equation}
c(M,z)=c_0\left(\frac{M}{10^{11}M_\odot}\right)^{-0.109}\left(\frac{4}{1+z}\right)~.
\end{equation}
The value of the parameter $\rho_0$ is found by requiring that the HI
mass obtained by integrating the density profile \ref{HI_profile_Paco}
is equal to this obtained from Eq. \ref{M_HI_Paco}. Given the HI
density profile and the function $M_{\rm HI}(M)$, the DLAs bias can be
computed using Eq.~\ref{biaseq}.
The DLA cross-section, $\sigma_{\rm DLA}=\pi s^2$, of
a given halo is calculated by finding the value $s$, for which
\begin{equation}
N_{\rm HI}=\frac{2}{m_H}\int_0^{\sqrt{R^2-s^2}}\rho_{\rm HI}(r=\sqrt{s^2+l^2})dl=10^{20.3}~{\rm cm}^{-2}~,
\label{N_HI_analytic}
\end{equation}
with $m_H$ being the mass of the Hydrogen atom. Similarly, the DLAs line density and the column density distribution can be obtained by computing:
\begin{eqnarray}
\frac{dN}{dX}&=&\frac{c}{H_0}\int_0^\infty n(M,z)\sigma_{\rm DLA}(M,z)dM~,\\
f_{\rm HI}(N_{\rm HI},X)&=&\frac{c}{H_0}\int_0^\infty n(M,z)\left|\frac{d\sigma(N_{\rm HI}|M,X)}{dN_{\rm HI}}\right|dM~,
\end{eqnarray}
with 
\begin{equation}
\sigma_{\rm DLA}(M,X)=\int_{10^{20.3}}^\infty \left|\frac{d\sigma(N_{\rm HI}|M,X)}{dN_{\rm HI}}\right| dN_{\rm HI}~.
\label{eqdlabias}
\end{equation}
$\sigma(N_{\rm HI})=\pi\tilde{s}^2$, with $\tilde{s}$ obtained by
using Eq. \ref{N_HI_analytic} but taking the actual value for $N_{\rm HI}$,
rather than $10^{20.3}$\, cm$^{-2}$. At $z=2.4$ we tune the values of the model free
parameters to reproduce both the DLAs bias \cite{Font_2012} and
the HI column density distribution \cite{Noterdaeme_2012}. At $z=3$
and $z=4$ we calibrate the value of the parameters to reproduce the
column density distribution, keeping the same value for the parameters
$v_c^0$ and $\beta$. At $z=3$ and $z=4$ the DLAs bias arises thus as a
prediction of the model. 

\begin{figure}
\begin{center}
\includegraphics[width=0.5\textwidth]{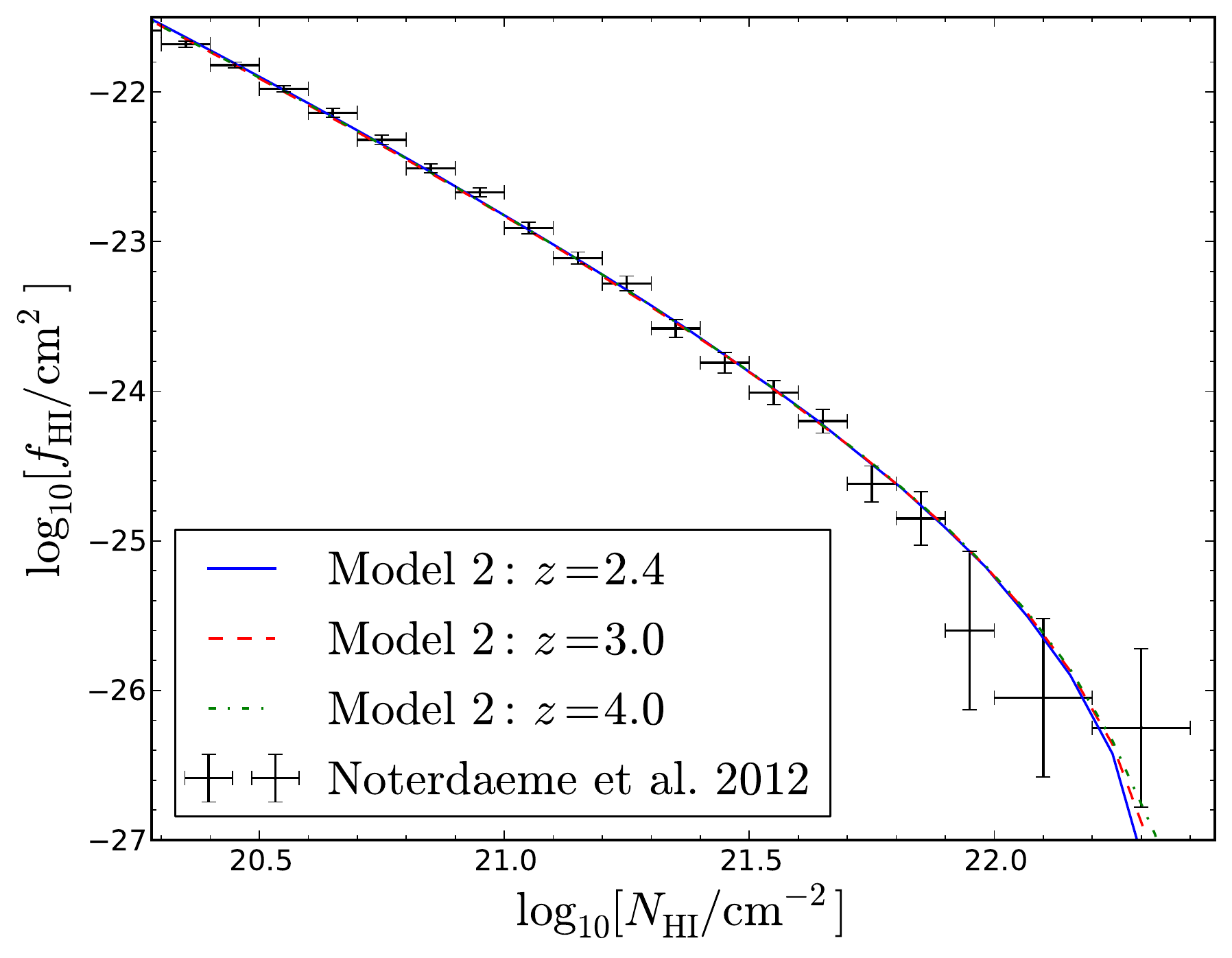}\\
\end{center}
\caption{Column density distribution from the analytic model 2 and its comparison with the observational measurements of Noterdaeme et al. 2012 \cite{Noterdaeme_2012}.}
\label{column_density_analytic_Paco}
\end{figure}

We have calibrated the free parameters of this simple analytic model using the
cosmological model of our N-body simulation. We have explicitly
checked that the values of the model parameters exhibit a weak
dependence with the cosmological parameters\footnote{Note that in
  \cite{Font_2012} the value of the cosmological parameters are
  slightly different to ours: $\Omega_{\rm m}=0.281$, $\Omega_{\rm
    b}=0.049$, $\Omega_\Lambda=0.719$, $\sigma_8=0.8$, $n_s=0.963$ and
  $h=0.71$.}. In table \ref{Paco_parameters} we show the value of the
model parameters together with the values of the DLAs line density,
$\Omega_{\rm HI}$ and the DLAs bias as a function of redshift. 
The value of $\Omega_{\rm HI}$ has been computed using Eq. 
\ref{Omega_HI_constrain}. We note that this model predicts a slightly
larger value for $\Omega_{\rm HI}$ than this obtained by observations.
We will see later that this happens because this model over-predicts the
abundance of LLS. In
Fig. \ref{column_density_analytic_Paco} we show the DLAs column
density distribution and its comparison with the observational
measurements. We remind the reader that in this paper 
we assume no redshift dependence in the observations of the
HI column density distribution. 

\begin{table}
\begin{center}
\resizebox{10cm}{!}{
\begin{tabular}{|c|c|c|c|c|c|c|c|}
\hline
z & $\alpha$ & $\beta$ & $v_c^0 (km/s)$ & $c_0$ & $dN/dX$ & $\Omega_{\rm HI}$ & $b_{\rm DLA}$\\ 
\hline
2.4 & 0.18 & 3 & 37 & 3.4 & 0.07 & $1.18\times10^{-3}$ & 2.15\\
\hline
3.0 & 0.222 & 3 & 37 & 3.2 & 0.07 & $1.12\times10^{-3}$ & 2.37\\ 
\hline
4.0 & 0.34 & 3 & 37 & 2.7 & 0.07 & $1.12\times10^{-3}$ & 2.76\\ 
\hline
\end{tabular}
}
\end{center}
\caption{Value of the parameters $\alpha$, $\beta$, $v_c^0$ and $c_0$ that best reproduce the DLAs observational properties for the analytic model 2. The values of the DLAs line density and bias, together with the value of $\Omega_{\rm HI}$ are also shown in the table.}
\label{Paco_parameters}
\end{table}

\begin{table}
\begin{center}
\resizebox{14cm}{!}{
\begin{tabular}{|c|c|c|c|c|c|}
\cline{1-6}
z & quantity & $\mathcal{B}120$ & $\mathcal{B}60$ & $\mathcal{B}30$ & $\mathcal{B}15$ \\ \cline{1-6}

\multicolumn{1}{ |c| }{\multirow{2}{*}{2.4} } &
\multicolumn{1}{ c| }{$\Omega_{\rm HI}$} & $0.886\times10^{-3}$ & $1.183\times10^{-3}$ & $1.208\times10^{-3}$ & $1.157\times10^{-3}$  \\ \cline{2-6}
\multicolumn{1}{ |c|  }{}   &
\multicolumn{1}{ c| }{$dN/dX$} & 0.057 & $0.067$ & $0.064$ & $0.064$ \\ \cline{1-6}

\multicolumn{1}{ |c| }{\multirow{2}{*}{3.0} } &
\multicolumn{1}{ c| }{$\Omega_{\rm HI}$} & $0.711\times10^{-3}$ & $1.067\times10^{-3}$ & $1.136\times10^{-3}$ & $1.012\times10^{-3}$  \\ \cline{2-6}
\multicolumn{1}{ |c|  }{}   &
\multicolumn{1}{ c| }{$dN/dX$} & $0.047$ & $0.066$ & $0.068$ & $0.043$ \\ \cline{1-6}

\multicolumn{1}{ |c| }{\multirow{2}{*}{4.0} } &
\multicolumn{1}{ c| }{$\Omega_{\rm HI}$} & $0.524\times10^{-3}$ & $0.991\times10^{-3}$ & $1.151\times10^{-3}$ & $1.115\times10^{-3}$  \\ \cline{2-6}
\multicolumn{1}{ |c|  }{}   &
\multicolumn{1}{ c| }{$dN/dX$} & $0.036$ & $0.066$ & $0.074$ & $0.073$ \\ \cline{1-6}

\end{tabular}
}
\end{center}
\caption{Values of the DLA line density, $dN/dX$ and $\Omega_{\rm HI}$ obtained by assigning the HI to the gas particles according to the halo-based model 2.}
\label{Omega_HI_tab_model2}
\end{table}

Motivated by the fact that the above simple model is able to reproduce
the most important properties of the DLAs, we now proceed to populate
the dark matter halos with HI. We compute the neutral hydrogen mass
that a dark matter halo of mass $M$ hosts by using Eq. \ref{M_HI_Paco},
with the calibrated parameter values at the corresponding redshift.
We then split the total HI mass
within a given dark matter halo among its gas particles. If
we distribute equally the overall HI mass among all the gas particles
we are not able to reproduce the observed column density
distribution: we have thus developed a simple recipe to distribute the HI
among the gas particles that is able to reproduce the HI density
profile of Eq. \ref{HI_profile_Paco}. The HI distribution proceeds as
follows: given the total HI that a given dark matter halo host,
$M_{\rm HI}$, we identify the gas particles belonging to it and assign a
HI mass to them equal to:
\begin{equation}
M_{\rm HI}^{p}=\frac{\left(\frac{{\rm HI}}{{\rm H}}\right)_p^{0.17}\left(r_{{\rm SPH,}p}\right)^{0.35}}{\sum_i \left(\frac{{\rm HI}}{{\rm H}}\right)_i^{0.17}\left(r_{{\rm SPH,}i}\right)^{0.35}}M_{\rm HI}~,
\label{M_HI_Paco_sim}
\end{equation}
where HI/H is the neutral hydrogen fraction that {\sc GADGET} assigns to each
gas particle, $r_{\rm SPH}$ is the gas particle SPH radius and $M_{\rm HI}$ is the
total HI mass within the halo. We denominate this method of assigning HI to the gas
particles belonging to dark matter halos as the \textit{halo-based model 2}.

\begin{figure}
\begin{center}
\includegraphics[width=1.0\textwidth]{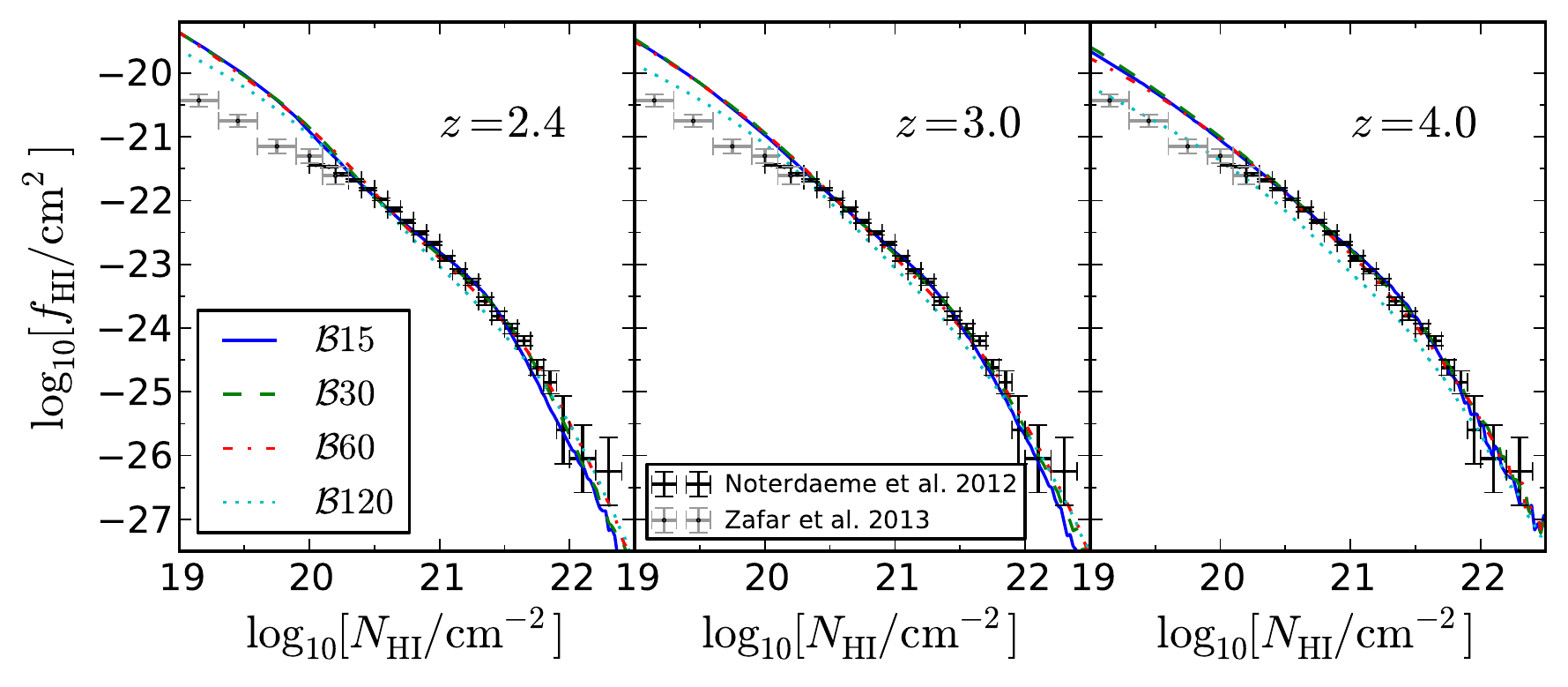}\\
\end{center}
\caption{Column density distribution obtained by assigning HI to the
  gas particles using the halo-based model 2 at $z=2.4$ (top), $z=3.0$
  (middle) and $z=4$ (bottom) for the simulations $\mathcal{B}15$
  (solid blue), $\mathcal{B}30$ (dashed green), $\mathcal{B}60$
  (dot-dashed red) and $\mathcal{B}120$ (dotted cyan). The
  observational measurements of Noterdaeme et al. 2012
  \cite{Noterdaeme_2012} are shown in black.}
\label{column_density_Paco}
\end{figure}

In table \ref{Omega_HI_tab_model2} we show the values of the DLAs line density
and $\Omega_{\rm HI}$, computed by summing the HI of all the gas particles,
obtained by assigning HI to the gas particles using the halo-based model 2
described above. The column density distribution and its comparison
with the observational measurements are displayed in
Fig. \ref{column_density_Paco}. We find an excellent agreement between
observations and our mock HI distribution in the DLA regime. 
However, this model, as the halo-based model 1, fails to reproduce the abundance of 
Lyman Limit systems, predicting a number significantly above the one
observed, which in turn produces an enhancement in the value of $\Omega_{\rm HI}$.
Moreover, we note that the results from the simulation $\mathcal{B}120$ are slightly
below the observational measurements: this is because this simulation
does not have resolution enough to resolve the smallest halos that
host DLAs. 

\begin{figure}
\begin{center}
\includegraphics[width=1.0\textwidth]{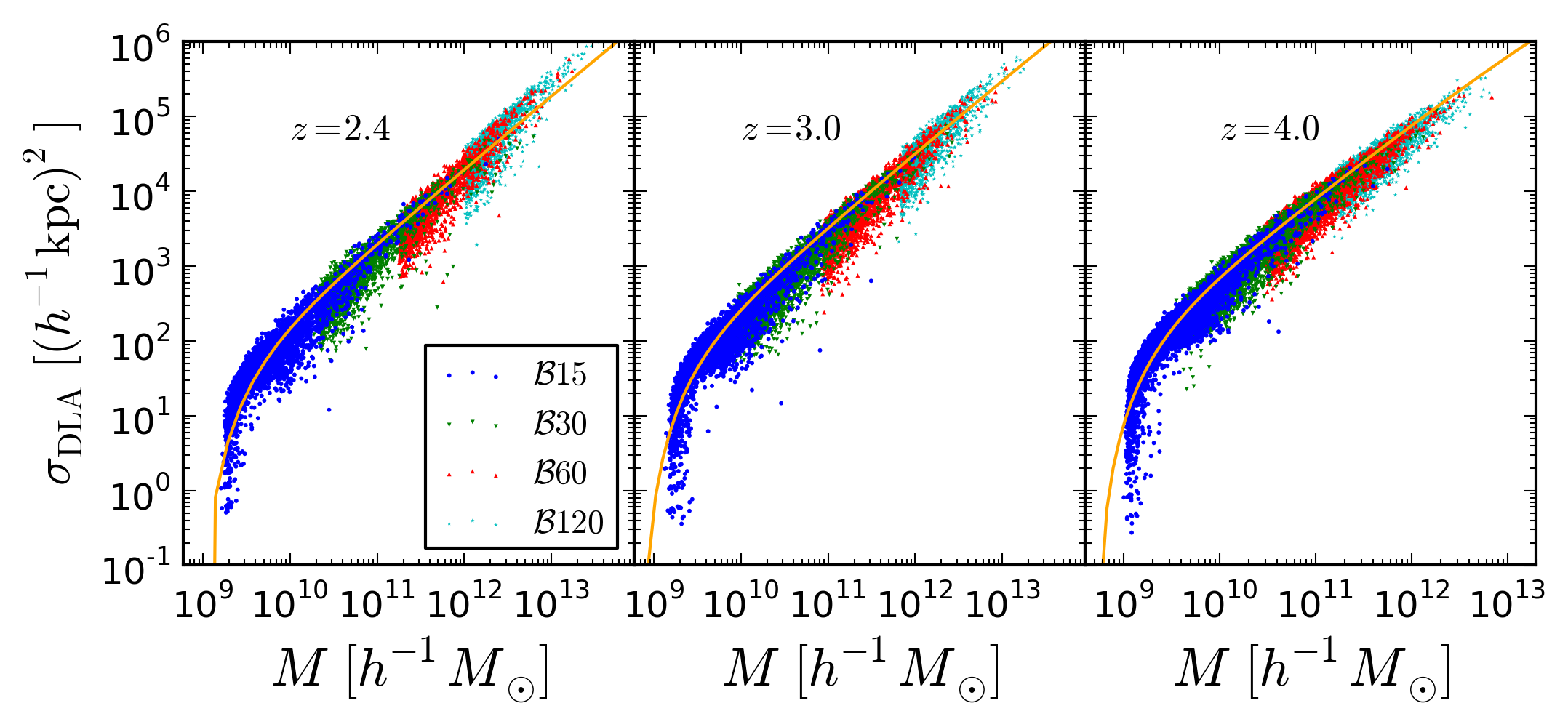}\\
\end{center}
\caption{DLAs comoving cross section for a subsample of halos at
  $z=2.4$ (left), $z=3$ (middle) and $z=4$ (right) from the
  simulations $\mathcal{B}120$ (cyan), $\mathcal{B}60$ (red),
  $\mathcal{B}30$ (green) and $\mathcal{B}15$ (blue) when the neutral
  hydrogen is assigned to the gas particles using the halo-based model
  2. The orange line represents the cross-section of the analytic
  model (see text for details).}
\label{cross_section_Paco}
\end{figure}

In order to verify that the HI density profile that we simulate
matches the one of the above analytic model we plot in
Fig. \ref{cross_section_Paco} the comoving cross-section of a subset
of dark matter halos from the different simulations. With a solid
orange line we show the cross-section obtained by using the above
analytic model. We find a good agreement between the analytic model
and cross-section of the halos populated with HI using the halo-based
model 2 at all redshifts. In other words, this model distributes the
HI in such a way that the spatial distribution of neutral hydrogen 
reproduces the measurements of both the DLAs 
column density distribution and bias (at $z=2.4$). We emphasize that the
orange line in Fig. \ref{cross_section_Paco} is not a fit to the DLAs
cross-section from the N-body but the prediction from the analytic model.

\begin{figure}
\begin{center}
\includegraphics[width=1.0\textwidth]{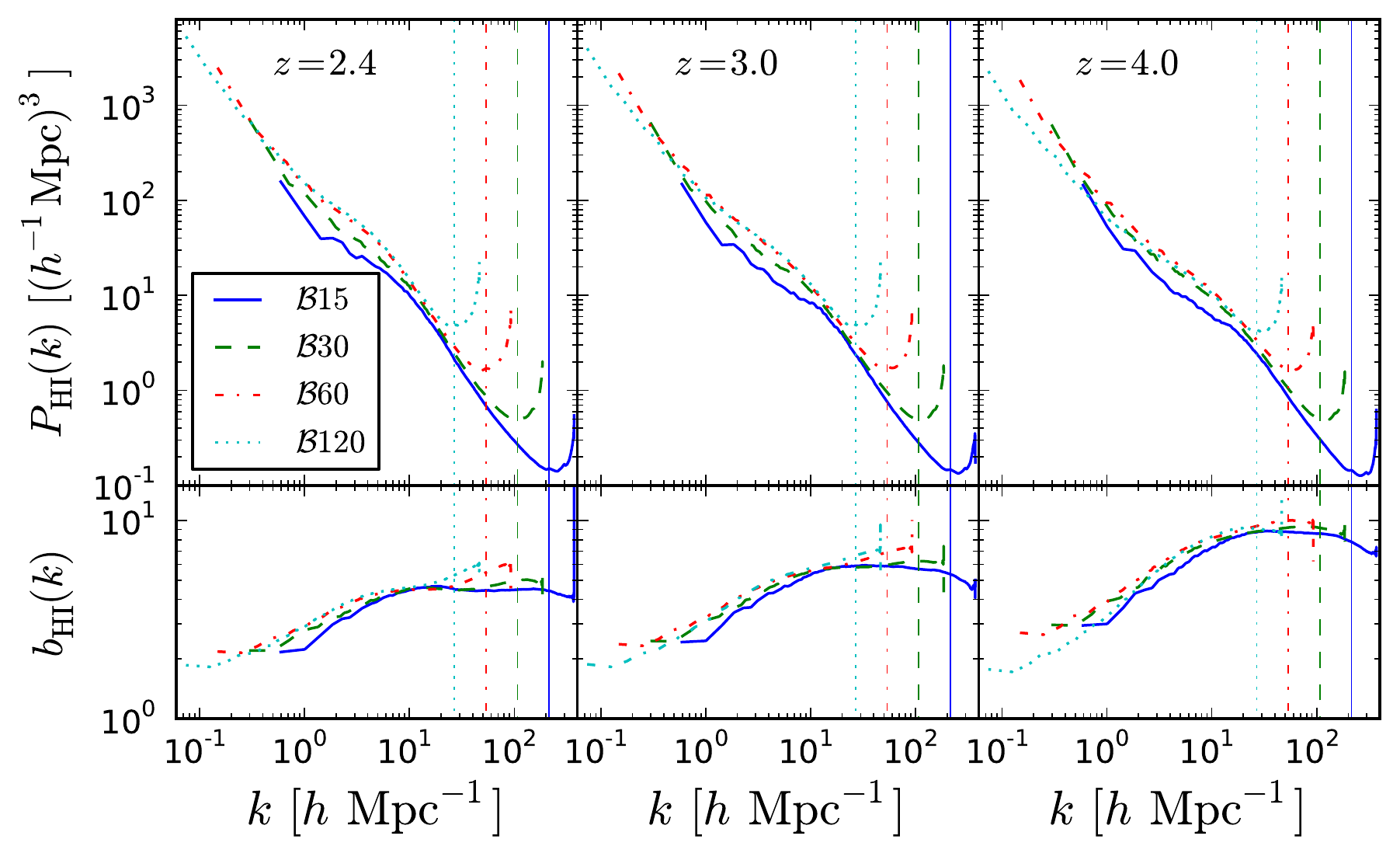}\\
\end{center}
\caption{HI power spectrum obtained by assigning the HI to the gas
  particles belonging to dark matter halos using the halo-based model
  2. The results are displayed at $z=2.4$ (top), $z=3$ (middle) and
  $z=4$ (bottom) for the simulations $\mathcal{B}120$ (dotted cyan),
  $\mathcal{B}60$ (dot-dashed red), $\mathcal{B}30$ (dashed green) and
  $\mathcal{B}15$ (solid blue). The value of the Nyquist frequency is
  displayed for each simulation with a vertical line.
  We show the bias between the distributions of HI and matter, 
  $b_{\rm HI}^2(k)=P_{\rm HI}(k)/P_{\rm m}(k)$, in the bottom panels.}
\label{Pk_HI_Paco}
\end{figure}

In Fig. \ref{Pk_HI_Paco} we plot the HI power spectrum when the HI is
assigned to the gas particles using the halo-based model 2. As
discussed at the beginning of this section, this model assigns a
constant $M_{\rm HI}/M$ fraction for very massive halos: this implies
that a strong contribution of the overall HI power spectrum is
expected to arise from very massive, and thus biased, dark matter
halos. We find that the HI power spectrum is converged on large and
intermediate scales only when using box sizes larger than $\sim
60~h^{-1}$ com. Mpc. This clearly shows the impact that massive halos have
in terms of power spectrum for this model. We note that the results from
the simulation $\mathcal{B}120$ should not be fully converged on all
scales, since that simulation is not able to resolve the smallest
halos that host HI (see from table \ref{Omega_HI_tab_model2} that the
value of $\Omega_{\rm HI}$ is below the expected one). For this scheme,
the $\Omega_{\rm HI}$ value used to compute the average HI mass
per grid cell (for the HI power spectrum computation) is the one
expected from the analytic model (see table \ref{Paco_parameters}),
not the value obtained from the simulations themselves.
To corroborate that the discrepancy in the HI power spectrum arises 
as a consequence of the different halo masses sampled by the different
simulations we have selected halos of the same mass range, converged
in terms of the halo mass function, and computed the power spectrum
of the HI residing within those halos. In this case we find that the HI power spectrum is 
almost the same among the different simulations, pointing out that the
discrepancies are related to the different mass ranges explored in the 
simulations.

The bottom panels of Fig. \ref{Pk_HI_Paco} display the HI bias. As expected, the
values of the HI bias are significantly above those obtained by using the halo-based
model 1 for the reasons depicted above. Notice that the HI bias at $z=2.4$ and
$z=3$ for the simulations $\mathcal{B}60$ and $\mathcal{B}120$ are in very good
agreement among themselves, since both the HI and the matter
power spectrum are very similar for those simulations.

In summary, whereas the distributions of HI obtained by using both halo-based
models can reproduce well the DLAs column density distribution, only the 
halo-based model 2 is capable of distribute the HI in such a way that it matches 
the recent measurements of the DLAs bias from SDSS-III/BOSS. This means that the HI in the halo-based
model 2 is more strongly clustered than in halo-based model 1. This is clearly
reflected in the amplitude of the HI power spectrum: on large scales the HI power
spectrum from the halo-based model 2 is a factor $\sim 3$ above this obtained
from the halo-based model 1.

\section{HI modeling: particle-based}
\label{HI_particles}

In this section we use a different approach to simulate the distribution of neutral hydrogen. The method, that we refer to as the \textit{particle-based} method, is based on assigning HI to all the gas particles of the N-body simulation on a particle-by-particle basis, 
according to the individual physical properties. 
Notice that one of the most important differences of this approach with respect to the halo-based method is that we do not make any assumption about the location of the HI (in the halo-based method we assume that all the HI reside within dark matter halos). 

\begin{table}
\begin{center}
\resizebox{15.5cm}{!}{
\begin{tabular}{|c|c|c|c|c|c|c|}
\cline{1-7}
z & quantity & $\mathcal{B}120$ & $\mathcal{B}60W$ & $\mathcal{B}60$ & $\mathcal{B}30$ & $\mathcal{B}15$ \\ \cline{1-7}

\multicolumn{1}{ |c| }{\multirow{2}{*}{2.4} } &
\multicolumn{1}{ c| }{$\Omega_{\rm HI}$ (total)} & $0.552\times10^{-3}$ & - & $0.647\times10^{-3}$ & $0.573\times10^{-3}$ & $0.510\times10^{-3}$  \\ \cline{2-7}
\multicolumn{1}{ |c|  }{}   &
\multicolumn{1}{ c| }{$\Omega_{\rm HI}$ (filaments)} & $7.265\times10^{-5}$ & - &$1.147\times10^{-5}$ & $3.231\times10^{-6}$ & $0.989\times10^{-6}$   \\ \cline{2-7}
\multicolumn{1}{ |c|  }{}   &
\multicolumn{1}{ c| }{$dN/dX$} & $0.033$ & - & $0.041$ & $0.042$ & $0.041$ \\ \cline{1-7}

\multicolumn{1}{ |c| }{\multirow{2}{*}{3.0} } &
\multicolumn{1}{ c| }{$\Omega_{\rm HI}$ (total)} & $0.524\times10^{-3}$ & $0.659\times10^{-3}$ & $0.667\times10^{-3}$ & $0.640\times10^{-3}$ & $0.619\times10^{-3}$  \\ \cline{2-7}
\multicolumn{1}{ |c|  }{}   &
\multicolumn{1}{ c| }{$\Omega_{\rm HI}$ (filaments)} & $9.163\times10^{-5}$ & $2.450\times10^{-5}$ &$2.154\times10^{-5}$ & $2.699\times10^{-6}$ & $1.884\times10^{-6}$   \\ \cline{2-7}
\multicolumn{1}{ |c|  }{}   &
\multicolumn{1}{ c| }{$dN/dX$} & $0.032$ & $0.045$ & $0.045$ & $0.047$ & $0.048$ \\ \cline{1-7}

\multicolumn{1}{ |c| }{\multirow{2}{*}{4.0} } &
\multicolumn{1}{ c| }{$\Omega_{\rm HI}$ (total)} & $0.351\times10^{-3}$ & $0.551\times10^{-3}$ & $0.536\times10^{-3}$ & $0.587\times10^{-3}$ & $0.619\times10^{-3}$ \\ \cline{2-7}
\multicolumn{1}{ |c|  }{}   &
\multicolumn{1}{ c| }{$\Omega_{\rm HI}$ (filaments)} & $8.104\times10^{-5}$ & $3.706\times10^{-5}$ & $3.440\times10^{-5}$ & $3.589\times10^{-6}$ & $2.879\times10^{-6}$   \\ \cline{2-7}
\multicolumn{1}{ |c|  }{}   &
\multicolumn{1}{ c| }{$dN/dX$} & $0.022$ & $0.038$ & $0.038$ & $0.044$ & $0.048$ \\ \cline{1-7}

\end{tabular}
}
\end{center}
\caption{Values of the DLAs line density, $dN/dX$, $\Omega_{\rm HI}$ over the whole simulation and the value of $\Omega_{\rm HI}$ for the gas particles residing outside dark matter halos (we refer to that environment as filaments, see section \ref{HI_outside_halos_section} for details) in each N-body simulation obtained by assigning the HI to the gas particles according to the particle-based method.}
\label{Omega_HI_tab_Dave}
\end{table}

We follow pretty closely the method recently presented by Dav\'e et al. 2013 \cite{Dave_2013}. We now briefly review the method here and refer the reader to \cite{Dave_2013} for further details. Firstly, HI is assigned to each gas particle assuming photo-ionization equilibrium with the external UV background. For star forming particles we assume that hydrogen residing in the cold phase is fully neutral, while hydrogen in the hot phase is fully ionized. Thus, we set the HI/H fraction of star forming particles equal to its multi-phase cold gas fraction (see \cite{Springel-Hernquist_2003, Nagamine_2004} for further details).
The value of the HI photo-ionization rate, $\Gamma_{\rm HI}$, is tuned such as the Ly$\alpha$ mean transmission flux is reproduced. Next, the method assumes that each gas particle has a density profile given by the SPH kernel $W(r,r_{\rm SPH})$:

\begin{equation}
W(r,r_{\rm SPH}) = \frac{8}{\pi r_{\rm SPH}^3} \left\{ 
  \begin{array}{l l}
    1-6\left(\frac{r}{r_{\rm SPH}}\right)^2+6\left(\frac{r}{r_{\rm SPH}}\right)^3 & \quad 0\leqslant\frac{r}{r_{\rm SPH}}\leqslant\frac{1}{2}\\
    \\
    2\left(1-\frac{r}{r_{\rm SPH}}\right)^3 & \quad \frac{1}{2}<\frac{r}{r_{\rm SPH}}\leqslant 1\\
    \\
    0 & \quad \frac{r}{r_{\rm SPH}}> 1\\
  \end{array} \right.
\label{SPH_kernel}
\end{equation}
where $r_{\rm SPH}$ is the SPH smoothing length of the particle and a correction to account for self-shielding is applied if a radius $r_{\lim}$, such as 

\begin{equation}
N_{\rm HI}(r_{\rm lim})=\frac{0.76m\left(\frac{\rm HI}{\rm H}\right)}{m_H}\int_{r_{\rm lim}}^{r_{\rm SPH}} W(r,r_{\rm SPH})dr=10^{17.2}~{\rm cm}^{-2}~,
\end{equation}
exists. In the above expression HI/H is the neutral hydrogen fraction obtained in the first step, $m$ is the gas particle mass and $m_H$ is the mass of the hydrogen atom. If such radius exists, then the method considers that the spherical shell from $r=0$ to $r=r_{\rm lim}$ is self-shielded againts the external radiation and assigns to it a HI/H fraction equal to 0.9. The HI/H fraction in the shell from $r=r_{\rm lim}$ to $r=r_{\rm SPH}$ is kept to the value obtained by assuming photo-ionization. Finally, the method takes into account the presence of molecular hydrogen, $H_2$, by assigning it to star forming particles using the observed ISM pressure relation found by the The HI Nearby Galaxy Survey (THINGS) \cite{Leroy_2008}

\begin{equation}
R_{\rm mol}=\frac{\Sigma_{H_2}}{\Sigma_{\rm HI}}=\left(\frac{P/k_B}{1.7\times10^{4}~{\rm cm^{-3}K}}\right)^{0.8}~.
\end{equation}
with $P$ being the pressure, $k_B$ the Boltzmann constant and $\Sigma_{\rm HI}$ and $\Sigma_{\rm H_2}$ are the HI and H$_2$ surface densities respectively.

We notice that our implementation of the above method only differs in the way in which the HI/H fractions are computed in photo-ionization equilibrium. Whereas Dav\'e et al. \cite{Dave_2013} uses a simplified hydrogen ionization balance computation, {\sc GADGET} uses a more refined method following \cite{Katz_1996}. We have however explicitly checked that differences are very small among the two methods.

In practice, we begin by extracting 5000 random QSO spectra from a given N-body simulation snapshot and by computing the mean transmitted flux. We tune the value of the photo-ionization rate such as we reproduce the observed mean transmitted flux from the Ly$\alpha$ forest \cite{Becker_2013} at the redshift of the N-body snapshot. 
We then calculate, in case it exits, the radius $r_{\rm lim}$ for every gas particle in the snapshot and correct the neutral hydrogen content of it to account for self-shielding. Finally, for star forming particles, we correct their self-shielded HI mass to take into account the presence of molecular hydrogen. We compute the pressure of the gas particles using their density and internal energy.

\begin{figure}
\begin{center}
\includegraphics[width=0.9\textwidth]{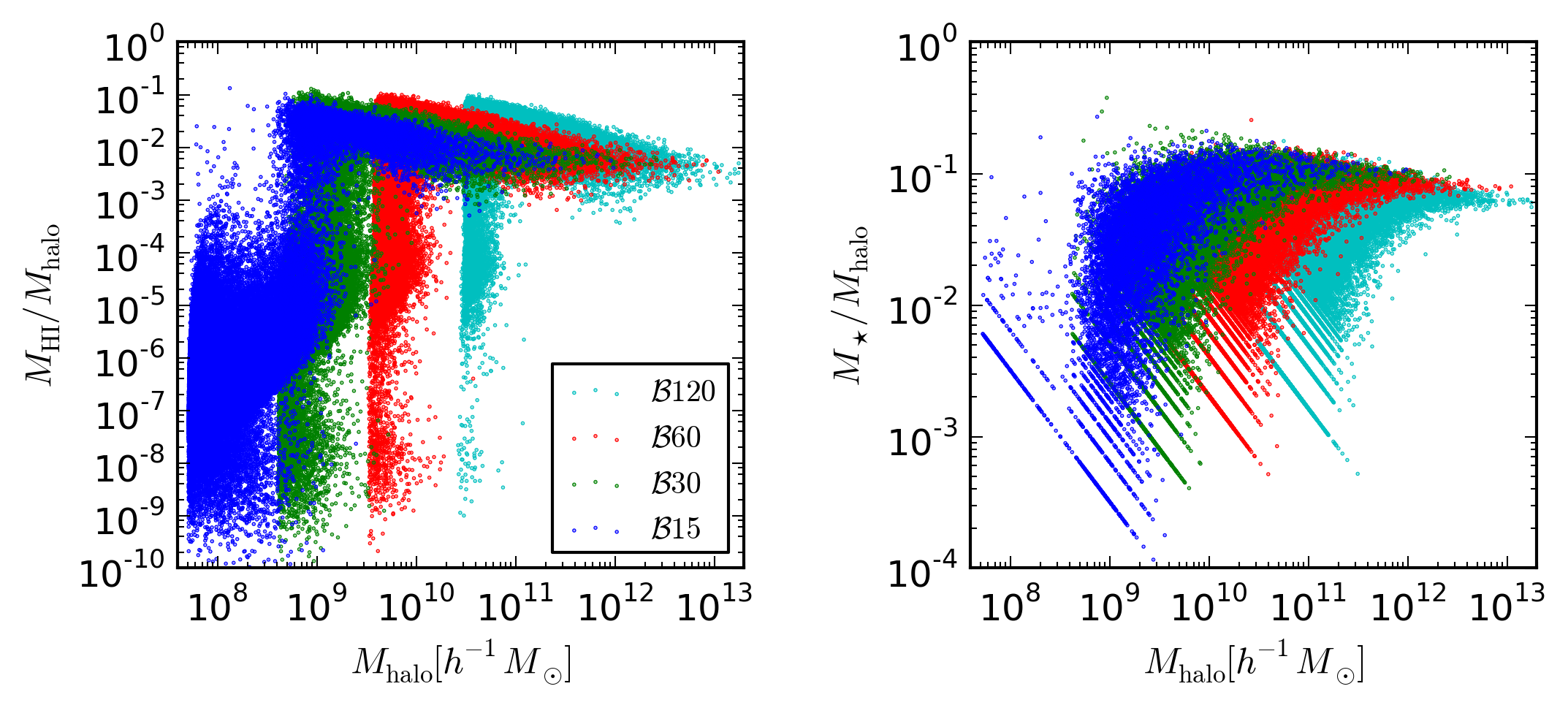}\\
\end{center}
\caption{For each dark matter halo of the simulation $\mathcal{B}120$ (cyan), $\mathcal{B}60$ (red), $\mathcal{B}30$ (green) and $\mathcal{B}15$ (blue), at $z=3$, we compute the neutral hydrogen mass, $M_{\rm HI}$, and the stellar mass $M_\star$. In the left panel we show the ratio $M_{\rm HI}/M_{\rm halo}$ as a function of the halo mass whereas the ratio $M_\star/M_{\rm halo}$ is displayed in the right panel.}
\label{M_HI_Dave}
\end{figure}  

We apply the above method to the simulations $\mathcal{B}120$, $\mathcal{B}60$, $\mathcal{B}60$W, $\mathcal{B}30$ and $\mathcal{B}15$ and show in table \ref{Omega_HI_tab_Dave} the values we obtain for the DLAs line density and the parameter $\Omega_{\rm HI}$. In Fig. \ref{M_HI_Dave} we show the HI mass fraction and the stellar mass fraction for each dark matter halo in each simulation (for clearness we do not show the results for the simulation with galactic winds). We find that as the resolution of the simulation increases, halos of the same mass contain larger fractions of stellar mass, i.e. star formation is more efficient in the high resolution simulations than in the low ones; note however that the results seem to begin converging for the simulations $\mathcal{B}30$ and $\mathcal{B}15$. The consequence of this is that in the highest resolution simulations there are fewer gas particles, and thus, the neutral hydrogen mass that a dark matter halo host critically depends on resolution as it can be seen in the left panel of Fig. \ref{M_HI_Dave}. We notice that this problem was already discussed in \cite{Dave_2013} and can be also understood in terms of semi-analytic models of galaxy formation \cite{DeLucia_2008,Guo_2011}.

As can be seen from table \ref{Omega_HI_tab_Dave} the value of $\Omega_{\rm HI}$ critically depends on the size of the simulation box. This unphysical behavior arises due to two competing effects. On one hand we have that large box size simulations can only resolve the most massive halos, and therefore, in those simulations the contribution to $\Omega_{\rm HI}$ from low mass halos is not accounted for. On the other hand, small box size simulations (higher resolution simulations) convert gas to stars more effectively, thus, the amount of gas left in those simulations capable of host HI decreases with resolution. These two effect compete and at $z=2.4$ and $z=3$ we find that the value of $\Omega_{\rm HI}$ peaks for the simulations $\mathcal{B}60$.

\begin{figure}
\begin{center}
\includegraphics[width=1.0\textwidth]{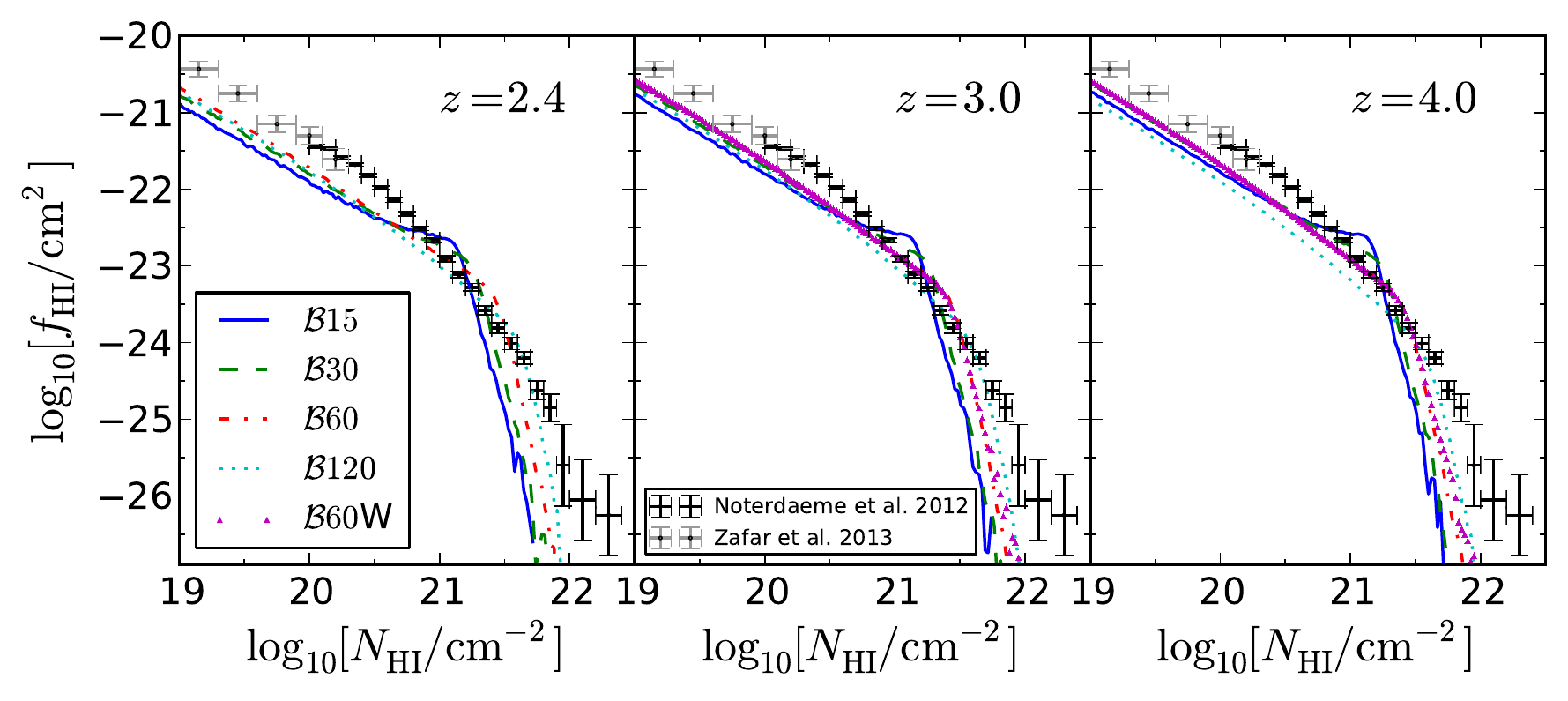}\\
\end{center}
\caption{Column density distribution obtained by using the particle-based method at $z=2.4$ (left), $z=3.0$ (middle) and $z=4$ (right) on top of the simulations $\mathcal{B}15$ (solid blue), $\mathcal{B}30$ (dashed green, $\mathcal{B}60$ (dot-dashed red), $\mathcal{B}120$ (dotted cyan) and $\mathcal{B}60$W (purple triangles). The observational measurements of Noterdaeme et al. 2012 \cite{Noterdaeme_2012} are shown in black whereas those by Zafar
  et al. 2013 \cite{Zafar_2013} are reported in gray.}
\label{f_HI_Dave}
\end{figure} 

In Fig. \ref{f_HI_Dave} we show the HI column density distribution at redshifts $z=2.4$, $z=3$ and $z=4$. We find that this method fails to reproduce the abundance of the systems with the highest column densities. As expected, the results are sensitive to the resolution of the simulation. The formation of H$_2$ is reflected by a sudden suppression of the abundance of absorbers with column densities higher than $N_{\rm HI}\sim10^{21}-10^{21.5}~{\rm cm^{-2}}$. This transition is quite abrupt for the simulation $\mathcal{B}15$, in which the number of star forming particles with high pressure is significantly larger than in the other simulations.

\begin{figure}
\begin{center}
\includegraphics[width=1.0\textwidth]{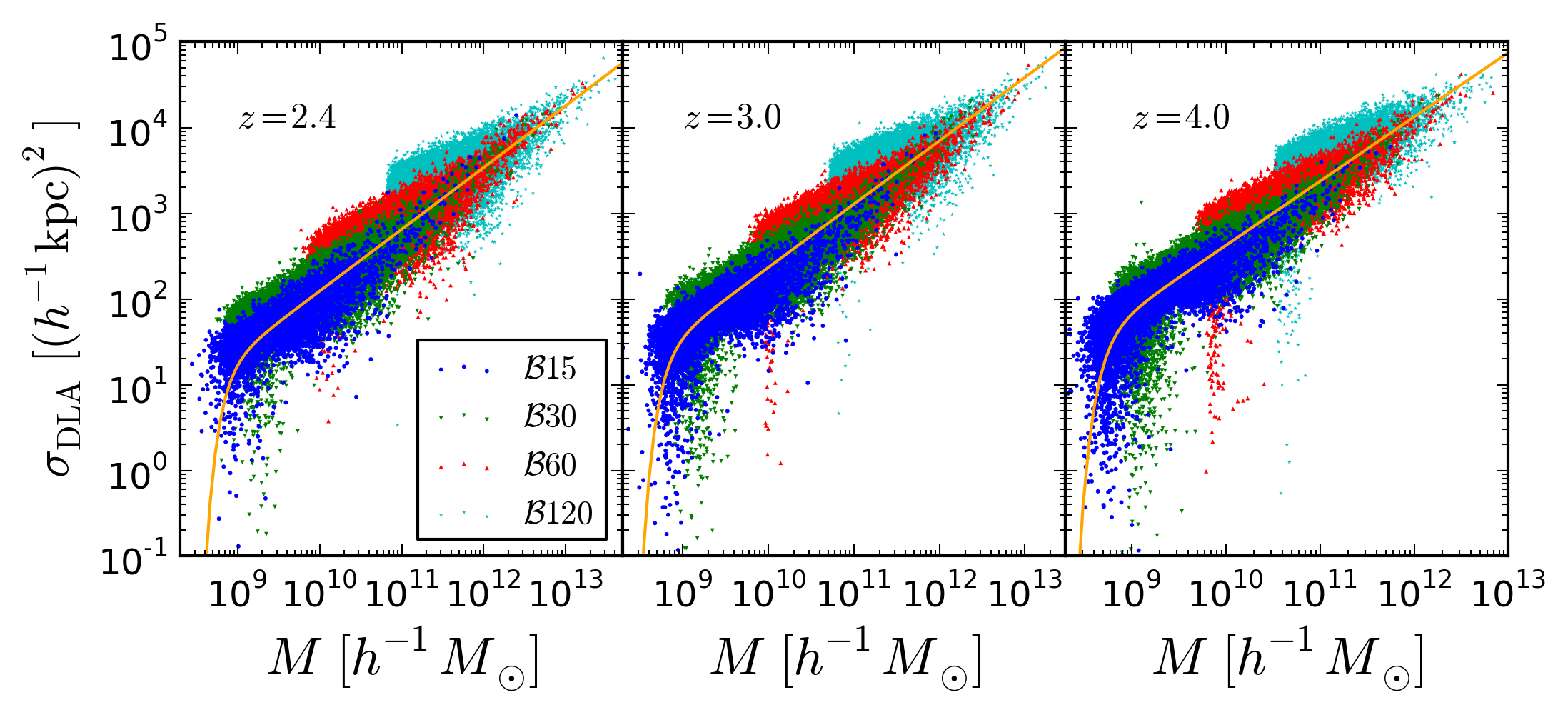}\\
\end{center}
\caption{DLA comoving cross section for a subsample of halos at $z=2.4$ (left), $z=3$ (middle) and $z=4$ (right) from the simulations $\mathcal{B}120$ (cyan), $\mathcal{B}60$ (red), $\mathcal{B}30$ (green) and $\mathcal{B}15$ (blue) when the neutral hydrogen is assigned to the gas particles using the particle-based method. A fit to the results is presented with a solid orange line.}
\label{cross_section_Dave}
\end{figure}

We show the DLAs comoving cross-section obtained by using the particle-based model in Fig. \ref{cross_section_Dave}. In contrast to the results obtained by using the halo-based model 1, 
where the DLA cross-section remains constant for very massive halos,
 we find that the DLA cross-section increases with halo mass. However, the amplitude of the cross-section is significantly lower in this model in comparison to the one obtained by employing the halo-based model 2. Our results suggest that the minimum mass that a dark matter halo should have to host DLAs is about $\sim5\times10^{8}~h^{-1}{\rm M}_\odot$, almost independently of redshift. We find that the mean DLA cross-section can be well described by the following fitting formula
\begin{equation}
\sigma_{\rm DLA}(M)=\sigma_0\left(\frac{M}{M_0}\right)^\beta\exp{\left[-(M_0/M)^3\right]}~.
\label{fitting_formula_Dave}
\end{equation}
In table \ref{cross-section_Dave_table} we show the values that best fit our results and show these in Fig. \ref{cross_section_Dave} with orange lines. By using the average DLA cross-section and Eq. \ref{biaseq} we find a value for the bias of the DLAs equal to 1.48, 1.69 and 2.07 at  $z=2.4, 3$ and 4 respectively. These values are very similar to those obtained by using the halo-based model 1, and thus, they are also in strong tension with the observational measurements by \cite{Font_2012}.

\begin{table}
\begin{center}
\resizebox{6cm}{!}{
\begin{tabular}{|c|c|c|c|}
\hline
$z$ & $\sigma_0$ & $M_0$ & $\beta$ \\ 
 & $(h^{-1}{\rm kpc})^2$ & $(h^{-1}M_\odot)$ & \\
\hline
$2.4$ & $18.1$ & $7\times10^{8}$  & $0.721$ \\
\hline
$3.0$ & $29.1$ & $6\times10^{8}$  & $0.739$ \\
\hline
$4.0$ & $44.4$ & $5\times10^{8}$  & $0.750$ \\
\hline
\end{tabular}
}
\end{center}
\caption{Best fit values of the fitting formula \ref{fitting_formula_Dave} used to reproduce the average cross-section obtained by assigning HI to the gas particles using the particle-based model.}
\label{cross-section_Dave_table}
\end{table}

\begin{figure}
\begin{center}
\includegraphics[width=1.0\textwidth]{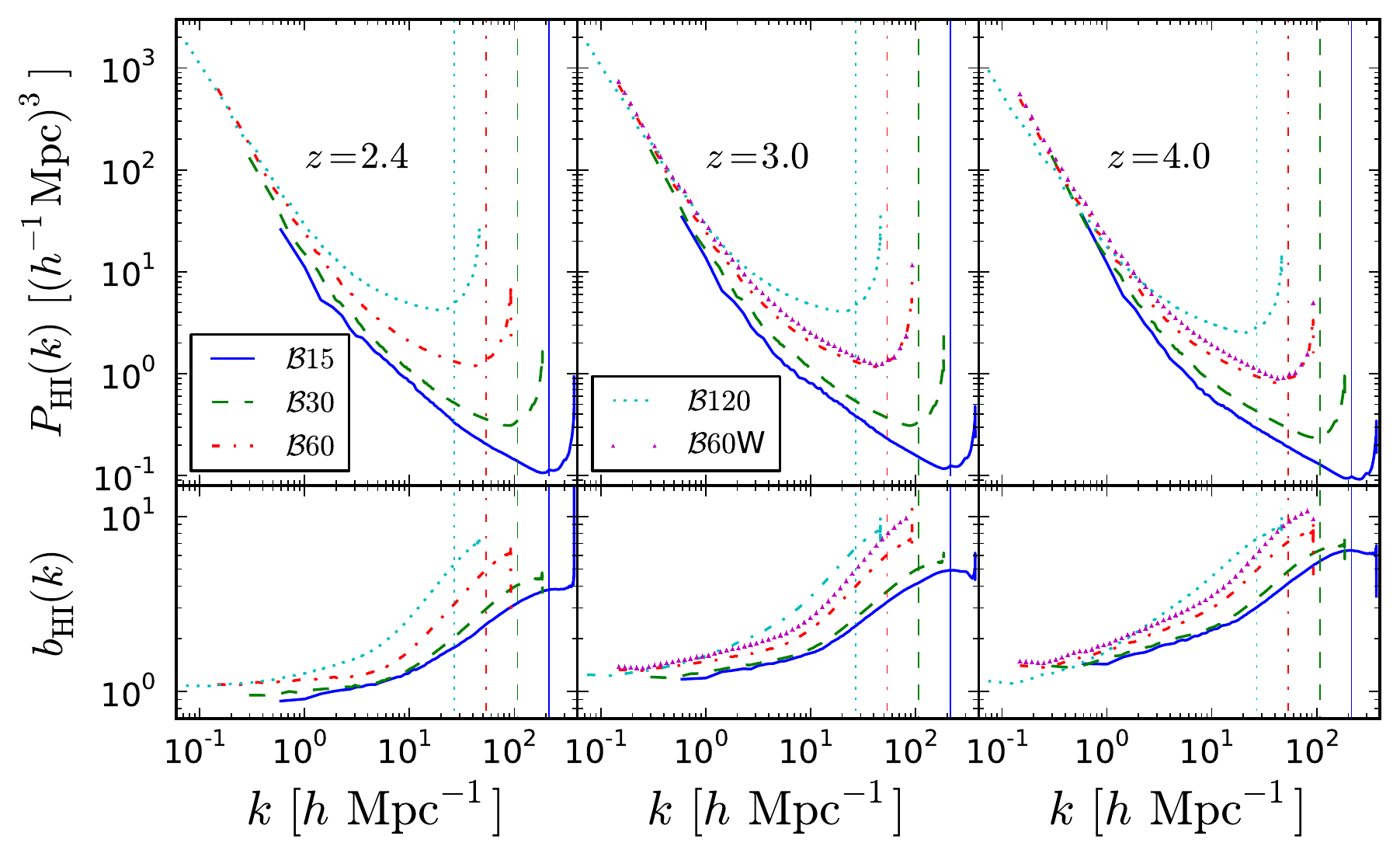}\\
\end{center}
\caption{HI power spectrum at redshift $z=2.4$ (left), $z=3$ (middle) and $z=4$ (right) obtained by assigning neutral hydrogen to the gas particles using the particle-based method.
Results are shown for the simulations $\mathcal{B}15$ (solid blue), $\mathcal{B}30$ (dashed green), $\mathcal{B}60$ (dot-dashed red), $\mathcal{B}120$ (dotted cyan) and $\mathcal{B}60$W (purple triangles). The vertical lines represent the value of the Nyquist frequency for the different simulations. We display the bias between the distributions of HI and matter, $b_{\rm HI}^2(k)=P_{\rm HI}(k)/P_{\rm m}(k)$, in the bottom panels.}
\label{Pk_HI_Dave}
\end{figure}

In Fig. \ref{Pk_HI_Dave} we display the power spectrum of the neutral hydrogen distribution for the different simulations at the redshifts $z=2.4$, $z=3$ and $z=4$. We stress that the mean HI mass per grid cell has been obtained by using $\Omega_{\rm HI}=10^{-3}$.
We find that the HI power spectrum depends on the simulation resolution, although results seem almost converged for the simulations $\mathcal{B}15$ and $\mathcal{B}30$. Our results indicate however that on large scales the HI power spectrum is converged even for the simulation $\mathcal{B}60$. The results of the simulation $\mathcal{B}120$ are slightly below those of the other simulations on large scales: this is due to the fact that this simulation can not resolve the smallest halos that can host HI, which is then reflected on the amplitude of the HI power spectrum on large scales. We show the HI bias in the bottom panels of Fig. \ref{Pk_HI_Dave}. The dependence of the results with resolution is clear from the plots, even though the HI bias seems almost converged for the simulations $\mathcal{B}15$ and $\mathcal{B}30$. Therefore, the differences we find in terms of the HI power spectrum for those simulations are mainly due to cosmic variance and box size effects rather than resolution. 

We now discuss the role played by the galactic winds in our results. As can be seen from table \ref{Omega_HI_tab_Dave}, the simulation $\mathcal{B}60$W contains a similar amount of HI as the simulation $\mathcal{B}60$ (the resolution is exactly the same in both simulations) at all redshifts. The reason for this is due to two effects going in opposite directions. On one hand, galactic winds suppress the star formation rate which means less star forming particles and thus, less HI. On the other hand the gas particles that experience  galactic winds are hydrodynamically decoupled for a certain time (or until their density is below some threshold). During that time, radiative cooling is very efficient, which means that the HI/H fraction computed assuming photo-ionization equilibrium is close to 1. Therefore, these two effects conspire to produce a similar total amount of HI. In terms of the HI column density distribution the simulations with and without winds produce very similar results. Finally, 
galactic winds also leave their signature on the HI power spectrum as can be seen from Fig. \ref{Pk_HI_Dave}. We find that the amplitude of the HI power spectrum is slightly higher in the simulation with galactic winds than in the simulation without winds.
This is due to the fact that galactic winds can escape from low-mass halos but are not able to leave the most massive ones. In the simulation with galactic winds, the most massive (and biased) halos will contribute more to the HI power spectrum than the low mass halos, producing an enhancement on the HI power spectrum as the one we obtain.

\section{HI outside dark matter halos}
\label{HI_outside_halos_section}

It is well known that in the post-reionization era the majority of the neutral hydrogen in the Universe is expected to reside in dense environments such as galaxies. We now investigate the importance, in terms of abundance and spatial clustering, of the HI residing outside dark matter halos, and we define this environment, i.e. all the gas outside dark matter halos (as identified by the FoF algorithm), as \textit{filaments}.
Our filamentary cosmic-web is realistic in the sense that it gives rise to a Lyman-$\alpha$ forest whose statistical properties are in agreement
with observations both in terms of continuous statistics like flux power and discrete statistics as properties of lines.

Since, by construction, the halo-based method does not assign any HI to gas particles outside halos, we use the particle-based method to investigate the properties of HI in filaments. We compute the value of $\Omega_{\rm HI}^{\rm filaments}$, i.e. the value of $\Omega_{\rm HI}$ using only the gas particles outside halos, and show the results in table \ref{Omega_HI_tab_Dave}. For the simulations with the highest resolution we find a typical value of $\Omega_{\rm HI}^{\rm filaments}\sim10^{-6}$. However, this value increases by more than an order of magnitude as resolution decreases.

\begin{figure}
\begin{center}
\includegraphics[width=0.8\textwidth]{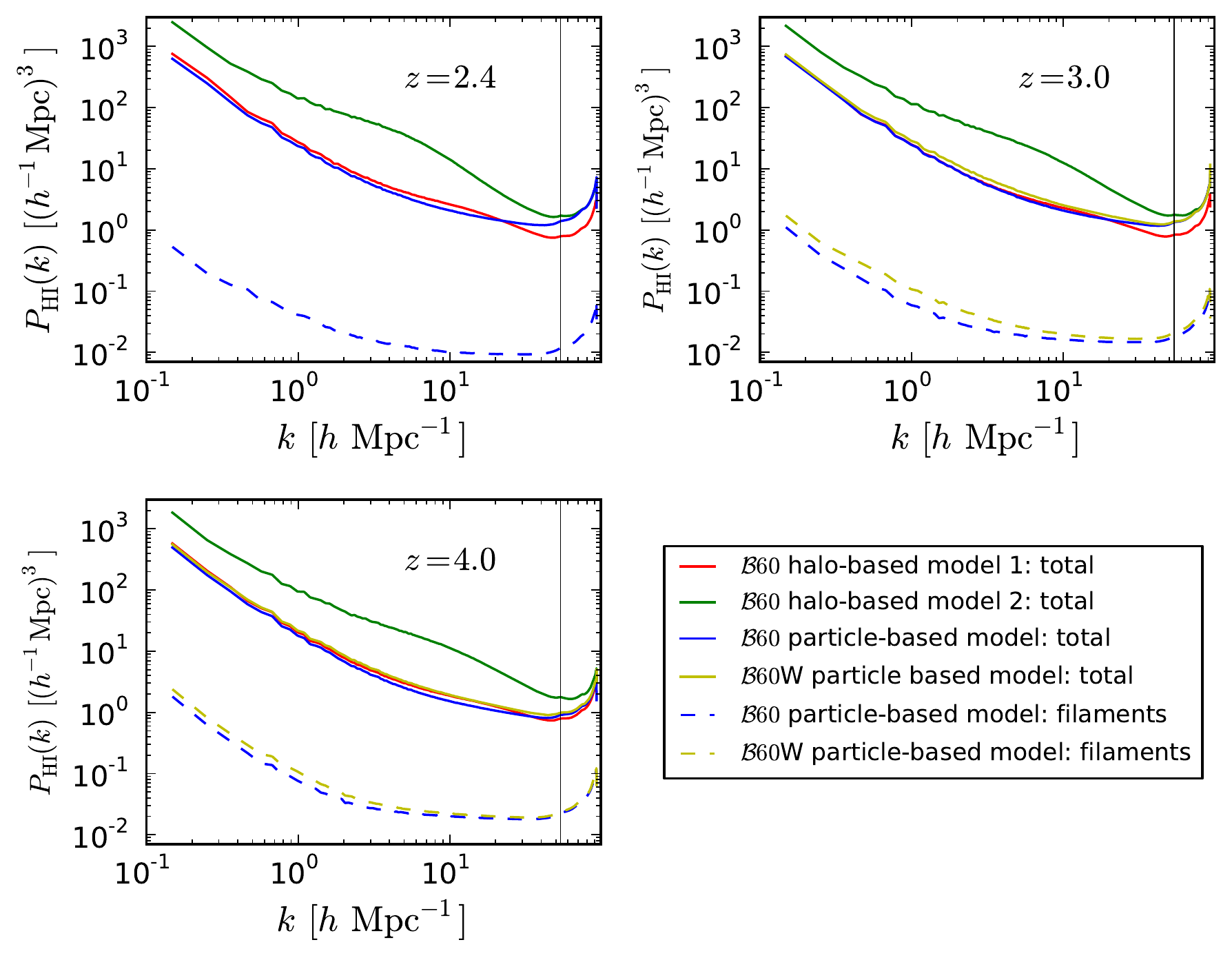}\\
\end{center}
\caption{Contribution of the HI in filaments, i.e. HI residing outside dark matter halos, to the total neutral hydrogen power spectrum. With solid lines we show the total HI power spectrum at $z=2.4$ (top-left), $z=3$ (top-right) and $z=4$ (bottom-left) when the HI is assigned to the gas particles of the simulation $\mathcal{B}60$ using the halo-based models 1 and 2 (red and green lines respectively) and the particle-based method (blue lines). The yellow lines represent the results obtained by using the simulation $\mathcal{B}60W$ (only applying the particle-based method). The dashed lines represent the power spectrum of the HI in filaments. The vertical lines display the value of the Nyquist frequency.}
\label{Pk_HI_filaments}
\end{figure}

We now quantify the contribution of the HI in filaments to the total neutral hydrogen power spectrum. In Fig. \ref{Pk_HI_filaments} we show with solid lines the total HI power spectrum from the simulation $\mathcal{B}60$ when the neutral hydrogen is assigned to the gas particles using the three different methods investigated in this paper. The dashed lines correspond to the power spectrum of the HI in filaments. We find that for all methods and  redshifts studied in this paper the contribution of the HI outside halos to the total HI power spectrum is negligible in all cases. However, we have checked that this is no longer the case as the redshift increases: for the simulation $\mathcal{B}60$, at $z=6$, about 30\% of the total HI content is contributed by the filaments. We expect that the contribution of HI in filaments to the overall HI content to increase with redshift since at higher redshift the abundance of dark matter halos decreases, leaving much more gas available for hosting HI outside dark matter halos.

\begin{figure}
\begin{center}
\includegraphics[width=0.6\textwidth]{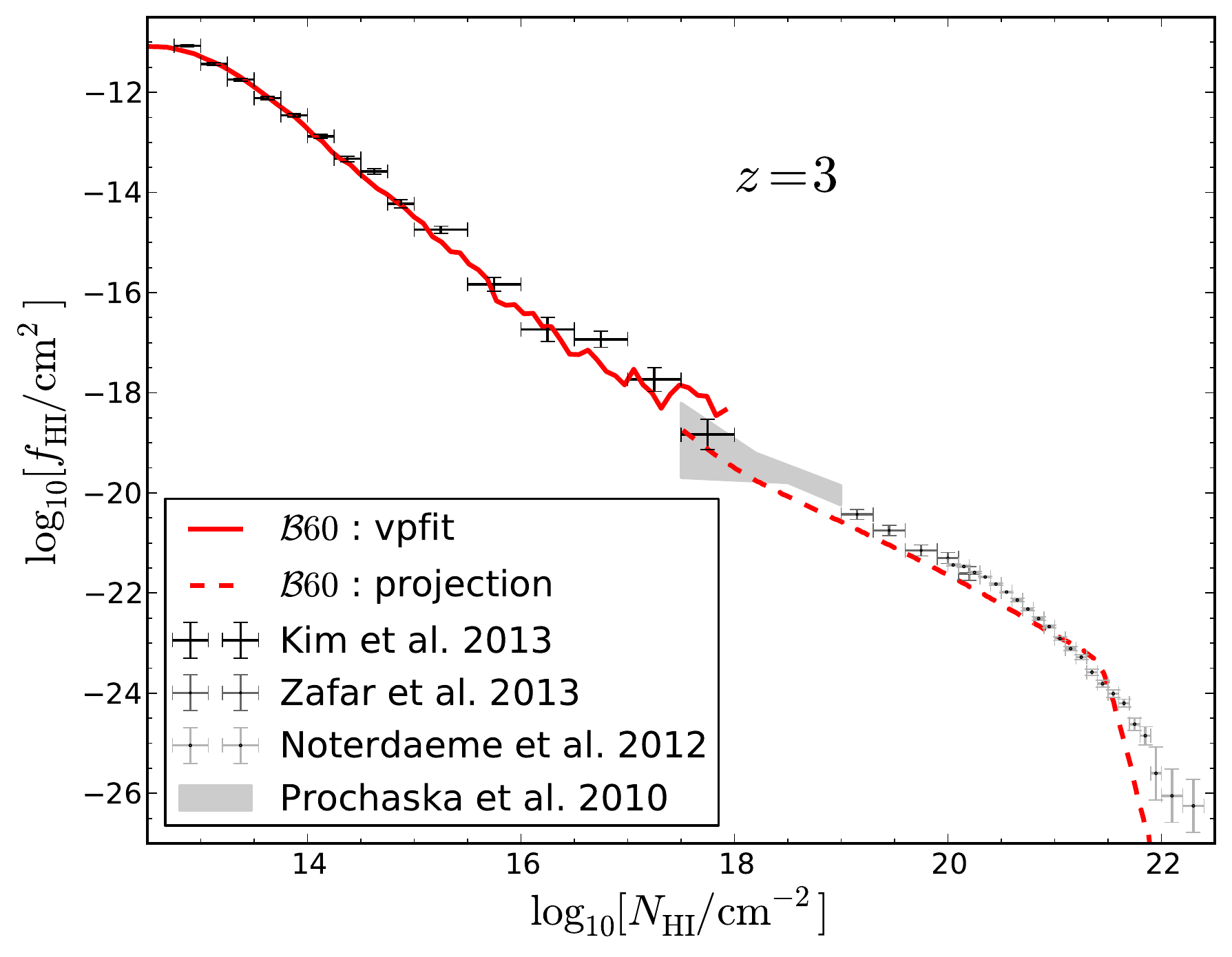}\\
\end{center}
\caption{HI column density distribution. The red solid line shows the HI column density distribution obtained by using VPFIT over the neutral hydrogen distribution obtained by assigning HI to the gas particles of the simulation $\mathcal{B}60$ at $z=3$ using the particle-based method. The red dashed line represents the HI column density distribution using our standard procedure of projecting the particles in a plane. The error points represent the measurements by Kim et al. 2013 \cite{Kim_2013} (black), Prochaska et al. 2010 \cite{Prochaska_2010} (gray area), Zafar et al. 2013 \cite{Zafar_2013} (dark gray) and Noterdaeme et al. 2012 \cite{Noterdaeme_2012} (light gray).}
\label{N_HI_vpfit}
\end{figure}

Finally, we investigate the properties of the HI in filaments in terms of the HI column density distribution. We assign HI to the gas particles of the simulation $\mathcal{B}60$  at $z=3$ using the particle-based method. We then generate 1000 QSO mock spectra and fit them using the VPFIT code \cite{Carswell_1987}. In Fig. ~\ref{N_HI_vpfit} we show with a solid red line the HI column density distribution obtained by using VPFIT together with the observational measurements by Kim et  al. \cite{Kim_2013}. The agreement between simulations and observations is remarkable. We also plot with a dashed red line the column density distribution for high column density values obtained using our standard procedure. As discussed in Sec. \ref{HI_particles}, the particle-based method produces reasonable results for the LLS systems but fails to reproduce the abundance of the highest column density absorbers. 

We emphasize that neither the value of $\Omega_{\rm HI}^{\rm filaments}$ nor the filaments HI power spectrum is converged and the results strongly depend on resolution, while the column density distribution function obtained with VPFIT is reasonably well converged. This happens because our definition of filaments not only includes the diffuse inter-galactic medium, but also the circum-galactic medium surrounding the dark matter halos and sub-resolution halos. Therefore, it is natural that some of the quantities shown exhibit such relatively large dependence with resolution. However, we can safely conclude that the amplitude of the HI power spectrum at $z<4$ is set by the HI residing in dark matter halos, with an almost negligible contribution coming from HI residing outside halos. These conclusions are no longer valid when considering higher redshift regimes.

\section{21 cm signal}
\label{21cm_section}

We now investigate the redshifted 21 cm signal generated by the distribution of neutral hydrogen and its detectability with the future telescopes like the SKA. We split this section into three subsections: In subsection \ref{21cm_Pk_subsection} we describe the method used to compute the 21 cm power spectrum. Our estimates of the noise expected in the 21 cm power are presented in subsection \ref{21cm_Pk_noise}. Finally, the detectability of the 21 cm power spectrum with the future SKA1-low and SKA1-mid telescopes is discussed in subsection \ref{detectability}.

\subsection{21 cm power spectrum from simulations}
\label{21cm_Pk_subsection}

The 21 cm power spectrum is computed from the spatial distribution of neutral hydrogen which in turn is obtained from the hydrodynamical simulations using the halo-based and particle-based methods. In real space, the brightness temperature excess due to the 21 cm emission from neutral hydrogen located in the real-space coordinate $\vec{r}$ and having a HI density $\rho_{\rm HI}(\vec{r})$ is given by \cite{Furlanetto_2006,Mao_2012}

\begin{equation}
\delta T_b(\nu)=\overline{\delta T_b}(z)\left(\frac{\rho_{\rm HI}(\vec{r})}{\bar{\rho}_{\rm HI}}\right)
\left[ 1-\frac{T_\gamma(z)}{T_s(\vec{r})}\right],
\end{equation}
where
\begin{equation}
\overline{\delta T_b}(z)=23.88~\bar{x}_{\rm HI}\left( \frac{\Omega_{\rm b}h^2}{0.02}\right)\sqrt{\frac{0.15}{\Omega_{\rm m}h^2}\frac{(1+z)}{10}}~{\rm mK},
\end{equation}
$\bar{\rho}_{\rm HI}$ is the mean density of neutral hydrogen and $\bar{x}_{\rm HI}=\bar{\rho}_{\rm HI}/\bar{\rho}_{\rm H}$ is the average neutral hydrogen fraction.
The quantity $T_\gamma(z)$ is the CMB temperature at redshift $z$ and $T_s$ is the spin temperature characterizing the relative population of HI atoms in different states.

Note that we have not accounted for the fact that the observed frequency not only depends on the cosmological redshift, but also on the peculiar velocity of the HI gas along the line of sight of observation. Since the radio telescope measurements will probe the signal in redshift-space, it is essential that the above equation is re-written in terms of the redshift space coordinates. If we assume the gas to be optically thin, and the spin temperature to be much larger than the CMB temperature ($T_s\gg T_\gamma$), the brightness temperature excess in redshift-space can be expressed as \cite{Mao_2012}
\begin{equation}
\delta T_b^s(\nu)=\overline{\delta T_b}(z)\left[\frac{\rho_{\rm HI}(\vec{s})}{\bar{\rho}_{\rm HI}}\right],
\label{delta_Tb}
\end{equation}
where the superscript $s$ in $\delta T_b$ indicates that the quantity is calculated in redshift-space and $\vec{s}$ is the redshift-space coordinate corresponding to $\vec{r}$. The relation between $\vec{s}$ and $\vec{r}$ is given by
\begin{equation}
\vec{s} = \vec{r} + \frac{1+z}{H(z)} \vec{v}_{\parallel}(\vec{r}),
\end{equation}
where $z$ should be interpreted as the redshift of observation and $\vec{v}_{\parallel}$ is the component of peculiar velocity along the line of sight. The 21 cm power spectrum in redshift space is then defined as
\begin{equation}
P^s_{\rm 21cm}(k)=\langle \delta T_b^s(\vec{k}) (\delta T_b^s)^*(\vec{k})  \rangle,
\label{Pk_21cm_s}
\end{equation}
where $\delta T_b^s(\vec{k})$ is simply the Fourier transform of the brightness temperature excess. 

We should mention here that the assumption $T_s \gg T_\gamma$ holds in the redshifts of our interest as the spin temperature is usually coupled to the gas kinetic temperature which, in turn, is expected to be much higher than $T_\gamma$. The other approximation we have made, i.e., the assumption that the gas is optically thin, is also a good approximation because of the forbidden nature of the hyperfine transition. There could be some regions (e.g., highly overdense regions) where the peculiar velocity effects can make the optical depth quite large in redshift space \cite{Mao_2012}. However, the fraction of such regions is expected to be quite small and hence we ignore them in this work.

Given the above formalism we compute the 21 cm power spectrum by assigning neutral hydrogen to the gas particles of simulation $\mathcal{B}60$ using the three different methods investigated in this paper. We choose simulation $\mathcal{B}60$ because it allows us to study the 21 cm power spectrum for sufficiently low values of $k$ (i.e., large scales) which can be accessed by telescopes like the SKA. We should, however, keep in mind that the HI power spectrum is converged only for the halo-based models (see Figs. \ref{Pk_HI_Bagla} and \ref{Pk_HI_Paco}), while the convergence is not satisfactory for the particle-based method (see Fig. \ref{Pk_HI_Dave}). 

\begin{figure}
\begin{center}
\includegraphics[width=1.0\textwidth]{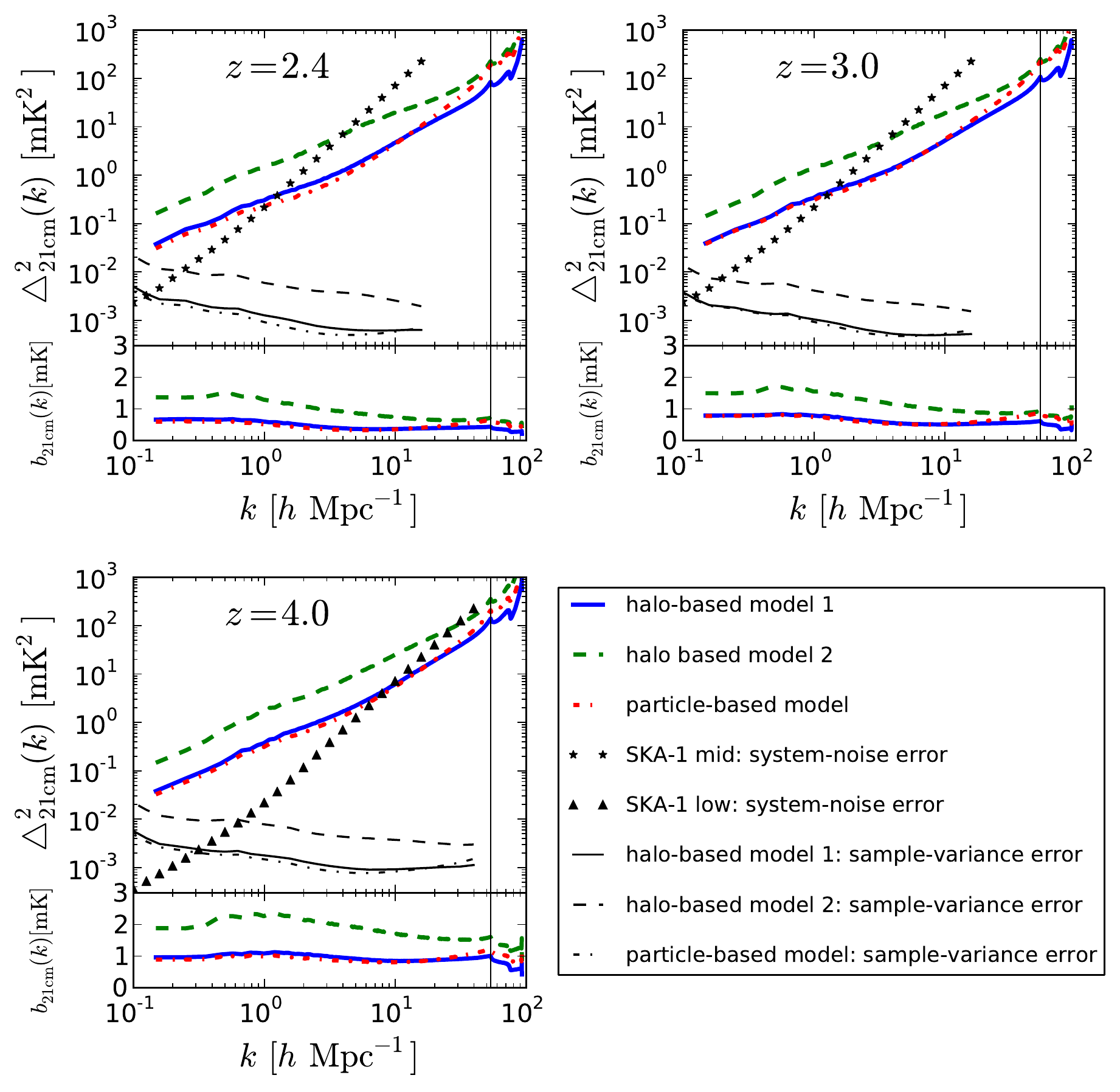}\\
\end{center}
\caption{21 cm power spectrum in redshift-space. We compute the dimensionless 21 cm power spectrum, $\bigtriangleup^2_{\rm 21cm}(k)=k^3P_{\rm 21cm}(k)/2\pi^2$, by
assigning HI to the gas particles of the simulation $\mathcal{B}60$ using the halo-based models 1 and 2 (solid blue and dashed green respectively) and the particle-based method (red dot-dashed) at $z=2.4$ (top-left), $z=3$ (top-right) and $z=4$ (bottom-left). The black stars and black triangles represents the expected SKA-1 mid and SKA-1 low system temperature noises for an observation time of 100 hours in a 32 MHz bandwidth and for intervals in $k$ with a width of $dk=k/5$ (see text for further details). The solid, dashed and dot-dashed black lines represent the sample variance noise for the halo-based model 1, the halo-based model 2 and the particle-based model, respectively. The vertical lines display the position of the Nyquist frequency. Below each panel we show the 21 cm bias defined as $b^2_{\rm 21cm}(k)=\bigtriangleup^2_{\rm 21cm}(k)/\bigtriangleup^2_{\rm m}(k)$ where $\bigtriangleup^2_{\rm m}(k)$ is the dimensionless matter power spectrum in real-space.}
\label{Pk_21cm}
\end{figure}

The plots for the 21 cm power spectrum are shown in Fig. \ref{Pk_21cm}. We find that the power spectrum obtained using the halo-based model 1 and the particle-based method are almost identical. This follows from the fact that the HI power spectra obtained from these two methods are quite similar as can be seen from Fig. \ref{Pk_HI_filaments}. We however note that the values of $\Omega_{\rm HI}$ from the two cases are different, with the particle-based method having a lower value. This implies that the HI in the particle-based method is more strongly clustered than that of the halo-based model 1. Should both methods be normalized to the same value of $\Omega_{\rm HI}$, the HI power spectrum from the particle-based method would have a higher amplitude. The 21 cm power spectrum obtained by using the halo-based model 2 has a significantly higher amplitude than those obtained by employing the other two methods. This is because of the much stronger HI clustering in that model, which is required to match the bias  measurements of DLAs \cite{Font_2012}. 

In the bottom of each panel of Fig. \ref{Pk_21cm} we display the 21 cm bias: $b^2_{\rm 21cm}(k)=P_{\rm 21cm}(k)/P_{\rm m}(k)$, where $P_{\rm m}(k)$ is the matter power spectrum in real-space whereas the 21 cm power spectrum is computed in redshift-space. Interestingly, on large scales ($k \lesssim 0.5 h~{\rm Mpc}^{-1}$), the slope of both the 21 cm power spectra and the 21 cm bias is the same in all the three models. This is because the HI power spectrum at large scales essentially traces the underlying dark matter fluctuations, the only difference being given in terms of the (scale-independent) linear bias parameter. Different prescriptions for generating the HI distribution only result in different values of this bias. Hence, it should be possible to constrain the dark matter power spectrum at large-scales using the 21 cm power spectrum signal. Surprisingly, we find that in the fully non-linear regime, the 21 cm bias exhibits a very weak scale dependence.

\subsection{Modeling the system noise}
\label{21cm_Pk_noise}

We now compute the sensitivity of 21 cm power spectrum that will possibly be achieved for the telescopes such as SKA1-low and SKA1-mid. We will assume that the only noise which contributes to the measurement is the system noise and sample variance, and we will ignore effects arising from astrophysical foregrounds, variations in the ionosphere, radio-frequency interference etc. Accounting for these effects is essential for detecting the 21 cm signal, however, we make the assumption that they can be separately identified and removed from the data before comparing with theoretical models. The $1\sigma$ error in the power spectrum for a single mode $\vec{k}$ arising from the system temperature can be written as (see Appendix A of \cite{geil} or \cite{mcquinn} for the detailed derivation) 
\be
\delta P_{N}(\vec{k},\nu)=\frac{T_{\rm sys}^2}{B t_0} \left ( \frac{\lambda^2}{A_e} \right)^2 \frac{r_{\nu}^2 L}{n_b(\vec{U},\nu)},
\e
where $\nu$ and $\lambda$ are the observing frequency and wavelength respectively. The quantities $T_{\rm sys}$, $B$ and $t_0$ denote the system temperature of the instrument, total frequency bandwidth and observation time, respectively. The effective collecting area of an individual antenna is denoted as $A_e$ which in turn can be written as $A_e=\epsilon A$, where $\epsilon$ and $A$ are the antenna efficiency and physical collecting area respectively. The comoving distance to an observer at redshift $z$, corresponding to an observing frequency $\nu$, is denoted as $r_{\nu}$, while $L$ is the comoving length associated to the bandwidth $B$. The number density of baselines $\vec{U}$ is written as $n_b(\vec{U},\nu)$. The baseline vector $\vec{U}$ is related to $\vec{k}$ by the relation   $\vec{U}=\vec{k}_{\perp}r_{\nu}/2 \pi$, where $\vec{k}_{\perp}$ is the component of $\vec{k}$ perpendicular to the line of sight. We write $n_b(\vec{U},\nu)$ as 
\be
n_b(\vec{U},\nu)=\frac{N(N-1)}{2}\rho_{2D}(\vec{U},\nu),
\e
where $N$ is the total number of antennae and $\rho_{2D}(\vec{U},\nu)$ is the two-dimensional normalized baseline distribution function which follows the condition $\int_0^{\infty} U d U \int_{0}^{\pi} d\phi ~\rho_{2D}(\vec{U},\nu)=1$. 

We now assume that the baseline distribution is circularly symmetric, i.e, it is only a function of $U=|\vec{U}|$. This assumption simplifies the calculations considerably and holds true for many cases of interest. The normalized baseline distribution $\rho_{2D}(U,\nu)$  can  be calculated for a given antenna distribution $\rho_{ant}(l)$ using the relation \cite{datta,petrovic,geil}
\be
\rho_{2D}(U,\nu)=B(\nu) \int_0^{\infty} 2 \pi l d l \, \rho_{\rm ant}(l) \int_0^{2\pi} d \phi~ \rho_{\rm ant}(|\vec{l}-\lambda \vec{U}|),
\label{eq:rhoant2rho2d}
\e
where $|\vec{l} - \lambda \vec{U}| = \sqrt{l^2 + \lambda^2 U^2 - 2 l \lambda U \cos \phi}$ and the constant $B(\nu)$ is determined from the above normalization condition for a given frequency $\nu$.

Now we consider a three-dimensional cell in momentum-space having coordinates between $k$ to $k+dk$ and $\theta$ to $\theta+d \theta$, with $\theta$ being the angle between $\vec{k}$ and the line of sight. Clearly, the perpendicular component is given by $k_{\perp}=k \sin \theta$ while the parallel component is $k_{\parallel}=k \cos \theta$. If we average the power spectrum  over all the modes which lie in the range $[k,k+dk]$ and $[\theta,\theta+d \theta]$,  the error in the power spectrum is reduced to 
\be
 \delta P_{N}(k,\theta)=\frac{T_{sys}^2}{B t_0} \left ( \frac{\lambda^2}{A_e}\right)^2 \frac{r_{\nu}^2 L}{[N(N-1)/2]\, \rho_{2D}(U)} \frac{1}{\sqrt{N_m(k,\theta)}}.
\label{eq:sigmap}
\e 
Note that we have omitted the frequency $\nu$ in the above expression, i.e., the frequency dependence is implicit. The quantity $N_m(k,\theta)$ is the number of independent modes between $[k, k+dk]$ and $[\theta, \theta+d \theta]$ and can be written as
\be
N_m(k,\theta)=\frac{2 \pi k^2 dk \sin \theta d \theta}{V_{\rm one-mode}},
\e
where 
\begin{equation}
V_{\rm one-mode}=\frac{(2 \pi)^3 A}{r_{\nu}^2 L \lambda^2}
\end{equation}
is the volume for a single independent mode in $k$-space.  

The noise error is a function of both $k$ and $\theta$, or equivalently, of both $k_{\perp}$ and $k_{\parallel}$. To calculate the noise error for the spherically averaged power spectrum we need to average over $\theta$ for a fixed $k$. Since  for a fixed $k$ the noise error varies with $\theta$, simple averaging is not optimum. We rather use an inverse-variance weighting scheme for averaging so as to minimize the noise error in the power spectrum. Under this scheme the noise error can be written as 
\be
\delta P_{N}(k)=\left [ \sum_{\theta}\frac{1}{\delta P^2_{N}(k, \theta)} \right]^{-1/2}~.
\label{eq:inv-variance}
\e

In the continuum limit, the system temperature error in the 21 cm power spectrum can be written as \cite{geil,mcquinn}
\be 
\delta P_{N}(k)=\frac{T_{sys}^2}{B t_0} \left ( \frac{\lambda^2}{A_e}\right)^2 \frac{r_{\nu}^2 L}{[N(N-1)/2]\, \rho_{3D}(k)} \frac{1}{\sqrt{N_k}},
\label{eq:sigmap2}
\e 
where $N_k$ is number of observable modes in the spherical shell between $k$ and $k+dk$ and can be calculated as
\be
N_k=\frac{2 \pi k^2 dk}{V_{\rm one-mode}}.
\e
The quantity $\rho_{3D}(k,\nu)$ is a distribution of measurements in 3D $k$-space which is related to the 2D normalized baseline distribution as
\be
 \rho_{3D}(k)=\left [ \int_{0}^{\pi/2} d\theta \sin \theta \, \rho_{2D}^2 \left ( \frac{r k}{2 \pi} \sin \theta \right ) \right ]^{1/2}.
\label{eq:2dto3d}
\e
In the above expression, we have replaced the baseline $U$ by $ r_{\nu} k \sin \theta/2 \pi$. Although Eq. \ref{eq:sigmap2} is an expression for the noise error in the continuum limit, it gives an accurate estimate of the noise error even when the measurements are in discrete points \cite{jensen13}.

Finally, the 21 cm power spectrum error arising from sample variance can be written as:
\be
\delta P_{SV}(k)=\left[ \sum_\theta \frac{N_m(k,\theta)}{P^2_{\rm 21cm}(k,\theta)} \right]^{-1/2}
\e
Notice that for convenience we have isolated both contributions to the total 21 cm power spectrum error (system temperature and sample variance) to investigate the scales at which each of them are most important.

\subsection{Detectability of the 21 cm signal}
\label{detectability}

We now address the detectability of the 21 cm power spectrum by the future SKA radio telescope. The SKA will have three different instruments working at different frequency bands and will be built in two phases. Here we consider two instruments called the SKA1-mid and SKA1-low which will be built in phase 1 in South Africa and Australia respectively. The SKA instrument specifications such as the antenna distribution, total number of antennae, total collecting area, frequency coverage, etc., have not been completely finalized yet and are possibly subject to change. Here we use the specifications  described in one of the most recent documents, i.e., the `Baseline Design Document'\footnote{\tt http://www.skatelescope.org/wp-content/uploads/2012/07/SKA-TEL-SKO-DD-001-1\_BaselineDesign1.pdf} which is available on the SKA website\footnote{\tt https://www.skatelescope.org/home/technicaldatainfo/key-documents/}. Let us briefly summarize the main properties of the two instruments which we have considered in this work:

{\bf SKA1-mid}: Based on the above document we assume that the SKA1-mid will cover a frequency range from $350$ MHz to $14$ GHz and will have a total of $250$ antennae of $15$ meters diameter each. This also includes $60$ antennae from the MeerKAT instrument. We use the baseline density given in the above document in (violet line in their Fig. 10) to calculate the normalized baseline distribution function $\rho_{2D}(U)$. Note that this baseline distribution is consistent with the proposed antenna distribution with $40\%$, $54\%$, $70\%$, $81\%$ and $100\%$ of the total antennae being within $0.4$ km, $1$ km, $2.5$ km, $4$ km and $100$ km radius respectively (see table 6 in the above document). The baseline density given in the document is essentially the total number of baselines measurements lying in the annulus, i.e., it is proportional to $U\rho_{2D}(U)$ for a fixed annulus size. We use the normalization condition $\int_0^{\infty} U d U \int_{0}^{\pi} d\phi  \rho_{2D}(\vec{U},\nu)=1$ to calculate the amplitude of the normalized  baseline distribution function $\rho_{2D}(U)$.  We then use Eq. \ref{eq:2dto3d} to compute $\rho_{3D}(U,\nu)$. We plug everything in Eq. \ref{eq:sigmap2} which gives the required noise error in the 21 cm power spectrum. We assume $T_{\rm sys}$ to be $30$K for the redshifts $z=2.4$ and $3$, corresponding to frequencies of $417.6$ MHz and $355$ MHz respectively. We also consider $100$ hours of observations over $32$ MHz bandwidth and a typical antenna efficiency equal to $0.7$. 

We show the system noise for SKA1-mid in the top panels of Fig. \ref{Pk_21cm}. On comparing the noise estimates with the amplitude of the 21 cm power spectra at $z = 2.4$ and $3.6$, we find that the SKA1-mid will be able to resolve the 21 cm power spectrum up to $k\sim 1 h~ {\rm Mpc}^{-1}$ with 100 hours of observations provided the HI distribution is modeled using the halo-based model 1 or the particle-based method. In case the HI distribution is modeled with the halo-based model 2 we find that the prospect of detecting the 21 cm power spectrum will be much better. One can detect it for relatively smaller scales: $k\sim 3 h~{\rm Mpc}^{-1}$, and even for larger scales, we expect to achieve a much better signal-to-noise ratio.

{\bf SKA1-low}: The proposed SKA1-low will operate in the frequency range $50-300$ MHz and will consist of $911$ stations with a diameter of $35$ meters each. The SKA1-low has total collecting area of $0.88 \, {\rm km^2}$ and thus is much more powerful compared to the SKA1-mid which has a collecting area of only $0.044 \, {\rm km^2}$. Although the maximum baseline for SKA1-low is expected to be $\sim 100$ km, most of the stations will be in the centre (see figure 3 \& 4 in the baseline design document). We find that the antenna distribution can be nicely approximated by a simple functional form $\rho_{ant}(l)=(A/l) \exp[-0.5(l/1000 \, m)^2]$ with no antenna within a $50$ m radius and $866$ stations within $5$ km radius. We ignore the larger baselines as their contribution to the sensitivity is negligible. We note however that the large baselines will be useful for accurate measurements of foreground sources and for removing them from the observed data. Given the antenna distribution, we can use Eq. \ref{eq:rhoant2rho2d} and the normalization condition to calculate the normalized baseline distribution function $\rho_{2D}(U)$. We then use the procedure described above to calculate the noise error in the power spectrum at redshift $z=4$ for SKA1-low. We assume $T_{\rm sys}$ to be $110$K for the redshift $z=4$, corresponding to a frequency of $284$ MHz. As before, we assume the antenna efficiency to be $0.7$. 

In the bottom-left panel of Fig. \ref{Pk_21cm} we show the SKA1-low system noise for $100$ hours of observations with a $32$ MHz bandwidth. We find that the 21 cm power spectrum will be detected by this telescope up to much smaller scales, i.e., $k\sim 5 h~{\rm Mpc}^{-1}$ for the halo-based 1 and particle-based models, while it can be detected up to $k\sim 20 h~{\rm Mpc}^{-1}$ for halo-based model 2. It is thus clear that SKA1-low, as per current specifications, would be quite sensitive for studying the HI power spectrum at $z \approx 4$. 

The different panels of Fig. \ref{Pk_21cm} also show the sample variance errors in the 21 cm power spectrum. On small scales, we find that errors arising from sample variance are much below those coming from the instrument noise, whereas they dominate on large scales. Although sample variance errors reduce the signal to noise ratio on which the 21 cm power spectrum can be determined on large scales, our conclusions do not change: the largest scales that we can probe with our simulations will be detected both by SKA1-mid and SKA1-low at the redshift studied here with 100 hours of observations.

Interestingly, for all the models studied in this paper, the 21 cm power spectra would be detectable at $k < 1 h~{\rm Mpc}^{-1}$ with $\sim 100$ hours of observations for a wide range of redshifts $2.4 < z < 4$ using the SKA1 telescopes. Since the signal at these scales essentially traces the dark matter fluctuations (see Fig. \ref{Pk_21cm}), it might be possible constrain cosmological parameters with future telescopes. We plan to study these issues in a different paper.

\section{Prospects of imaging}
\label{imaging}

In this section we investigate the prospects of imaging the HI distribution using radio telescopes. In general, the 21 cm signal in an average region of the Universe may not be strong enough for imaging, hence we focus on regions of very high densities where the concentration of HI is expected to be considerably larger than average. In particular we investigate whether SKA1-low and SKA1-mid might be able to detect a few individual bright HI peaks which we find in our simulations. 

We begin by creating brightness temperature maps from our simulated distribution of neutral hydrogen. The procedure used to produce these maps is as follows: Given a frequency channel $[\nu_0-\triangle \nu/2,~\nu_0+\triangle \nu/2]$ of width $\triangle \nu$, with $\nu_0=1420/(1+z)$ MHz, we take a slice of the $N$-body snapshot at redshift $z$ with a width equal to $L=r_{\nu_0-\triangle \nu/2}-r_{\nu_0+\triangle \nu/2}$, where $r_{\nu}$ is the comoving distance to redshift $z=(1420~{\rm MHz})/\nu-1$. We then divide the slice into $N_{\rm pixels}\times N_{\rm pixels}$ cells of equal volume, where $N_{\rm pixels}$ determines the resolution of the map. For each cell we compute the HI density and use Eq. \ref{delta_Tb} to calculate the brightness temperature excess within it. Next, we compute the specific intensity excess from the brightness temperature excess using the relation
\begin{equation}
I_\nu = \frac{2\nu^2}{c^2}k_B\delta{T_b},
\end{equation}
where $k_B$ is the Boltzmann constant and $c$ is the light speed. In order to account for instrumental resolution, we smooth the $I_\nu$ field with  a Gaussian window of angular radius $\theta_{\rm beam}$, with $\theta_{\rm beam}$ being the synthesized beam width. Since radio-interferometers are not sensitive to the mean value of the specific intensity and can only measure deviations from the mean, we correct the value of the specific intensity in each pixel by making the transformation $I_\nu \rightarrow I_\nu - \langle I_\nu \rangle$. We then multiply the $I_\nu$ map with the beam solid angle $\Delta \Omega_{\rm beam}=\theta^2_{\rm beam}$ which essentially gives the total flux within a single beam, i.e., the resultant image maps will be in units of flux per beam. We repeat the above procedure with different slices taken from the simulation box until we find the slice containing the highest value of $I_\nu$. 

\begin{figure}
\begin{center}
\includegraphics[width=1.0\textwidth]{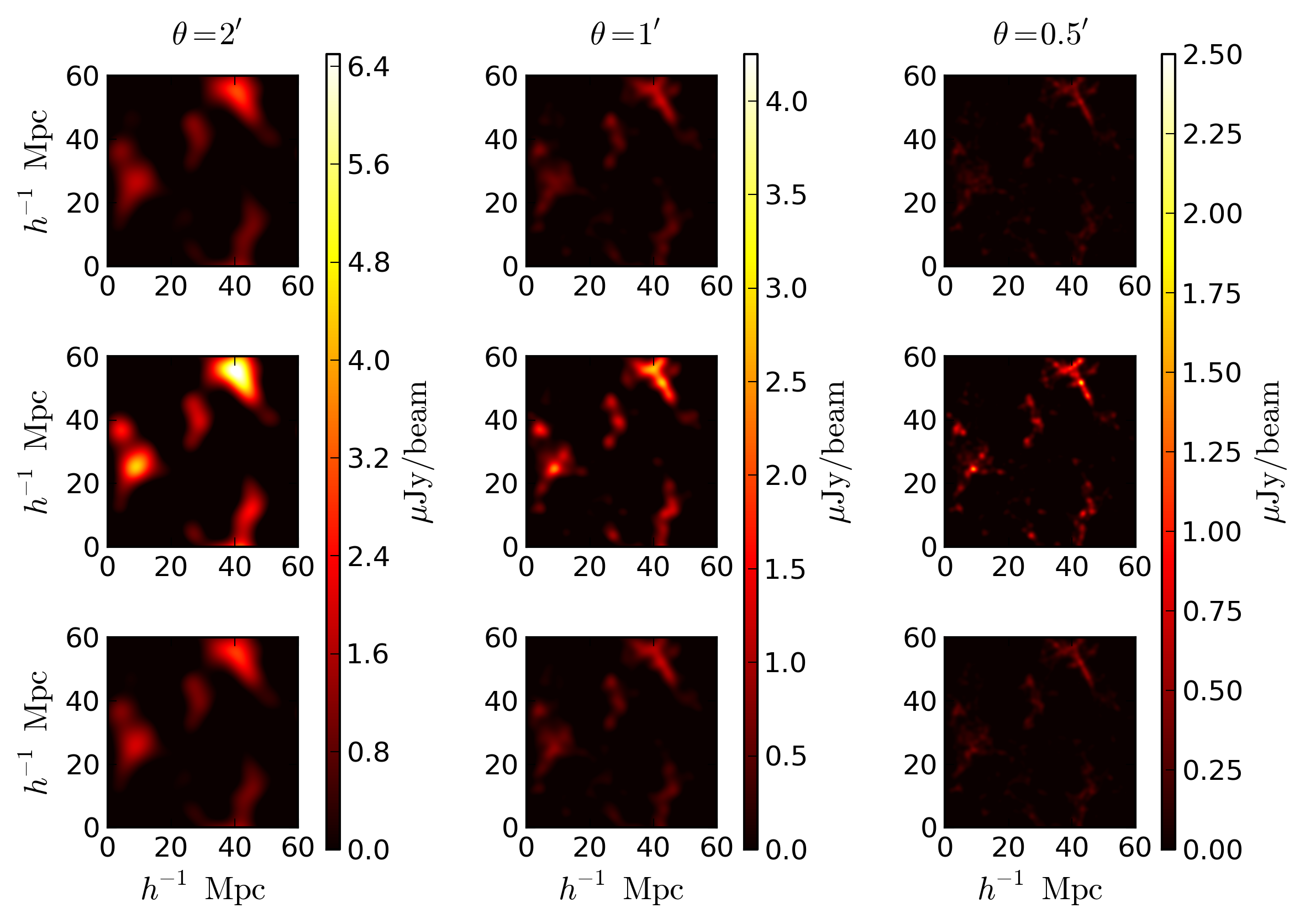}\\
\end{center}
\caption{Mock radio-maps, for a frequency channel of 500 kHz, created from the HI distribution obtained by using the halo-based model 1 (top row), the halo-based model 2 (middle row) and the particle-based method (bottom row) on top of the simulation $\mathcal{B}60$ at $z=3$. Three different synthesized beam widths has been used: $\theta_{\rm beam}=2'$ (left column), $\theta_{\rm beam}=1'$ (middle column) and $\theta_{\rm beam}=0.5'$ (right column). The maps do not include the system noise associated to radio-telescopes.}
\label{radio_maps}
\end{figure}

In Fig. \ref{radio_maps} we show mock radio-maps created using the above procedure where the HI distributions have been obtained using the halo-based and the particle-based methods on top of the simulation $\mathcal{B}60$ at $z=3$. The simulated maps are for a frequency channel width of 500 kHz and for three different synthesized beam widths of $2'$, $1'$ and $0.5'$. We should mention that these are not proper radio-maps as would be observed in telescopes, since we have used a very large number of pixels\footnote{Notice that the purpose of using a large number of pixels is just to show a smooth image rather than the pixelated one.} ($N_{\rm pixels}=1024$) and we have not included the system noise. In this figure, we only aim to show the specific intensity field that will be sampled by the radio-telescopes.

We find that the maps, for a given $\theta_{\rm beam}$, look quite similar. However, the peak amplitude obtained in the halo-based model 2 is significantly higher than the one found by using the other two methods (typically a factor $\sim2-3$). This is because the HI fraction in the large halos for halo-based model 2 is much higher than in the other two models. 

We now investigate whether isolated peaks can be directly imaged by the future SKA1-low and SKA1-mid radio-telescopes, i.e, whether there will be enough sensitivity in the maps to identify the cosmological HI. The noise r.m.s. per synthesized beam in radio images for two polarizations can be written as
\be
\Delta S_{\nu}= \frac{\sqrt{2}k_BT_{\rm sys}}{A_e\sqrt{N_b ~\triangle \nu~ t_0}},
\label{eq:noise-beam}
\e
where $N_b=N(N-1)/2$ is the total number of instantaneous baselines. We have already defined $\triangle \nu$ as the frequency channel width, $T_{\rm sys}$ as the instrument temperature and $t_0$ as the observation time. When the number of antennae is large ($N>>1$) the above equation can be written as
\be
\Delta S_{\nu}= 65.7 \, {\mu \rm Jy}\left (\frac{T_{\rm sys}}{100 \, K}\right)\left (\frac{0.7}{\epsilon}\right)\left (\frac{100 \, {\rm m}^2}{A}\right)\left (\frac{100}{N}\right)\left (\frac{1 \, {\rm MHz}}{\triangle\nu}\right)^{0.5}\left (\frac{100 \, {\rm hrs}}{t_0}\right)^{0.5}.
\label{eq:noise-beam1}
\e

\begin{figure}
\begin{center}
\includegraphics[width=0.7\textwidth]{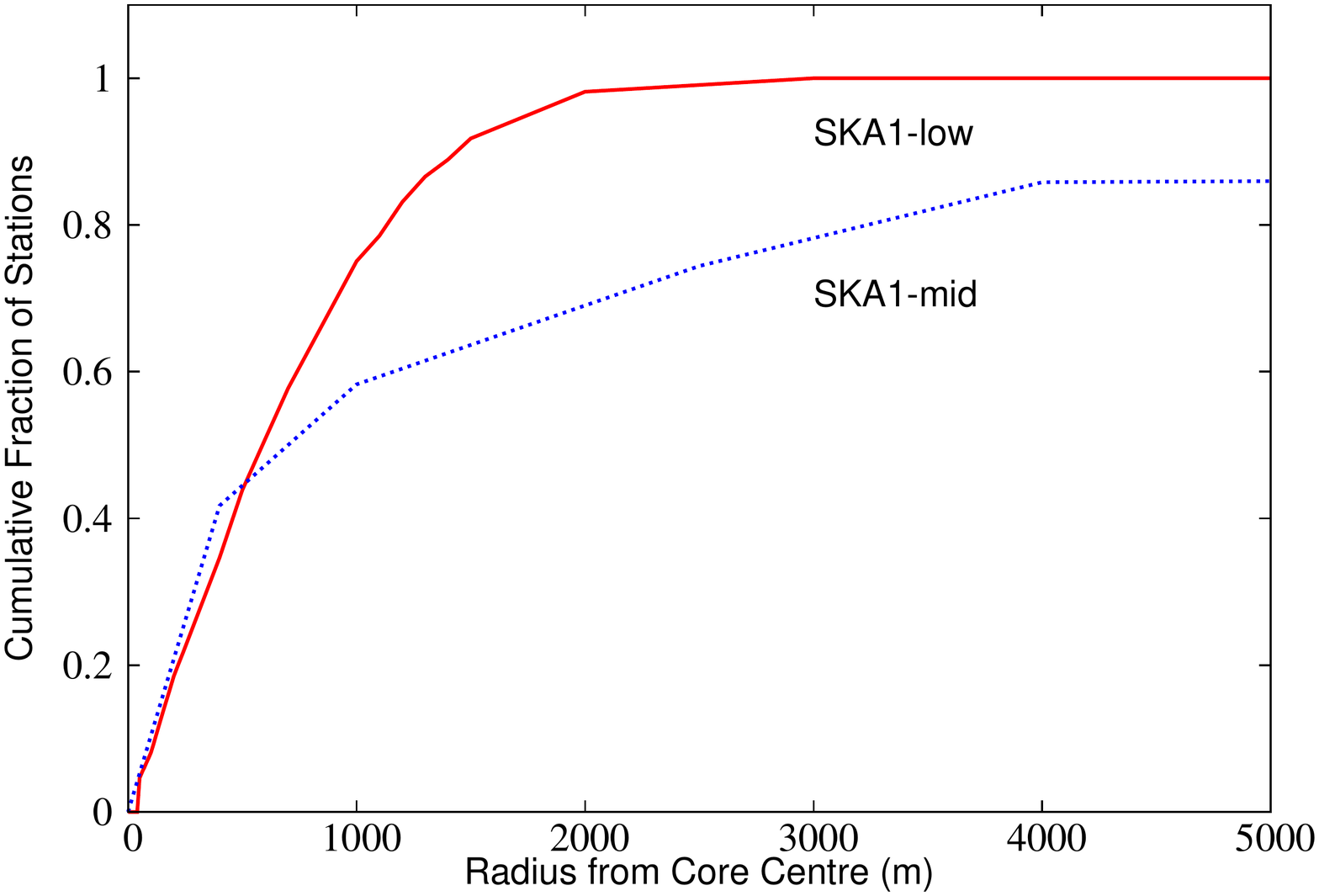}\\
\end{center}
\caption{The cumulative fraction of antennae (stations) as a function of distance from the core centre for the SKA1-mid and SKA1-low.}
\label{fig:cum_frac}
\end{figure}

We calculate the noise r.m.s. values for three different synthesized beams of $0.5'$, $1'$ and $2'$. As we will see later, the sensitivity for beam sizes smaller than the $0.5'$ is very poor and therefore we do not consider smaller beams. Higher beam sizes smooth out the HI signal in peaks, hence we avoid using higher beam sizes. To achieve a desired synthesized beam width, $\theta_{\rm beam}$, we consider antennae only from the central region which will provide baselines up to $U_{\rm max}=1/\theta_{\rm beam}$. We simply discard antennae outside the central region which do not contribute to baselines $U \leq U_{\rm max}$. For example, we need a maximum baseline $U_{\rm max} \sim 6879, \, 3440 $ and $1720$ to achieve the synthesized beam sizes of $0.5'$, $1'$ and $2'$ respectively.  Thus, at redshift $z=4$ (corresponding to the observing wavelength $\lambda=1.056$m) we take into account the antennae which are within  a circular region of diameter $7264$ m, $3632$ m and $1816$ m from the core centre. Fig. \ref{fig:cum_frac}  shows the cumulative fraction of antennae (stations) as a function of distance from the core centre for the SKA1-mid and SKA1-low. The cumulative fraction of antennae we use here is consistent with the antenna distribution given in the SKA baseline design document mentioned in the previous section (see table 6 and figure 3 in the document). We can calculate the total number of antennae lying within a given radius using Fig. \ref{fig:cum_frac}. For SKA1-low, at frequencies corresponding to $z=4$, there are $866$, $830$ and $600$ antennae within distances corresponding to synthesized beam widths of $0.5'$, $1'$ and $2'$ respectively. We note that we do not necessarily lose a large amount of sensitivity in the images while discarding antennae at large distances from the core. This is because the antenna distribution is highly condensed in the central region. The advantage in discarding antennae at large distances is that the noise calculation in images becomes considerably simpler. The same argument holds for the SKA1-mid at redshifts $z=2.4$ and $z=3$.  

\begin{table}[ht]
\caption{The brighest HI peak at redshift $z=4$ detection prospects for the SKA1-low with $1000$ hrs of observations} 
\centering 
\vspace{.2in}
\begin{tabular}{c rrrrrrrrr} 
\hline
\hline 
 Frequency  & \multicolumn{3}{c}{$\theta_{\rm beam}=0.5'$} & \multicolumn{3}{c}{$\theta_{\rm beam}=1'$} & \multicolumn{3}{c}{$\theta_{\rm beam}=2'$} \\ 
Channel& Peak flux & noise & SNR & Peak flux & noise & SNR & Peak flux & noise & SNR \\
(kHz)&($\mu$Jy) & ($\mu$Jy)  &  &($\mu$Jy) &($\mu$Jy)& &($\mu$Jy)&($\mu$Jy)&  \\
\\ [0.5ex]
\hline 
125 & 1.79 & 0.77 & 2.32 &    3.20 & 0.81 & 3.95 &    5.20 & 1.12 & 4.64 \\
\hline
500 & 1.34 & 0.39 & 3.44 &    2.57 & 0.40 & 6.42 &    4.19 & 0.56 & 7.48 \\
\hline
1000 & 0.71 & 0.27 & 2.63 &   1.40 & 0.29 & 4.83 &    2.47 & 0.40 & 6.18 \\
\hline 
\end{tabular}
\label{tab:ska1-low}
\end{table}

\begin{table}[ht]
\caption{The brighest HI peak at redshift $z=2.4$ detection prospects for the SKA1-mid with $1000$ hrs of observations} 
\centering 
\vspace{.2in}
\begin{tabular}{c rrrrrrrrr} 
\hline
\hline 
 Frequency  & \multicolumn{3}{c}{$\theta_{\rm beam}=0.5'$} & \multicolumn{3}{c}{$\theta_{\rm beam}=1'$} & \multicolumn{3}{c}{$\theta_{\rm beam}=2'$} \\ 
Channel& Peak flux & noise & SNR & Peak flux & noise & SNR & Peak flux & noise & SNR \\
(kHz)&($\mu$Jy) & ($\mu$Jy)  &  &($\mu$Jy) &($\mu$Jy)& &($\mu$Jy)&($\mu$Jy)&  \\
\\ [0.5ex]
\hline 
125 & 5.82 & 5.25 & 1.11 &    9.48 & 6.44 & 1.47 &    14.3 & 8.22 & 1.74 \\
\hline
500 & 4.79 & 2.62 & 1.83 &    7.93 & 3.22 & 2.46 &    12.1 & 4.11 & 2.94 \\
\hline
1000 & 3.32 & 1.85 & 1.79 &    5.43 & 2.28 & 2.38 &    8.57 & 2.90 & 2.54 \\
\hline 
\end{tabular}
\label{tab:ska1-mid}
\end{table}

Tables \ref{tab:ska1-low} and \ref{tab:ska1-mid} show the noise r.m.s. values for the SKA1-low at redshift $z=4$ and the SKA1-mid at $z=2.4$ for three different frequency channels for $1000$ hours of observations. Note that the time of observation is 10 times higher than what was used for calculating the noise in power spectra measurements. For the values of the parameters $T_{\rm sys}$, $A$, $\epsilon$, $N$ we simply use those quoted in the previous section.  We also quote the peak HI flux obtained from our simulations using the halo-based model 2 in tables \ref{tab:ska1-low} and \ref{tab:ska1-mid}. As we have seen above, the peak flux would be a factor $\sim 2-3$ lower for the other two methods.

We find that at $z=4$, for synthesized beam widths of $\theta_{\rm beam}=1'$ and $2'$, the SKA1-low will be able to detect the brightest HI peaks with a high signal to noise ratio (SNR$> 4.5$), while the SNR is relatively smaller $\lesssim 3$ for $\theta_{\rm beam} = 0.5'$. For the parameter values we have explored, it seems that SKA1-low will achieve the highest SNR for a $500$ kHz channel with a $2'$ synthesized beam, resulting in a SNR higher than 7. For higher frequency channel width the SNR starts to decline as the peak flux drops rapidly. This is because of the fact that the HI density around high density peaks follow the density profile of the halo and thus does not extend very far. 

\begin{figure}
\begin{center}
\includegraphics[width=1.0\textwidth]{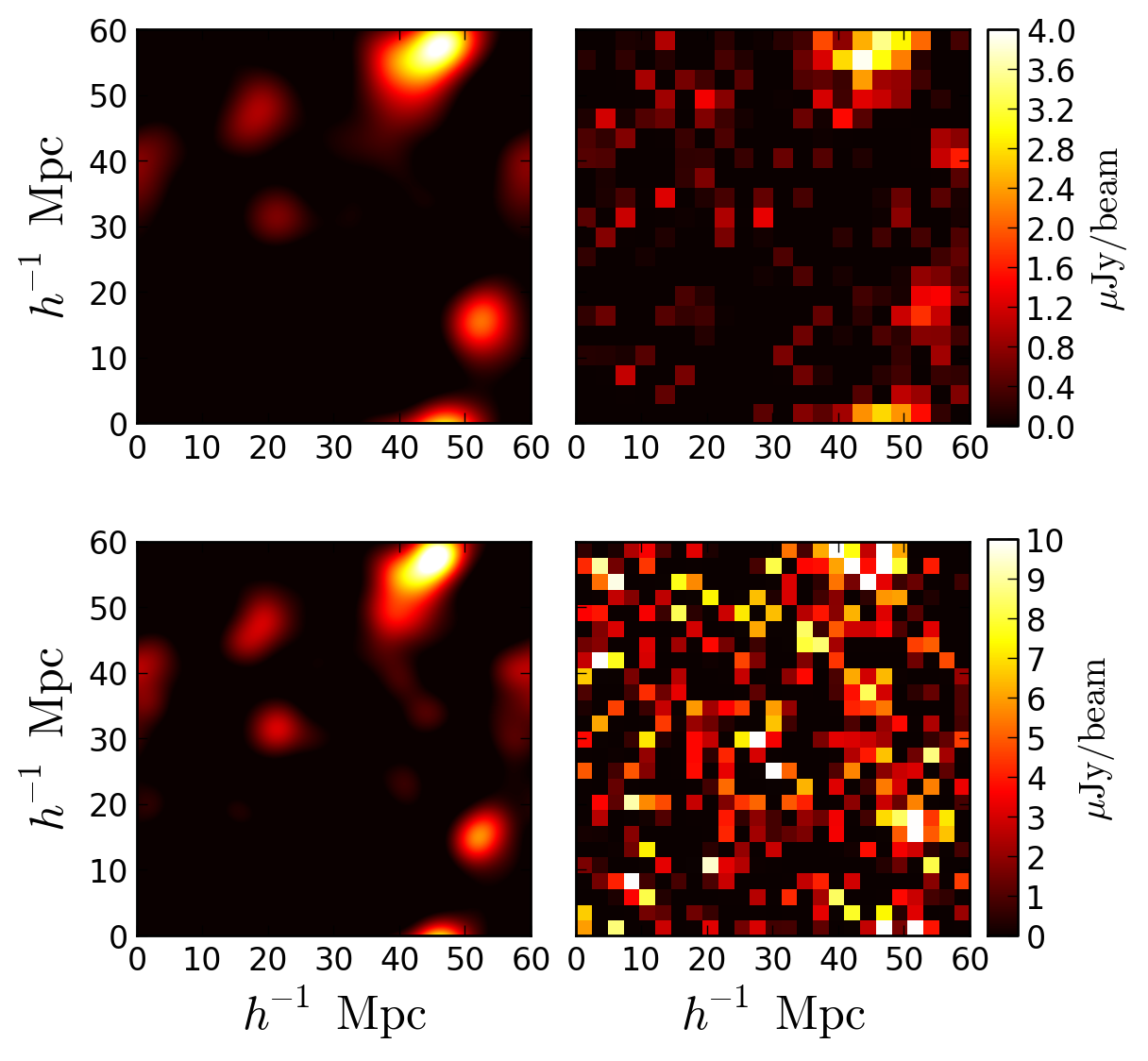}\\
\end{center}
\caption{Detectability of HI peaks in radio-maps. The top-left panel shows a radio-maps for a frequency channel of 500 kHz and for a synthesized beam width of $2'$ at z=4 obtained from the HI distribution generated by using the halo-based model 2 on top of the simulation $\mathcal{B}60$. In the top-right panel we show the same map but with a pixel size in correspondence with $\theta_{\rm beam}$ and with noise added for each pixel according to the noise r.m.s. The bottom panels display the same but at $z=2.4$.}
\label{radio_maps_noise}
\end{figure}

At lower redshift the SKA1-mid may be able to detect only the brightest peaks, albeit with a very low SNR. From our results we find that the most promising case would be to detect a peak using a synthesized beam width of $2'$ with a frequency channel width of 500 kHz. In this case, the signal to noise ratio would be close to 3. Hence, with the current specifications, the prospect of imaging cosmological HI with SKA1-mid at the redshifts studied in this paper is not very promising.

In order to visualize the difficulties in detecting peaks, we show in Fig. \ref{radio_maps_noise} radio-maps with instrumental noise included. The left hand panels in the figure show maps smoothed over the beam size, but using $1024 \times 1024$ pixels and without adding  any noise (i.e, these maps are similar to those presented in Fig. \ref{radio_maps}). In the right column panels show the same maps made using a pixel size in correspondence with the synthesized beam width and with noise added according to the noise r.m.s. value. The top panels are for $z = 4$ (i.e., SKA1-low) while the bottom panels are for $z = 2.4$ (SKA1-mid). The regions of peak flux, as is clear from the left panels, lie in the top-right corner of the maps. From the figure it seems that we can possibly detect the HI in the density peaks for SKA1-low as it looks quite distinct from the noise fluctuations. On the other hand, it will be extremely difficult to image the bright HI peak with SKA1-mid as the signal is virtually indistinguishable from the noise fluctuations.

We should remind the reader that the results presented in tables \ref{tab:ska1-low} and \ref{tab:ska1-mid} have been obtained by using the halo-based model 2. The other two models, i.e., the halo-based model 1 and the particle-based model, predict lower peak flux values and hence are much more difficult to image. On the other hand it is expected that the mass of the most massive halos at those redshifts would be significantly higher than the one used in this analysis, due to the small simulation volumes of our simulations. An interesting consequence of this calculation is that the detection of isolated bright HI peaks by SKA can provide some indication on the amount of HI present in very high density peaks.

\section{Summary and conclusions}
\label{Conclusions}

The aim of this paper is to model the neutral hydrogen content of the Universe and investigate its detectability by future radio telescopes like the SKA. We have run high-resolution hydrodynamical N-body simulations using the code {\sc GADGET-III} and we have modeled the distribution of HI using two different techniques: the \textit{halo-based method} and the \textit{particle-based method}.

The halo-based method is built on the assumption that all the HI in the Universe resides within dark matter halos. According to this scheme, the neutral hydrogen mass assigned to a particular gas particle residing in a given dark matter halo depends on the halo HI mass, $M_{\rm HI}(M)$ and on the density profile of the HI in the halo $\rho_{\rm HI}(r|M)$. For simplicity, we have neglected any dependence of the above quantities on the environment.
We have investigated two different models that rely on this methodology: the halo-based model 1 and the halo-based model 2. Each model uses a different $M_{\rm HI}(M)$ and $\rho_{\rm HI}(r|M)$ functions. In particular, the halo-model 1 uses a simple prescription for the function $M_{\rm HI}(M)$ \cite{Bagla_2010} based on the observations whereas the halo-based model 2 is constructed to reproduce the recent estimate of DLAs bias obtained by SDSS-III/BOSS \cite{Font_2012}. We find that both models reproduce very well the DLAs column density distribution, although they over-predict the abundance of Lyman Limit Systems. In terms of the HI power spectrum our results indicate that the amplitude of the HI power spectrum is significantly higher in the halo-based model 2. The reason is that model has been constructed to reproduce the DLAs bias measurements \cite{Font_2012} and therefore the HI in that model is much more strongly clustered than the HI spatial distribution obtained by employing the halo-based model 1. Overall, these two models, whose primarily difference is in the bias between DLAs and matter, bracket a conservative and physical range for assigning neutral hydrogen to dark matter halos.

In the particle-based method, we instead do not make any assumption on the location of the neutral hydrogen: HI (and H$_2$ for star forming particles) is assigned to all gas particles in the simulation according to their physical properties. In this scheme, we first compute the HI/H fraction associated to each gas particle assuming photo-ionization equilibrium with the external UV background. Next, the HI/H fraction is corrected to account for
self-shielding effects and finally, for star forming particles, a further correction to the HI/H fraction is carried out to account for molecular hydrogen. This method not only predicts the spatial distribution of HI but also its amount, i.e. $\Omega_{\rm HI}$: we find a typical value of $\Omega_{\rm HI}$ equal to $0.6\times10^{-3}$, about a $40\%$ smaller than the observational measurements ($\Omega_{\rm HI}\sim10^{-3}$). This method reproduces fairly well the abundance of LLS and DLAs even though it fails to reproduce the abundance of the absorbers with the highest column densities. In terms of the HI power spectrum the predictions of this model are very similar to those obtained by using the halo-based model 1. However, since the value of $\Omega_{\rm HI}$ predicted by this method is below the observational measurements, we conclude that the HI is more clustered in this model than in the halo-based model 1. We notice that this model is not able to reproduce the DLAs bias measurements of  \cite{Font_2012}  (bias is lower than in observations) and therefore the amplitude of the HI power spectrum in this model is below the one computed from the halo-based model 2.

By using the particle-based method we have investigated the contribution of HI outside dark matter halos (an environment that we denominate \textit{filaments}) to both the total amount of HI in the Universe and to the HI power spectrum. We find that the amount of HI in filaments contributes to a very small fraction of the overall HI content, $\Omega_{\rm HI}^{\rm filaments}\sim10^{-6}$, with that fraction increasing significantly with redshift. Our results also point out that the contribution of HI in filaments to the total HI power spectrum is negligible for the redshifts studied in this paper. 
We stress that our HI modeling  reproduces extremely well the abundance of absorbers in the Lyman-$\alpha$ forest.

From our simulated HI distribution we have computed the 21 cm power spectrum. We find that the same features we observe in the HI power spectrum are also present in the 21 cm power spectrum, i.e. the amplitude and shape of the 21 cm power spectrum extracted from the HI distribution obtained by using the halo-based model 1 and the particle-based, are very similar. However, the amplitude of the 21 cm power spectrum calculated from the HI distribution found by employing the halo-based model 2 is much higher than those predicted by two other methods, reflecting the fact that the HI is much strongly clustered in this model in comparison to the other two. In order to asses the detectability of such quantity,
we have computed the system noise for the future SKA1-mid and SKA1-low radio-telescopes. Our results indicate that with 100 hours of observations the 21 cm power spectrum will be detected by these instrument up to very small scales: $k\sim1-3~h{\rm Mpc}^{-1}$ at redshifts $z=2.4$ and $z=3$ and $k\sim5-20~h{\rm Mpc}^{-1}$ at redshift $z=4$, depending on the particular model used to simulate the HI distribution. Since the 21 cm power spectra for all the three models have the same slope at large scale $k \lesssim 1 h~{\rm Mpc}^{-1}$, it should also be possible to constrain the underlying dark matter power spectrum at large scales using the 21 cm observations, though detailed investigations on this aspect will be done in a future work.

Furthermore, we have investigated the possibility of directly imaging rare peaks in the HI distribution. We find that SKA1-low may detect the brightest peaks with a SNR higher than $\sim5$ for a synthesized beam width of $2'$ at $z=4$ for 1000 hours of observation, if the HI distribution is described by the halo-based model 2. In case the HI distribution is modeled using the other two methods the SNR drops by a factor of $\sim2-3$ due to the fact that the HI content in halos is less those models. At redshifts $z=2.4$ and $z=3$ our results suggest that directly imaging large HI peaks with SKA1-mid would be challenging. By using the halo-based model 2 we find that the most massive HI peaks can be detected with a SNR close to 3 at $z=2.4$ for 1000 hours of observations and for a synthesized beam width of $2'$. 

This work constitutes a first step in modeling the neutral hydrogen content in and outside halos by using semi-analytical recipes applied on top of hydrodynamical simulations and at the same time reproducing most of the relevant observational constraints.

\section*{Acknowledgements}
We thank Paramita Barai, Aritra Basu, Simeon Bird, Jayaram Changalur, Federica Govoni, Tae-sun Kim, 
and Narendra Nath Patra for useful discussions.  We thank Jamie Bolton for help with VPFIT and for having provided
us with the reference thermal IGM model.
Calculations were performed on SOM2 and SOM3 at IFIC and 
on the COSMOS Consortium supercomputer within the DiRAC 
Facility jointly funded by STFC, the Large Facilities Capital Fund of BIS and the
University of Cambridge, as well as the Darwin Supercomputer of the
University of Cambridge High Performance Computing Service (http://
www.hpc.cam.ac.uk/), provided by Dell Inc. using Strategic Research
Infrastructure Funding from the Higher Education Funding Council for
England.  FVN and MV are supported by the ERC Starting Grant
``cosmoIGM'' and partially supported by INFN IS PD51 "INDARK".
KKD thanks the Department of Science \& Technology (DST), India for the research grant SR/FTP/PS-119/2012 under the Fast Track Scheme for Young Scientist.
We acknowledge partial support from "Consorzio per la Fisica - Trieste".

\appendix

\section{Visual comparison of the HI distribution}
\label{HI_distribution_appendix}

\begin{figure}
\begin{center}
\includegraphics[width=1.0\textwidth]{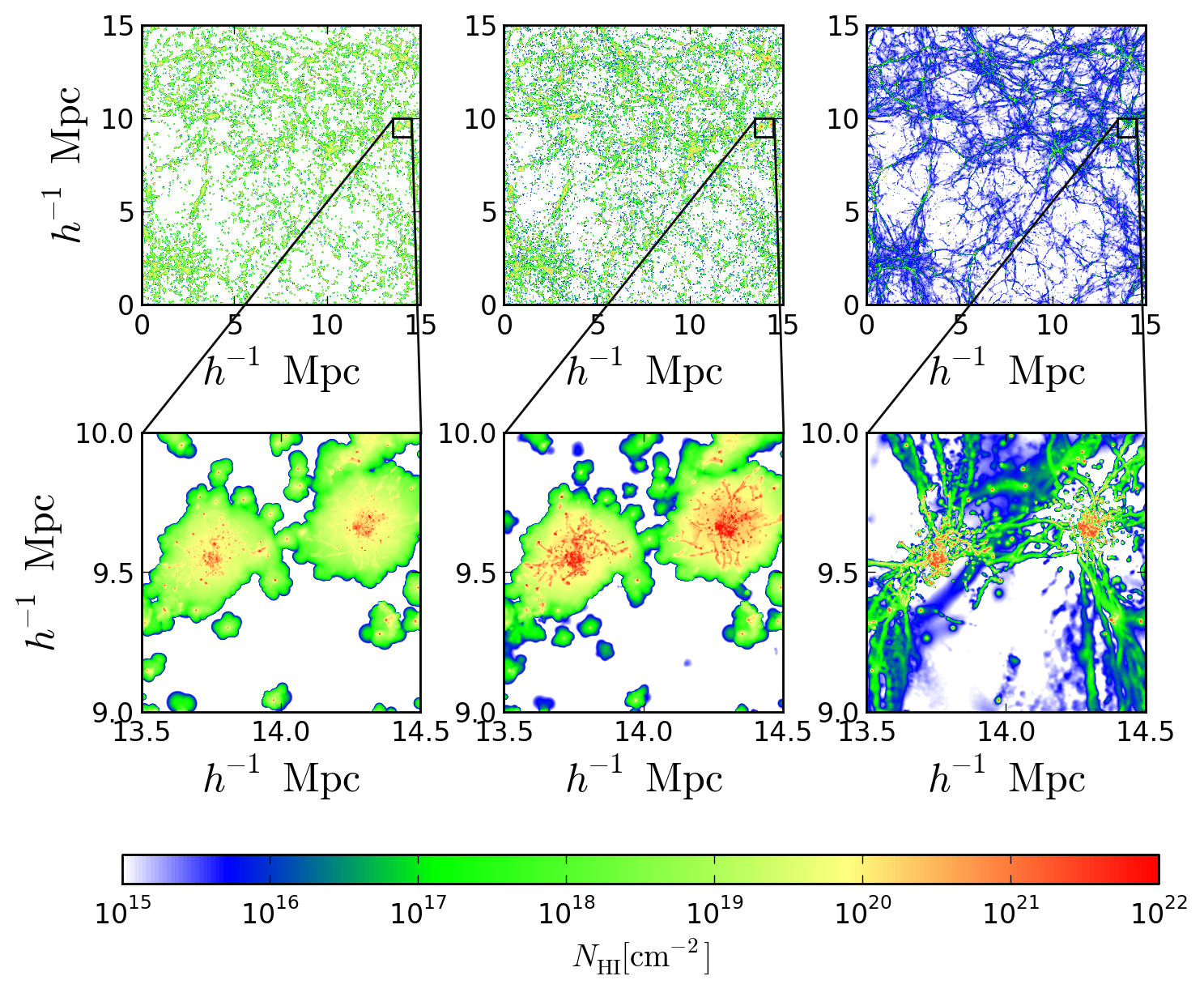}\\
\end{center}
\caption{Distribution of the HI column densities, along lines of sights that expand along the whole simulation box, obtained by using the halo-based model 1 (left column), the halo-based model 2 (middle column) and the particle-based model (right column) on top of the simulation $\mathcal{B}60$ at $z=3$. The top row shows the distribution of HI on large scales whereas the bottom row represents a zoom into the region marked with a black square into each panel of the top row.}
\label{N_HI_map}
\end{figure}

Here we show the distribution of neutral hydrogen, obtained by assigning HI to the gas particles of the simulation $\mathcal{B}60$ at $z=3$,  with the three methods investigated in this paper. In Fig. \ref{N_HI_map} we display the distribution of the column densities, computed along lines of sights that span
along the whole simulation box, when the halo-based model 1 (left column), the halo-based model 2 (middle column) and the particle based method is used. We find that the halo-based models over-predict the abundance of Lyman Limit Systems, whereas those are slightly under-predicted in the particle-based model. This is in agreement with the distribution of column densities from Figs. \ref{f_HI_Bagla}, \ref{column_density_Paco} and \ref{f_HI_Dave}.
The distribution of the DLAs is quite different among the models. Whereas the halo-based model 2 predicts a large cross-section for massive halos, this is saturated for massive halos in the halo-based model 1. The particle-based model is however in the intermediate situation, with a growing DLA cross-section with halo mass but with a lower amplitude than the one obtained from the halo-based model 2. We also stress the fact that the particle-based method also assign HI to particles outside halos (which give rise to the Lyman-$\alpha$ forest), that it is not present in the HI distribution generated by the other two models.

\section{Column density distribution: computation}
\label{column_density_appendix}

Here we describe in detail the method used to compute the column density distribution, $f_{\rm HI}$, from the N-body simulations. Once the neutral hydrogen has been assigned to the gas particles, using any of the methods investigated in this paper, the value of the column density along an arbitrary line of sight can be computed using the physical properties of the gas particles: mass, HI/H fraction and SPH smoothing length. 

We start by projecting all the gas particle positions onto the XY plane (we have checked that results do not change if the projection is performed onto a different plane). We then draw lines of sights
(LOS) from a regular grid in the XY plane that go from $Z=0$ to $Z=L$, where $Z$ is the cartesian coordinate, perpendicular to the XY plane, and $L$ is the size of the simulation box. For any given LOS we compute the minimum distance between any gas particle and the LOS, $b$. If that distance is smaller than the gas particle smoothing length, $h^i$, then we integrate its density along the path that the LOS intersects the physical size of the gas particle (see Fig. \ref{integration}):
\begin{figure}
\begin{center}
\includegraphics[width=0.3\textwidth]{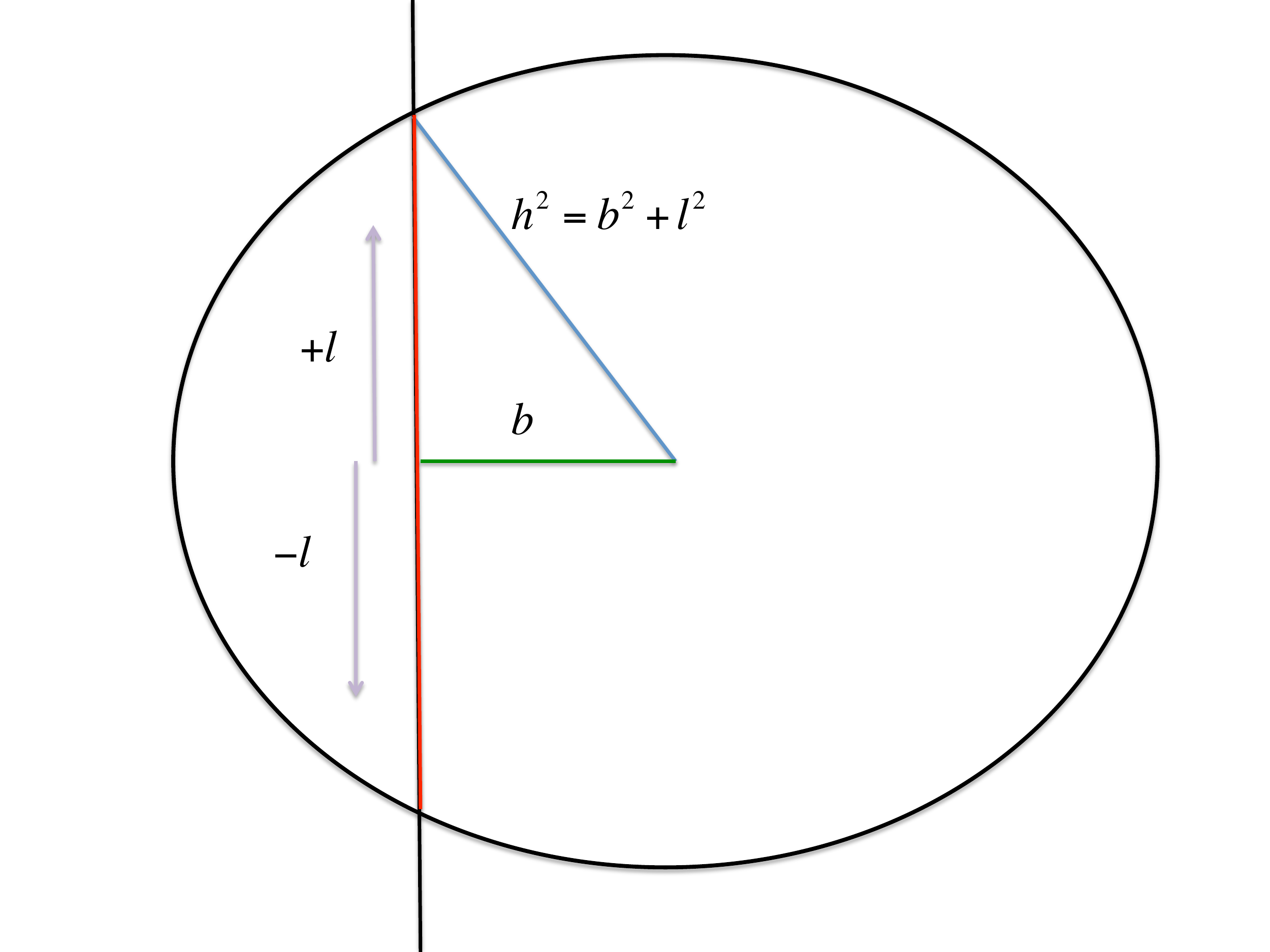}\\
\end{center}
\caption{Scheme showing a gas particle together with its boundary (circle black line). The SPH density has to be integrated along the orange line.}
\label{integration}
\end{figure}
\begin{equation}
N_{\rm HI}^i=\frac{0.76\left(\frac{\rm HI}{H}\right)^i}{m_H}\int_{-l_{max}}^{+l_{max}}\rho^i(r)dl=2\frac{0.76\left(\frac{\rm HI}{H}\right)^i}{m_H}m^i\int_{0}^{l_{max}}W(r,h^i)dl\, ,
\end{equation}
where $N_{\rm HI}^i$ is the column density due to the particle $i$, having mass $m^i$, neutral hydrogen fraction ${(\rm HI/H})^i$ and SPH smoothing length $h^i$. $m_H$ is the mass of the hydrogen atom. The relation between the integration variable $l$ and the radius $r$ is given by $r^2=b^2+l^2$, with $l_{max}^2=(h^i)^2-b^2$. For each LOS, we sum the $N_{\rm HI}$ of all the gas particles contributing to it. We repeat the procedure for all the LOS and finally we compute HI column density distribution function as:
\begin{equation}
f_{\rm HI}(N_{\rm HI})=\frac{d^2n(N_{\rm HI})}{dN_{\rm HI}dX}\,,
\end{equation}
where $n(N_{\rm HI})$ is the number of lines with column densities equal to $N_{\rm HI}$ and $dX=H_0(1+z)^2/H(z)dz$ is the absorption distance. 

Since we compute the column density along a LOS that spans along the whole box, by using this method we are implicitly assuming that the column density of a given LOS is due to a single absorber. We have tested the validity of this assumption by dividing the simulation box into $N$ slices of width $L/N$ and computing the column densities along a grid of LOS that spans within any of those slices. In other words, for given LOS that goes from $Z=0$ to $Z=L$, we compute the column density for $N$ LOS that span among $[Z=0,Z=L/N]$, $[Z=L/N,Z=2L/N]$ and so on.

\begin{figure}
\begin{center}
\includegraphics[width=0.49\textwidth]{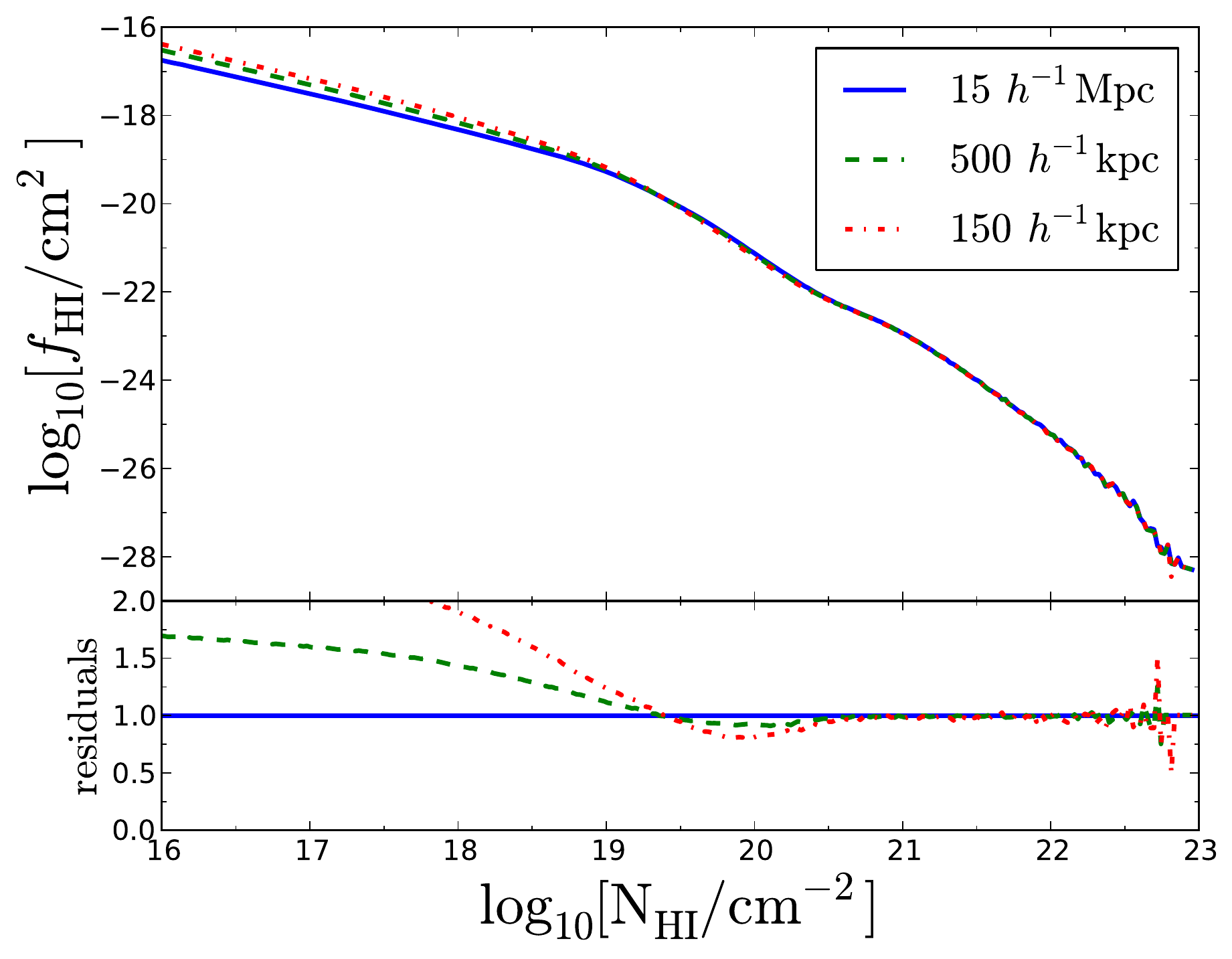}
\includegraphics[width=0.49\textwidth]{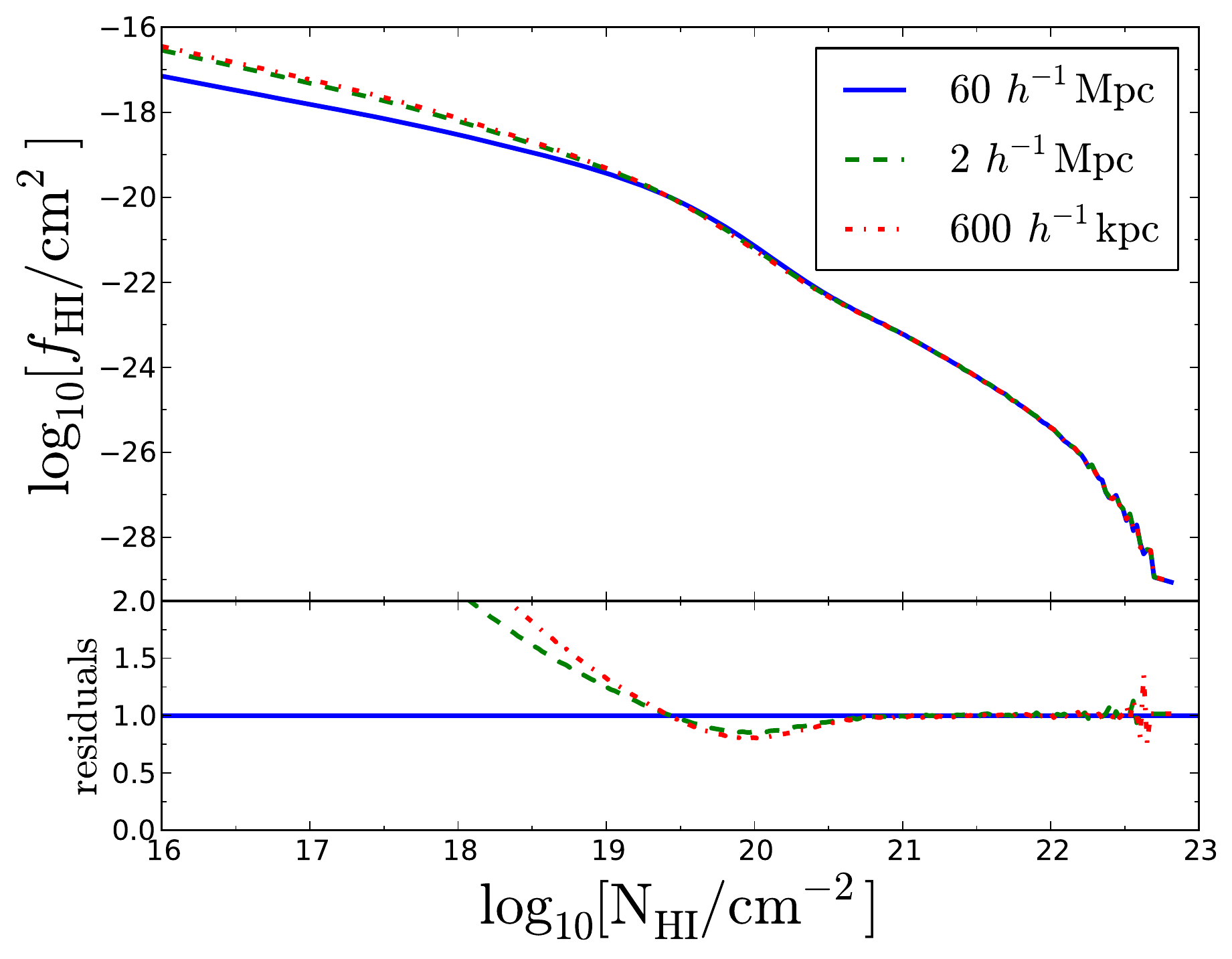}\\
\end{center}
\caption{Impact of the absorption distance on the column density distribution. We show the results of computing the column density distribution using lines of sights that span along the entire simulation box (solid blue lines) and line of sights that span along $1/30$ (dashed green) and $1/100$ (dash-dotted red) of the box size for the simulations $\mathcal{B}15$ (left) and $\mathcal{B}60$ (right) at $z=3$. The residuals are shown in the bottom panels. We have used the halo-based model 1 to assign the neutral hydrogen to the gas particles.}
\label{f_HI_convergence_test}
\end{figure}

The distribution of the column densities for the different values of $N$ are shown in Fig. \ref{f_HI_convergence_test} when the HI is assigned used the halo-based model 1. We find that for column densities larger than $\sim~10^{19}$ cm$^{-2}$ the column density distribution is insensitive to the absorption distance of the LOS used to compute the values of the column densities. We have explicitly checked that by assigning the HI using the other two methods the column density distribution is also converged above $\sim~10^{19}$ cm$^{-2}$.

We thus conclude that this method produces converged results for high column densities values ($N_{\rm HI}\gtrsim10^{19}$cm$^{-2}$). Notice that the advantage of this way of computing the column density distribution is that the calculations are very fast, allowing us to compute the column density of a very large number of LOS.

\section{Power spectrum: method}
\label{power_spectrum_method}

The power spectrum measurements presented in the paper have been obtained through the standard procedure of assigning the particle positions (or any other quantity associated to them as the HI mass) to a regular cubic grid using the Cloud-in-Cell (CIC) interpolation technique. By doing this, it is  implicitly assumed that each particle represents a cube of size $L/N_{\rm grid}$, with $L$ being the size of the simulation box and $N_{\rm part}$ the number of points along one axis in the grid, with a uniform interior density. However, the actual density profile of the gas particles it is not given by a uniform cube, but it is instead described by the SPH kernel (Eq. \ref{SPH_kernel}). In order to investigate how much our results are affected by the fact of having computed the power spectrum using the CIC, i.e. by ignoring the internal structure of the gas particles, we have calculated the power spectrum taking into account SPH kernel of
 the gas particles. 

\begin{figure}
\begin{center}
\includegraphics[width=0.6\textwidth]{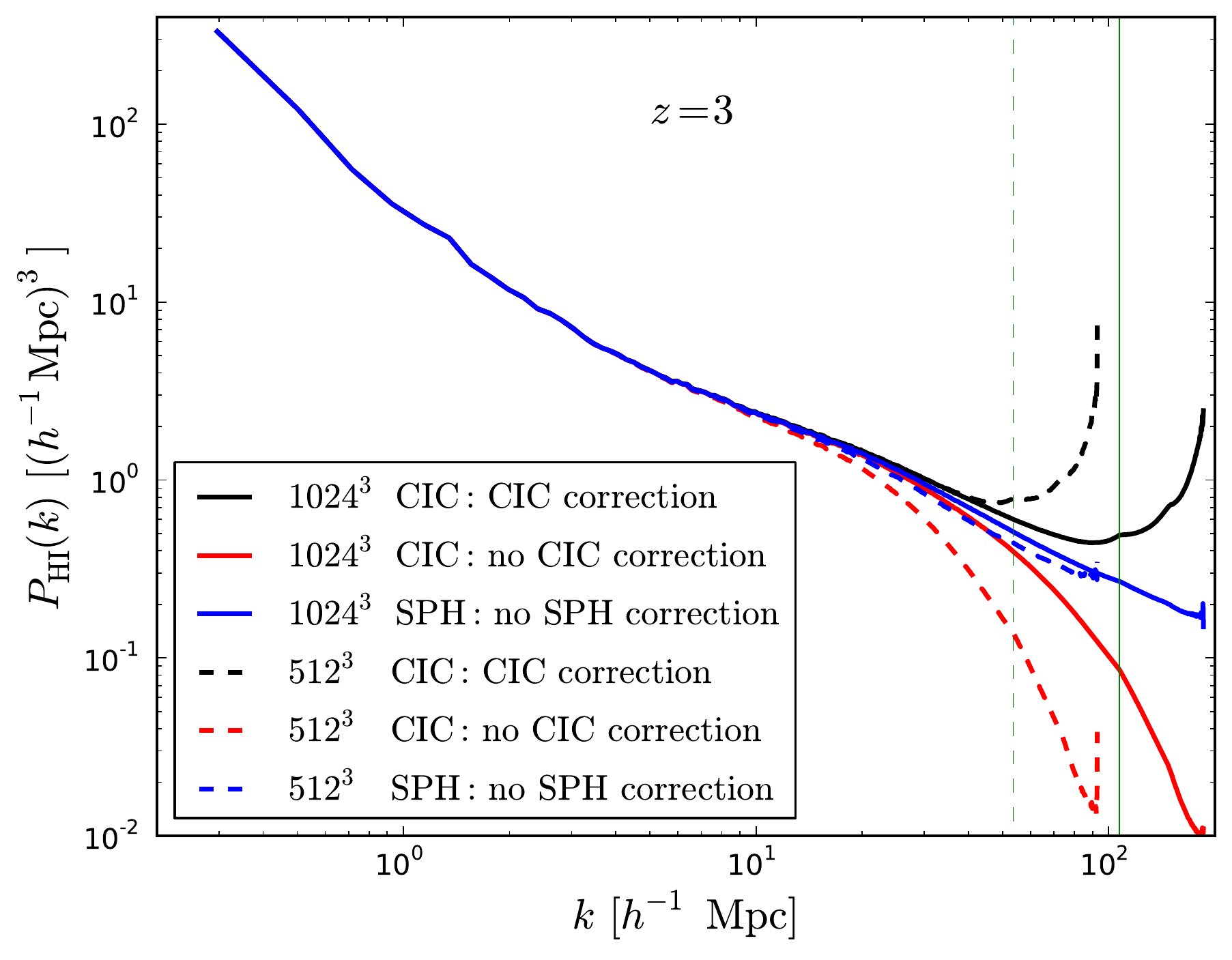}\\
\end{center}
\caption{HI power spectrum, obtained by using the halo-based model 1, for the simulation $\mathcal{B}30$ at $z=3$ using different methods. The black lines represent the results obtained using the standard procedure: interpolation through CIC, FFT, mode correction and measurement of the P(k). The red lines display the results of using the above procedure but without correct the modes amplitude to take into account the CIC assignment to the grid. The blue lines show the results of computing the power spectrum taking into account the SPH kernel of each gas particle (without mode correction). Two grids have been used: $1024^3$ (solid lines) and $512^3$ (dashed lines) points. The value of the Nyquist frequency for the two grids are represented by vertical lines.}
\label{Pk_HI_test}
\end{figure}

The procedure used is as follows: for each gas particle we select $N^3$ points (we have verified that with $N>7$ our results are converged) within the SPH smoothing length in such a way that each of them represent the same interior volume. We then assign each gas particle interior point to the grid cell that it belongs to. Finally, we follow the standard procedure of computing the FFT of the field obtained in that way and calculating the power spectrum $P(k)$. We have assigned HI to the gas particles of the simulation $\mathcal{B}30$ at $z=3$ using the halo-based model 1 and computed the HI power spectrum using the standard procedure (CIC) and the above one which takes into account the gas particles SPH kernel. We show the results in Fig. \ref{Pk_HI_test}. We find that the at large scales both procedures yield to identical results while at  smaller scales the results, as expected, differ. We notice that whereas we have corrected the amplitude of the modes to take into account the CIC assignment, we have not implemented that correction when using the SPH kernel of the gas particles (for details about this correction see \cite{Jing_2005}). In Fig. \ref{Pk_HI_test} the results obtained by using the CIC interpolation procedure but not correcting the modes amplitude are represented by the red curves.  It is clear that the mode correction plays a very important role when computing the power spectrum and thus, in order to quantify the scale at which both procedures begin to diverge it is needed to implement the mode correction to account for the SPH kernel. Unfortunately, the implementation of the amplitude mode correction to account for the SPH kernel is beyond the scope of this paper. We have explicitly checked that similar results are obtained by using different simulations and different HI assignments.

In summary, both procedures yield equivalent results on large scales. On small scales some differences show up, but in order to quantify the discrepancy a correction to the modes amplitude needs to be implemented. For the scales of interest in the present work, we can safely interpolate with CIC.

\bibliographystyle{JHEP}
\bibliography{Bibliography}

\providecommand{\href}[2]{#2}\begingroup\raggedright\begin{thebibliography}{10}

\bibitem{Peroux_2003}
C.~{P{\'e}roux}, R.~G. {McMahon}, L.~J. {Storrie-Lombardi}, and M.~J. {Irwin},
  {\it {The evolution of {$\Omega$}$_{HI}$ and the epoch of formation of damped
  Lyman {$\alpha$} absorbers}},  {\em \mnras} {\bf 346} (Dec., 2003)
  1103--1115, [\href{http://xxx.lanl.gov/abs/astro-ph/0107045}{{\tt
  astro-ph/0107045}}].

\bibitem{Zwaan_2005}
M.~A. {Zwaan}, M.~J. {Meyer}, L.~{Staveley-Smith}, and R.~L. {Webster}, {\it
  {The HIPASS catalogue: {$\Omega$}$_{HI}$ and environmental effects on the HI
  mass function of galaxies}},  {\em \mnras} {\bf 359} (May, 2005) L30--L34,
  [\href{http://xxx.lanl.gov/abs/astro-ph/0502257}{{\tt astro-ph/0502257}}].

\bibitem{Rao_2006}
S.~M. {Rao}, D.~A. {Turnshek}, and D.~B. {Nestor}, {\it {Damped Ly{$\alpha$}
  Systems at z{$\textless$}1.65: The Expanded Sloan Digital Sky Survey Hubble
  Space Telescope Sample}},  {\em \apj} {\bf 636} (Jan., 2006) 610--630,
  [\href{http://xxx.lanl.gov/abs/astro-ph/0509469}{{\tt astro-ph/0509469}}].

\bibitem{Lah_2007}
P.~{Lah}, J.~N. {Chengalur}, F.~H. {Briggs}, M.~{Colless}, R.~{de Propris},
  M.~B. {Pracy}, W.~J.~G. {de Blok}, S.~S. {Fujita}, M.~{Ajiki}, Y.~{Shioya},
  T.~{Nagao}, T.~{Murayama}, Y.~{Taniguchi}, M.~{Yagi}, and S.~{Okamura}, {\it
  {The HI content of star-forming galaxies at z = 0.24}},  {\em \mnras} {\bf
  376} (Apr., 2007) 1357--1366,
  [\href{http://xxx.lanl.gov/abs/astro-ph/0701668}{{\tt astro-ph/0701668}}].

\bibitem{Martin_2010}
A.~M. {Martin}, E.~{Papastergis}, R.~{Giovanelli}, M.~P. {Haynes}, C.~M.
  {Springob}, and S.~{Stierwalt}, {\it {The Arecibo Legacy Fast ALFA Survey. X.
  The H I Mass Function and {$\Omega$}\_H I from the 40\% ALFALFA Survey}},
  {\em \apj} {\bf 723} (Nov., 2010) 1359--1374,
  [\href{http://xxx.lanl.gov/abs/1008.5107}{{\tt arXiv:1008.5107}}].

\bibitem{Braun_2012}
R.~{Braun}, {\it {Cosmological Evolution of Atomic Gas and Implications for 21
  cm H I Absorption}},  {\em \apj} {\bf 749} (Apr., 2012) 87,
  [\href{http://xxx.lanl.gov/abs/1202.1840}{{\tt arXiv:1202.1840}}].

\bibitem{Noterdaeme_2012}
P.~{Noterdaeme}, P.~{Petitjean}, W.~C. {Carithers}, I.~{P{\^a}ris},
  A.~{Font-Ribera}, S.~{Bailey}, E.~{Aubourg}, D.~{Bizyaev}, G.~{Ebelke},
  H.~{Finley}, J.~{Ge}, E.~{Malanushenko}, V.~{Malanushenko},
  J.~{Miralda-Escud{\'e}}, A.~D. {Myers}, D.~{Oravetz}, K.~{Pan}, M.~M.
  {Pieri}, N.~P. {Ross}, D.~P. {Schneider}, A.~{Simmons}, and D.~G. {York},
  {\it {Column density distribution and cosmological mass density of neutral
  gas: Sloan Digital Sky Survey-III Data Release 9}},  {\em \aap} {\bf 547}
  (Nov., 2012) L1, [\href{http://xxx.lanl.gov/abs/1210.1213}{{\tt
  arXiv:1210.1213}}].

\bibitem{Zafar_2013}
T.~{Zafar}, C.~{P{\'e}roux}, A.~{Popping}, B.~{Milliard}, J.-M. {Deharveng},
  and S.~{Frank}, {\it {The ESO UVES advanced data products quasar sample. II.
  Cosmological evolution of the neutral gas mass density}},  {\em \aap} {\bf
  556} (Aug., 2013) A141, [\href{http://xxx.lanl.gov/abs/1307.0602}{{\tt
  arXiv:1307.0602}}].

\bibitem{Furlanetto_2006}
S.~R. {Furlanetto}, S.~P. {Oh}, and F.~H. {Briggs}, {\it {Cosmology at low
  frequencies: The 21 cm transition and the high-redshift Universe}},  {\em
  \physrep} {\bf 433} (Oct., 2006) 181--301,
  [\href{http://xxx.lanl.gov/abs/astro-ph/0608032}{{\tt astro-ph/0608032}}].

\bibitem{Morales_2009}
M.~F. {Morales} and J.~S.~B. {Wyithe}, {\it {Reionization and Cosmology with
  21-cm Fluctuations}},  {\em \araa} {\bf 48} (Sept., 2010) 127--171,
  [\href{http://xxx.lanl.gov/abs/0910.3010}{{\tt arXiv:0910.3010}}].

\bibitem{Bharadwaj_2001A}
S.~{Bharadwaj}, B.~B. {Nath}, and S.~K. {Sethi}, {\it {Using HI to Probe Large
  Scale Structures at z \~{} 3}},  {\em Journal of Astrophysics and Astronomy}
  {\bf 22} (Mar., 2001) 21,
  [\href{http://xxx.lanl.gov/abs/astro-ph/0003200}{{\tt astro-ph/0003200}}].

\bibitem{Bharadwaj_2001B}
S.~{Bharadwaj} and S.~K. {Sethi}, {\it {HI Fluctuations at Large Redshifts:
  I--Visibility correlation}},  {\em Journal of Astrophysics and Astronomy}
  {\bf 22} (Dec., 2001) 293--307,
  [\href{http://xxx.lanl.gov/abs/astro-ph/0203269}{{\tt astro-ph/0203269}}].

\bibitem{Chang_2008}
T.-C. {Chang}, U.-L. {Pen}, J.~B. {Peterson}, and P.~{McDonald}, {\it {Baryon
  Acoustic Oscillation Intensity Mapping of Dark Energy}},  {\em Physical
  Review Letters} {\bf 100} (Mar., 2008) 091303,
  [\href{http://xxx.lanl.gov/abs/0709.3672}{{\tt arXiv:0709.3672}}].

\bibitem{Loeb_Wyithe_2008}
A.~{Loeb} and J.~S.~B. {Wyithe}, {\it {Possibility of Precise Measurement of
  the Cosmological Power Spectrum with a Dedicated Survey of 21cm Emission
  after Reionization}},  {\em Physical Review Letters} {\bf 100} (Apr., 2008)
  161301, [\href{http://xxx.lanl.gov/abs/0801.1677}{{\tt arXiv:0801.1677}}].

\bibitem{Zaldarriaga_2004}
M.~{Zaldarriaga}, S.~R. {Furlanetto}, and L.~{Hernquist}, {\it {21 Centimeter
  Fluctuations from Cosmic Gas at High Redshifts}},  {\em \apj} {\bf 608}
  (June, 2004) 622--635, [\href{http://xxx.lanl.gov/abs/astro-ph/0311514}{{\tt
  astro-ph/0311514}}].

\bibitem{Datta_2007}
K.~K. {Datta}, T.~R. {Choudhury}, and S.~{Bharadwaj}, {\it {The multifrequency
  angular power spectrum of the epoch of reionization 21-cm signal}},  {\em
  \mnras} {\bf 378} (June, 2007) 119--128,
  [\href{http://xxx.lanl.gov/abs/astro-ph/0605546}{{\tt astro-ph/0605546}}].

\bibitem{Pritchard_2011}
J.~R. {Pritchard} and A.~{Loeb}, {\it {21 cm cosmology in the 21st century}},
  {\em Reports on Progress in Physics} {\bf 75} (Aug., 2012) 086901,
  [\href{http://xxx.lanl.gov/abs/1109.6012}{{\tt arXiv:1109.6012}}].

\bibitem{Wyithe_2008}
J.~S.~B. {Wyithe}, A.~{Loeb}, and P.~M. {Geil}, {\it {Baryonic acoustic
  oscillations in 21-cm emission: a probe of dark energy out to high
  redshifts}},  {\em \mnras} {\bf 383} (Jan., 2008) 1195--1209.

\bibitem{Camera_2013}
S.~{Camera}, M.~G. {Santos}, P.~G. {Ferreira}, and L.~{Ferramacho}, {\it
  {Cosmology on Ultralarge Scales with Intensity Mapping of the Neutral
  Hydrogen 21 cm Emission: Limits on Primordial Non-Gaussianity}},  {\em
  Physical Review Letters} {\bf 111} (Oct., 2013) 171302,
  [\href{http://xxx.lanl.gov/abs/1305.6928}{{\tt arXiv:1305.6928}}].

\bibitem{Bull_2014}
P.~{Bull}, P.~G. {Ferreira}, P.~{Patel}, and M.~G. {Santos}, {\it {Late-time
  cosmology with 21cm intensity mapping experiments}},  {\em ArXiv e-prints}
  (May, 2014) [\href{http://xxx.lanl.gov/abs/1405.1452}{{\tt
  arXiv:1405.1452}}].

\bibitem{Ooty}
S.~{Saiyad Ali} and S.~{Bharadwaj}, {\it {Prospects for detecting the 326.5 MHz
  redshifted 21 cm HI signal with the Ooty Radio Telescope (ORT)}},  {\em ArXiv
  e-prints} (Oct., 2013) [\href{http://xxx.lanl.gov/abs/1310.1707}{{\tt
  arXiv:1310.1707}}].

\bibitem{Chang_2010}
T.-C. {Chang}, U.-L. {Pen}, K.~{Bandura}, and J.~B. {Peterson}, {\it {Hydrogen
  21-cm Intensity Mapping at redshift 0.8}},  {\em ArXiv e-prints} (July, 2010)
  [\href{http://xxx.lanl.gov/abs/1007.3709}{{\tt arXiv:1007.3709}}].

\bibitem{Masui_2013}
K.~W. {Masui}, E.~R. {Switzer}, N.~{Banavar}, K.~{Bandura}, C.~{Blake}, L.-M.
  {Calin}, T.-C. {Chang}, X.~{Chen}, Y.-C. {Li}, Y.-W. {Liao}, A.~{Natarajan},
  U.-L. {Pen}, J.~B. {Peterson}, J.~R. {Shaw}, and T.~C. {Voytek}, {\it
  {Measurement of 21 cm Brightness Fluctuations at z \~{} 0.8 in
  Cross-correlation}},  {\em \apjl} {\bf 763} (Jan., 2013) L20,
  [\href{http://xxx.lanl.gov/abs/1208.0331}{{\tt arXiv:1208.0331}}].

\bibitem{Ghosh_2010}
A.~{Ghosh}, S.~{Bharadwaj}, S.~S. {Ali}, and J.~N. {Chengalur}, {\it {GMRT
  observation towards detecting the post-reionization 21-cm signal}},  {\em
  \mnras} {\bf 411} (Mar., 2011) 2426--2438,
  [\href{http://xxx.lanl.gov/abs/1010.4489}{{\tt arXiv:1010.4489}}].

\bibitem{Ghosh_2011}
A.~{Ghosh}, S.~{Bharadwaj}, S.~S. {Ali}, and J.~N. {Chengalur}, {\it {Improved
  foreground removal in GMRT 610 MHz observations towards redshifted 21-cm
  tomography}},  {\em \mnras} {\bf 418} (Dec., 2011) 2584--2589,
  [\href{http://xxx.lanl.gov/abs/1108.3707}{{\tt arXiv:1108.3707}}].

\bibitem{Bharadwaj_2004}
S.~{Bharadwaj} and P.~S. {Srikant}, {\it {HI Fluctuations at Large Redshifts:
  III - Simulating the Signal Expected at GMRT}},  {\em Journal of Astrophysics
  and Astronomy} {\bf 25} (Mar., 2004) 67,
  [\href{http://xxx.lanl.gov/abs/astro-ph/0402262}{{\tt astro-ph/0402262}}].

\bibitem{Nagamine_2004}
K.~{Nagamine}, V.~{Springel}, and L.~{Hernquist}, {\it {Abundance of damped
  Lyman {$\alpha$} absorbers in cosmological smoothed particle hydrodynamics
  simulations}},  {\em \mnras} {\bf 348} (Feb., 2004) 421--434,
  [\href{http://xxx.lanl.gov/abs/astro-ph/0302187}{{\tt astro-ph/0302187}}].

\bibitem{Kohler_2006}
K.~{Kohler} and N.~Y. {Gnedin}, {\it {Lyman Limit Systems in Cosmological
  Simulations}},  {\em \apj} {\bf 655} (Feb., 2007) 685--690,
  [\href{http://xxx.lanl.gov/abs/astro-ph/0605032}{{\tt astro-ph/0605032}}].

\bibitem{Pontzen_2008}
A.~{Pontzen}, F.~{Governato}, M.~{Pettini}, C.~M. {Booth}, G.~{Stinson},
  J.~{Wadsley}, A.~{Brooks}, T.~{Quinn}, and M.~{Haehnelt}, {\it {Damped Lyman
  {$\alpha$} systems in galaxy formation simulations}},  {\em \mnras} {\bf 390}
  (Nov., 2008) 1349--1371, [\href{http://xxx.lanl.gov/abs/0804.4474}{{\tt
  arXiv:0804.4474}}].

\bibitem{Razoumov_2008}
A.~O. {Razoumov}, M.~L. {Norman}, J.~X. {Prochaska}, J.~{Sommer-Larsen}, A.~M.
  {Wolfe}, and Y.-J. {Yang}, {\it {Can Gravitational Infall Energy Lead to the
  Observed Velocity Dispersion in DLAs?}},  {\em \apj} {\bf 683} (Aug., 2008)
  149--160, [\href{http://xxx.lanl.gov/abs/0710.4137}{{\tt arXiv:0710.4137}}].

\bibitem{Tescari_2009}
E.~{Tescari}, M.~{Viel}, L.~{Tornatore}, and S.~{Borgani}, {\it {Damped Lyman
  {$\alpha$} systems in high-resolution hydrodynamical simulations}},  {\em
  \mnras} {\bf 397} (July, 2009) 411--430,
  [\href{http://xxx.lanl.gov/abs/0904.3545}{{\tt arXiv:0904.3545}}].

\bibitem{Altay_2010}
G.~{Altay}, T.~{Theuns}, J.~{Schaye}, N.~H.~M. {Crighton}, and C.~{Dalla
  Vecchia}, {\it {Through Thick and Thin-H I Absorption in Cosmological
  Simulations}},  {\em \apjl} {\bf 737} (Aug., 2011) L37,
  [\href{http://xxx.lanl.gov/abs/1012.4014}{{\tt arXiv:1012.4014}}].

\bibitem{Popping_2009}
A.~{Popping}, R.~{Dav{\'e}}, R.~{Braun}, and B.~D. {Oppenheimer}, {\it {The
  simulated H I sky at low redshift}},  {\em \aap} {\bf 504} (Sept., 2009)
  15--32, [\href{http://xxx.lanl.gov/abs/0906.3067}{{\tt arXiv:0906.3067}}].

\bibitem{Bagla_2010}
J.~S. {Bagla}, N.~{Khandai}, and K.~K. {Datta}, {\it {HI as a probe of the
  large-scale structure in the post-reionization universe}},  {\em \mnras} {\bf
  407} (Sept., 2010) 567--580, [\href{http://xxx.lanl.gov/abs/0908.3796}{{\tt
  arXiv:0908.3796}}].

\bibitem{Nagamine_2010}
K.~{Nagamine}, J.-H. {Choi}, and H.~{Yajima}, {\it {Effects of Ultraviolet
  Background and Local Stellar Radiation on the H I Column Density
  Distribution}},  {\em \apjl} {\bf 725} (Dec., 2010) L219--L222,
  [\href{http://xxx.lanl.gov/abs/1006.5345}{{\tt arXiv:1006.5345}}].

\bibitem{McQuinn_2011}
M.~{McQuinn}, S.~P. {Oh}, and C.-A. {Faucher-Gigu{\`e}re}, {\it {On Lyman-limit
  Systems and the Evolution of the Intergalactic Ionizing Background}},  {\em
  \apj} {\bf 743} (Dec., 2011) 82,
  [\href{http://xxx.lanl.gov/abs/1101.1964}{{\tt arXiv:1101.1964}}].

\bibitem{Fumagalli_2011}
M.~{Fumagalli}, J.~X. {Prochaska}, D.~{Kasen}, A.~{Dekel}, D.~{Ceverino}, and
  J.~R. {Primack}, {\it {Absorption-line systems in simulated galaxies fed by
  cold streams}},  {\em \mnras} {\bf 418} (Dec., 2011) 1796--1821,
  [\href{http://xxx.lanl.gov/abs/1103.2130}{{\tt arXiv:1103.2130}}].

\bibitem{Duffy_2012}
A.~R. {Duffy}, S.~T. {Kay}, R.~A. {Battye}, C.~M. {Booth}, C.~{Dalla Vecchia},
  and J.~{Schaye}, {\it {Modelling neutral hydrogen in galaxies using
  cosmological hydrodynamical simulations}},  {\em \mnras} {\bf 420} (Mar.,
  2012) 2799--2818, [\href{http://xxx.lanl.gov/abs/1107.3720}{{\tt
  arXiv:1107.3720}}].

\bibitem{Bird_2013}
S.~{Bird}, M.~{Vogelsberger}, D.~{Sijacki}, M.~{Zaldarriaga}, V.~{Springel},
  and L.~{Hernquist}, {\it {Moving-mesh cosmology: properties of neutral
  hydrogen in absorption}},  {\em \mnras} {\bf 429} (Mar., 2013) 3341--3352,
  [\href{http://xxx.lanl.gov/abs/1209.2118}{{\tt arXiv:1209.2118}}].

\bibitem{Rahmati_2013}
A.~{Rahmati}, A.~H. {Pawlik}, M.~{Raicevic}, and J.~{Schaye}, {\it {On the
  evolution of the H I column density distribution in cosmological
  simulations}},  {\em \mnras} {\bf 430} (Apr., 2013) 2427--2445,
  [\href{http://xxx.lanl.gov/abs/1210.7808}{{\tt arXiv:1210.7808}}].

\bibitem{Dave_2013}
R.~{Dav{\'e}}, N.~{Katz}, B.~D. {Oppenheimer}, J.~A. {Kollmeier}, and D.~H.
  {Weinberg}, {\it {The neutral hydrogen content of galaxies in cosmological
  hydrodynamic simulations}},  {\em \mnras} {\bf 434} (Sept., 2013) 2645--2663,
  [\href{http://xxx.lanl.gov/abs/1302.3631}{{\tt arXiv:1302.3631}}].

\bibitem{Marinacci_2014}
F.~{Marinacci}, R.~{Pakmor}, V.~{Springel}, and C.~M. {Simpson}, {\it {Diffuse
  gas properties and stellar metallicities in cosmological simulations of disc
  galaxy formation}},  {\em ArXiv e-prints} (Mar., 2014)
  [\href{http://xxx.lanl.gov/abs/1403.4934}{{\tt arXiv:1403.4934}}].

\bibitem{Bird_2014}
S.~{Bird}, M.~{Vogelsberger}, M.~{Haehnelt}, D.~{Sijacki}, S.~{Genel},
  P.~{Torrey}, V.~{Springel}, and L.~{Hernquist}, {\it {Damped Lyman-alpha
  absorbers as a probe of stellar feedback}},  {\em ArXiv e-prints} (May, 2014)
  [\href{http://xxx.lanl.gov/abs/1405.3994}{{\tt arXiv:1405.3994}}].

\bibitem{Font_2012}
A.~{Font-Ribera}, J.~{Miralda-Escud{\'e}}, E.~{Arnau}, B.~{Carithers}, K.-G.
  {Lee}, P.~{Noterdaeme}, I.~{P{\^a}ris}, P.~{Petitjean}, J.~{Rich},
  E.~{Rollinde}, N.~P. {Ross}, D.~P. {Schneider}, M.~{White}, and D.~G. {York},
  {\it {The large-scale cross-correlation of Damped Lyman alpha systems with
  the Lyman alpha forest: first measurements from BOSS}},  {\em \jcap} {\bf 11}
  (Nov., 2012) 59, [\href{http://xxx.lanl.gov/abs/1209.4596}{{\tt
  arXiv:1209.4596}}].

\bibitem{Meyer_2004}
M.~J. {Meyer}, M.~A. {Zwaan}, R.~L. {Webster}, L.~{Staveley-Smith},
  E.~{Ryan-Weber}, M.~J. {Drinkwater}, D.~G. {Barnes}, M.~{Howlett}, V.~A.
  {Kilborn}, J.~{Stevens}, M.~{Waugh}, M.~J. {Pierce}, R.~{Bhathal}, W.~J.~G.
  {de Blok}, M.~J. {Disney}, R.~D. {Ekers}, K.~C. {Freeman}, D.~A. {Garcia},
  B.~K. {Gibson}, J.~{Harnett}, P.~A. {Henning}, H.~{Jerjen}, M.~J. {Kesteven},
  P.~M. {Knezek}, B.~S. {Koribalski}, S.~{Mader}, M.~{Marquarding}, R.~F.
  {Minchin}, J.~{O'Brien}, T.~{Oosterloo}, R.~M. {Price}, M.~E. {Putman}, S.~D.
  {Ryder}, E.~M. {Sadler}, I.~M. {Stewart}, F.~{Stootman}, and A.~E. {Wright},
  {\it {The HIPASS catalogue - I. Data presentation}},  {\em \mnras} {\bf 350}
  (June, 2004) 1195--1209,
  [\href{http://xxx.lanl.gov/abs/astro-ph/0406384}{{\tt astro-ph/0406384}}].

\bibitem{Giovanelli_2005}
R.~{Giovanelli}, M.~P. {Haynes}, B.~R. {Kent}, P.~{Perillat}, A.~{Saintonge},
  N.~{Brosch}, B.~{Catinella}, G.~L. {Hoffman}, S.~{Stierwalt}, K.~{Spekkens},
  M.~S. {Lerner}, K.~L. {Masters}, E.~{Momjian}, J.~L. {Rosenberg}, C.~M.
  {Springob}, A.~{Boselli}, V.~{Charmandaris}, J.~K. {Darling}, J.~{Davies},
  D.~{Garcia Lambas}, G.~{Gavazzi}, C.~{Giovanardi}, E.~{Hardy}, L.~K. {Hunt},
  A.~{Iovino}, I.~D. {Karachentsev}, V.~E. {Karachentseva}, R.~A. {Koopmann},
  C.~{Marinoni}, R.~{Minchin}, E.~{Muller}, M.~{Putman}, C.~{Pantoja}, J.~J.
  {Salzer}, M.~{Scodeggio}, E.~{Skillman}, J.~M. {Solanes}, C.~{Valotto},
  W.~{van Driel}, and L.~{van Zee}, {\it {The Arecibo Legacy Fast ALFA Survey.
  I. Science Goals, Survey Design, and Strategy}},  {\em \aj} {\bf 130} (Dec.,
  2005) 2598--2612, [\href{http://xxx.lanl.gov/abs/astro-ph/0508301}{{\tt
  astro-ph/0508301}}].

\bibitem{Catinella_2010}
B.~{Catinella}, D.~{Schiminovich}, G.~{Kauffmann}, S.~{Fabello}, J.~{Wang},
  C.~{Hummels}, J.~{Lemonias}, S.~M. {Moran}, R.~{Wu}, R.~{Giovanelli}, M.~P.
  {Haynes}, T.~M. {Heckman}, A.~R. {Basu-Zych}, M.~R. {Blanton},
  J.~{Brinchmann}, T.~{Budav{\'a}ri}, T.~{Gon{\c c}alves}, B.~D. {Johnson},
  R.~C. {Kennicutt}, B.~F. {Madore}, C.~D. {Martin}, M.~R. {Rich}, L.~J.
  {Tacconi}, D.~A. {Thilker}, V.~{Wild}, and T.~K. {Wyder}, {\it {The GALEX
  Arecibo SDSS Survey - I. Gas fraction scaling relations of massive galaxies
  and first data release}},  {\em \mnras} {\bf 403} (Apr., 2010) 683--708,
  [\href{http://xxx.lanl.gov/abs/0912.1610}{{\tt arXiv:0912.1610}}].

\bibitem{Springel_2005}
V.~{Springel}, {\it {The cosmological simulation code GADGET-2}},  {\em \mnras}
  {\bf 364} (Dec., 2005) 1105--1134,
  [\href{http://xxx.lanl.gov/abs/astro-ph/}{{\tt astro-ph/}}].

\bibitem{CAMB}
A.~{Lewis}, A.~{Challinor}, and A.~{Lasenby}, {\it {Efficient Computation of
  Cosmic Microwave Background Anisotropies in Closed Friedmann-Robertson-Walker
  Models}},  {\em \apj} {\bf 538} (Aug., 2000) 473--476,
  [\href{http://xxx.lanl.gov/abs/astro-ph/}{{\tt astro-ph/}}].

\bibitem{Planck_2013}
{Planck Collaboration}, P.~A.~R. {Ade}, N.~{Aghanim}, C.~{Armitage-Caplan},
  M.~{Arnaud}, M.~{Ashdown}, F.~{Atrio-Barandela}, J.~{Aumont},
  C.~{Baccigalupi}, A.~J. {Banday}, and et~al., {\it {Planck 2013 results. XVI.
  Cosmological parameters}},  {\em ArXiv e-prints} (Mar., 2013)
  [\href{http://xxx.lanl.gov/abs/1303.5076}{{\tt arXiv:1303.5076}}].

\bibitem{Springel-Hernquist_2003}
V.~{Springel} and L.~{Hernquist}, {\it {Cosmological smoothed particle
  hydrodynamics simulations: a hybrid multiphase model for star formation}},
  {\em \mnras} {\bf 339} (Feb., 2003) 289--311,
  [\href{http://xxx.lanl.gov/abs/astro-ph/0206393}{{\tt astro-ph/0206393}}].

\bibitem{viel13}
M.~{Viel}, G.~D. {Becker}, J.~S. {Bolton}, and M.~G. {Haehnelt}, {\it {Warm
  dark matter as a solution to the small scale crisis: New constraints from
  high redshift Lyman-{$\alpha$} forest data}},  {\em \prd} {\bf 88} (Aug.,
  2013) 043502, [\href{http://xxx.lanl.gov/abs/1306.2314}{{\tt
  arXiv:1306.2314}}].

\bibitem{Barai_2013}
P.~{Barai}, M.~{Viel}, S.~{Borgani}, E.~{Tescari}, L.~{Tornatore}, K.~{Dolag},
  M.~{Killedar}, P.~{Monaco}, V.~{D'Odorico}, and S.~{Cristiani}, {\it
  {Galactic winds in cosmological simulations of the circumgalactic medium}},
  {\em \mnras} {\bf 430} (Apr., 2013) 3213--3234,
  [\href{http://xxx.lanl.gov/abs/1210.3582}{{\tt arXiv:1210.3582}}].

\bibitem{FoF}
M.~{Davis}, G.~{Efstathiou}, C.~S. {Frenk}, and S.~D.~M. {White}, {\it {The
  evolution of large-scale structure in a universe dominated by cold dark
  matter}},  {\em \apj} {\bf 292} (May, 1985) 371--394.

\bibitem{Katz_1996}
N.~{Katz}, D.~H. {Weinberg}, and L.~{Hernquist}, {\it {Cosmological Simulations
  with TreeSPH}},  {\em \apjs} {\bf 105} (July, 1996) 19,
  [\href{http://xxx.lanl.gov/abs/astro-ph/9509107}{{\tt astro-ph/9509107}}].

\bibitem{Sheth-Tormen}
R.~K. {Sheth} and G.~{Tormen}, {\it {An excursion set model of hierarchical
  clustering: ellipsoidal collapse and the moving barrier}},  {\em \mnras} {\bf
  329} (Jan., 2002) 61--75, [\href{http://xxx.lanl.gov/abs/astro-ph/}{{\tt
  astro-ph/}}].

\bibitem{Weiguang_2014}
W.~{Cui}, S.~{Borgani}, and G.~{Murante}, {\it {The effect of AGN feedback on
  the halo mass function}},  {\em ArXiv e-prints} (Feb., 2014)
  [\href{http://xxx.lanl.gov/abs/1402.1493}{{\tt arXiv:1402.1493}}].

\bibitem{SMT}
R.~K. {Sheth}, H.~J. {Mo}, and G.~{Tormen}, {\it {Ellipsoidal collapse and an
  improved model for the number and spatial distribution of dark matter
  haloes}},  {\em \mnras} {\bf 323} (May, 2001) 1--12,
  [\href{http://xxx.lanl.gov/abs/astro-ph/9907024}{{\tt astro-ph/9907024}}].

\bibitem{Guha_2012}
T.~{Guha Sarkar}, S.~{Mitra}, S.~{Majumdar}, and T.~R. {Choudhury}, {\it
  {Constraining large-scale H I bias using redshifted 21-cm signal from the
  post-reionization epoch}},  {\em \mnras} {\bf 421} (Apr., 2012) 3570--3578,
  [\href{http://xxx.lanl.gov/abs/1109.5552}{{\tt arXiv:1109.5552}}].

\bibitem{Barnes_2010}
L.~A. {Barnes} and M.~G. {Haehnelt}, {\it {Faint extended Ly{$\alpha$} emission
  due to star formation at the centre of high column density QSO absorption
  systems}},  {\em \mnras} {\bf 403} (Apr., 2010) 870--885,
  [\href{http://xxx.lanl.gov/abs/0912.1345}{{\tt arXiv:0912.1345}}].

\bibitem{Barnes_2014}
L.~A. {Barnes} and M.~G. {Haehnelt}, {\it {The bias of DLAs at z \~{} 2.3:
  Evidence for very strong stellar feedback in shallow potential wells}},  {\em
  ArXiv e-prints} (Mar., 2014) [\href{http://xxx.lanl.gov/abs/1403.1873}{{\tt
  arXiv:1403.1873}}].

\bibitem{Bullock_2001}
J.~S. {Bullock}, T.~S. {Kolatt}, Y.~{Sigad}, R.~S. {Somerville}, A.~V.
  {Kravtsov}, A.~A. {Klypin}, J.~R. {Primack}, and A.~{Dekel}, {\it {Profiles
  of dark haloes: evolution, scatter and environment}},  {\em \mnras} {\bf 321}
  (Mar., 2001) 559--575, [\href{http://xxx.lanl.gov/abs/astro-ph/9908159}{{\tt
  astro-ph/9908159}}].

\bibitem{Maccio_2007}
A.~V. {Macci{\`o}}, A.~A. {Dutton}, F.~C. {van den Bosch}, B.~{Moore},
  D.~{Potter}, and J.~{Stadel}, {\it {Concentration, spin and shape of dark
  matter haloes: scatter and the dependence on mass and environment}},  {\em
  \mnras} {\bf 378} (June, 2007) 55--71,
  [\href{http://xxx.lanl.gov/abs/astro-ph/0608157}{{\tt astro-ph/0608157}}].

\bibitem{Leroy_2008}
A.~K. {Leroy}, F.~{Walter}, E.~{Brinks}, F.~{Bigiel}, W.~J.~G. {de Blok},
  B.~{Madore}, and M.~D. {Thornley}, {\it {The Star Formation Efficiency in
  Nearby Galaxies: Measuring Where Gas Forms Stars Effectively}},  {\em \aj}
  {\bf 136} (Dec., 2008) 2782--2845,
  [\href{http://xxx.lanl.gov/abs/0810.2556}{{\tt arXiv:0810.2556}}].

\bibitem{Becker_2013}
G.~D. {Becker}, P.~C. {Hewett}, G.~{Worseck}, and J.~X. {Prochaska}, {\it {A
  refined measurement of the mean transmitted flux in the Ly{$\alpha$} forest
  over 2 {\textless} z {\textless} 5 using composite quasar spectra}},  {\em
  \mnras} {\bf 430} (Apr., 2013) 2067--2081,
  [\href{http://xxx.lanl.gov/abs/1208.2584}{{\tt arXiv:1208.2584}}].

\bibitem{DeLucia_2008}
G.~{De Lucia} and A.~{Helmi}, {\it {The Galaxy and its stellar halo: insights
  on their formation from a hybrid cosmological approach}},  {\em \mnras} {\bf
  391} (Nov., 2008) 14--31, [\href{http://xxx.lanl.gov/abs/0804.2465}{{\tt
  arXiv:0804.2465}}].

\bibitem{Guo_2011}
Q.~{Guo}, S.~{White}, M.~{Boylan-Kolchin}, G.~{De Lucia}, G.~{Kauffmann},
  G.~{Lemson}, C.~{Li}, V.~{Springel}, and S.~{Weinmann}, {\it {From dwarf
  spheroidals to cD galaxies: simulating the galaxy population in a
  {$\Lambda$}CDM cosmology}},  {\em \mnras} {\bf 413} (May, 2011) 101--131,
  [\href{http://xxx.lanl.gov/abs/1006.0106}{{\tt arXiv:1006.0106}}].

\bibitem{Kim_2013}
T.-S. {Kim}, A.~M. {Partl}, R.~F. {Carswell}, and V.~{M{\"u}ller}, {\it {The
  evolution of H I and C IV quasar absorption line systems at 1.9 {\textless} z
  {\textless} 3.2}},  {\em \aap} {\bf 552} (Apr., 2013) A77,
  [\href{http://xxx.lanl.gov/abs/1302.6622}{{\tt arXiv:1302.6622}}].

\bibitem{Prochaska_2010}
J.~X. {Prochaska}, J.~M. {O'Meara}, and G.~{Worseck}, {\it {A Definitive Survey
  for Lyman Limit Systems at z \~{} 3.5 with the Sloan Digital Sky Survey}},
  {\em \apj} {\bf 718} (July, 2010) 392--416,
  [\href{http://xxx.lanl.gov/abs/0912.0292}{{\tt arXiv:0912.0292}}].

\bibitem{Carswell_1987}
R.~F. {Carswell}, J.~K. {Webb}, J.~A. {Baldwin}, and B.~{Atwood}, {\it
  {High-redshift QSO absorbing clouds and the background ionizing source}},
  {\em \apj} {\bf 319} (Aug., 1987) 709--722.

\bibitem{Mao_2012}
Y.~{Mao}, P.~R. {Shapiro}, G.~{Mellema}, I.~T. {Iliev}, J.~{Koda}, and
  K.~{Ahn}, {\it {Redshift-space distortion of the 21-cm background from the
  epoch of reionization - I. Methodology re-examined}},  {\em \mnras} {\bf 422}
  (May, 2012) 926--954, [\href{http://xxx.lanl.gov/abs/1104.2094}{{\tt
  arXiv:1104.2094}}].

\bibitem{geil}
P.~M. {Geil}, B.~M. {Gaensler}, and J.~S.~B. {Wyithe}, {\it {Polarized
  foreground removal at low radio frequencies using rotation measure synthesis:
  uncovering the signature of hydrogen reionization}},  {\em \mnras} {\bf 418}
  (Nov., 2011) 516--535, [\href{http://xxx.lanl.gov/abs/1011.2321}{{\tt
  arXiv:1011.2321}}].

\bibitem{mcquinn}
M.~{McQuinn}, O.~{Zahn}, M.~{Zaldarriaga}, L.~{Hernquist}, and S.~R.
  {Furlanetto}, {\it {Cosmological Parameter Estimation Using 21 cm Radiation
  from the Epoch of Reionization}},  {\em \apj} {\bf 653} (Dec., 2006)
  815--834, [\href{http://xxx.lanl.gov/abs/astro-ph/0512263}{{\tt
  astro-ph/0512263}}].

\bibitem{datta}
K.~K. {Datta}, S.~{Bharadwaj}, and T.~R. {Choudhury}, {\it {Detecting ionized
  bubbles in redshifted 21-cm maps}},  {\em \mnras} {\bf 382} (Dec., 2007)
  809--818, [\href{http://xxx.lanl.gov/abs/astro-ph/0703677}{{\tt
  astro-ph/0703677}}].

\bibitem{petrovic}
N.~{Petrovic} and S.~P. {Oh}, {\it {Systematic effects of foreground removal in
  21-cm surveys of reionization}},  {\em \mnras} {\bf 413} (May, 2011)
  2103--2120, [\href{http://xxx.lanl.gov/abs/1010.4109}{{\tt
  arXiv:1010.4109}}].

\bibitem{jensen13}
H.~{Jensen}, K.~K. {Datta}, G.~{Mellema}, E.~{Chapman}, F.~B. {Abdalla}, I.~T.
  {Iliev}, Y.~{Mao}, M.~G. {Santos}, P.~R. {Shapiro}, S.~{Zaroubi},
  G.~{Bernardi}, M.~A. {Brentjens}, A.~G. {de Bruyn}, B.~{Ciardi}, G.~J.~A.
  {Harker}, V.~{Jeli{\'c}}, S.~{Kazemi}, L.~V.~E. {Koopmans}, P.~{Labropoulos},
  O.~{Martinez}, A.~R. {Offringa}, V.~N. {Pandey}, J.~{Schaye}, R.~M. {Thomas},
  V.~{Veligatla}, H.~{Vedantham}, and S.~{Yatawatta}, {\it {Probing
  reionization with LOFAR using 21-cm redshift space distortions}},  {\em
  \mnras} {\bf 435} (Oct., 2013) 460--474,
  [\href{http://xxx.lanl.gov/abs/1303.5627}{{\tt arXiv:1303.5627}}].

\bibitem{Jing_2005}
Y.~P. {Jing}, {\it {Correcting for the Alias Effect When Measuring the Power
  Spectrum Using a Fast Fourier Transform}},  {\em \apj} {\bf 620} (Feb., 2005)
  559--563, [\href{http://xxx.lanl.gov/abs/astro-ph/}{{\tt astro-ph/}}].

\end{thebibliography}\endgroup

\end{document}